\documentclass[preprint,preprintnumbers,amsmath,amssymb,superscriptaddress,nofootinbib]{revtex4}

\usepackage{graphicx,color}
\usepackage{amsmath,amssymb}
\usepackage{url}
\usepackage{epstopdf}

\newcommand{\hs}{\hspace*{0.5cm}}

\newcommand{\eq}[1]{Eq.~(\ref{#1})}

\newcommand{\be}{\begin{equation}}
\newcommand{\ee}{\end{equation}}
\newcommand{\bea}{\begin{eqnarray}}
\newcommand{\eea}{\end{eqnarray}}
\newcommand{\nn}{\nonumber}
\newcommand{\crn}{\nonumber \\}

\newcommand{\fr}{\frac}
\newcommand{\bc}{\begin{center}}
\newcommand{\ec}{\end{center}}

\newcommand {\ba}{\begin{array}}
\newcommand {\ea}{\end{array}}
\newcommand{\ben}{\begin{enumerate}}
\newcommand{\een}{\end{enumerate}}

\usepackage{epsfig,graphicx}
\usepackage{bm}
\usepackage{dcolumn}

\begin{document}

\title{Neutral Higgs  decays $H \rightarrow Z \gamma,\gamma\gamma$ in 3-3-1 models}

\author{H.~T.~Hung}\email{hathanhhung@hpu2.edu.vn}
\affiliation{Department of Physics, Hanoi Pedagogical University 2, Phuc Yen,  Vinh Phuc 280000, Vietnam}
\author{T.~T.~Hong}\email{tthong@agu.edu.vn}
\affiliation{Department of Physics, An Giang University, Vietnam National University HCMC, 	Ung Van Khiem Street, Long Xuyen,  An Giang 880000, Vietnam}
\affiliation{Department of Physics, Hanoi Pedagogical University 2, Phuc Yen,  Vinh Phuc 280000, Vietnam}
\author{H.~H.~Phuong}\email{trongnghia.hd@gmail.com}
\affiliation{Department of Physics, Hanoi Pedagogical University 2, Phuc Yen,  Vinh Phuc 280000, Vietnam}

\author{H.~L.~T.~Mai}\email{huynhmaict1509@gmail.com }
\affiliation{Faculty of Physics Science, Can Tho Medical College, Nguyen Van Cu Street, Can Tho 900000, Vietnam}
\author{L.~T.~Hue\footnote{Corresponding author}} \email{lethohue@duytan.edu.vn}
\affiliation{Institute of Research and Development, Duy Tan University,  Da Nang 550000, Vietnam}
\affiliation{Institute of Physics, Vietnam Academy of Science and Technology, 10 Dao Tan, Ba
	Dinh, Hanoi 100000, Vietnam} 

\begin{abstract}
The significance of new physics appearing in the loop-induced decays of neutral Higgs bosons into pairs of dibosons $\gamma\gamma$ and $Z\gamma$ will be discussed in the framework of the 3-3-1 models based on a recent work~\cite{Okada:2016whh}, where the Higgs sector  becomes effectively the same as that in the two Higgs doublet models (2HDM) after the first symmetry breaking from $SU(3)_L$ scale into the electroweak scale. For large $SU(3)_L$ scale $v_3\simeq10$ TeV,  dominant one-loop  contributions to the two decay amplitudes  arise from only the single charged Higgs boson predicted by the 2HDM, leading to that experimental constraint on the signal strength $\mu^{331}_{\gamma\gamma}$ of the Standard Model-like  Higgs boson decay $h\rightarrow \gamma\gamma$ will result in a strict upper bound on the signal strength $\mu^{331}_{Z\gamma}$ of the decay $h\rightarrow\, Z\gamma$. For a particular model with lower  $v_3$ around 3 TeV, contributions  from heavy charged gauge and Higgs bosons may have the same order, therefore may give strong destructive or constructive correlations.  As a by-product,  a deviation from the SM prediction $|\mu^{331}_{\gamma\gamma}-1| \le 0.04$ still allows $|\mu^{331}_{Z\gamma}-1|$ to reach values near 0.1.  We also show that there exists an $CP$-even neutral Higgs boson $h^0_3$ predicted by the 3-3-1 models, but beyond the 2HDM, has an  interesting property that the branching ratio Br$(h^0_3\rightarrow \gamma\gamma)$ is very sensitive to the parameter $\beta$ used to  distinguish different 3-3-1 models.
	
\end{abstract}
\pacs{ 
 }

\maketitle
\allowdisplaybreaks
\section{Introduction}

One of the most important channels confirming the existence of the  Standard Model-like (SM-like) Higgs boson is the loop-induced decay channel  $h\rightarrow \gamma\gamma$. Experimentally,  the respective signal strength $\mu_{\gamma\gamma}\equiv (\sigma. \mathrm{Br})_{\mathrm{obs}}/(\sigma. \mathrm{Br})_{\mathrm{SM}}$,  which is the observed product of the Higgs boson production cross section ($\sigma$) and its branching ratio (Br) in units of the corresponding values predicted by the standard model (SM)~\cite{Tanabashi:2018oca}, has been updated  recently by  ATLAS and CMS  \cite{Aaboud:2018ezd, Sirunyan:2018ouh, Aaboud:2018xdt}.  There is another loop-induced decay $h\rightarrow\,Z\gamma$, which the branching ratio (Br) predicted by the SM is $\mathrm{Br}(h\rightarrow\,Z\gamma)=1.54\times 10^{-3}$  $(\pm 5.7\%)$ corresponding to the Higgs boson mass $m_h=125.09$ GeV~\cite{Heinemeyer:2013tqa, deFlorian:2016spz}. This decay channel has not been observed experimentally. The recent upper constraints of the signal strength are  $\mu_{Z\gamma}<6.6$ and  $\mu_{Z\gamma}<3.9$ from ATLAS and CMS ~\cite{Aaboud:2017uhw,Sirunyan:2018tbk}, respectively. In the future project from LHC with its High Luminosity (HL-LHC) and High Energy (HE-LHC),  precision measurements for  the signal  strengths of the two decays $h\rightarrow Z\gamma$ and $h\rightarrow \gamma\gamma$  can reach the respective values of $\mu_{Z\gamma}=1\pm0.23$ and $\mu_{\gamma \gamma}=1\pm 0.04$ for both ATLAS and CMS~\cite{Cepeda:2019klc}. In addition, the ATLAS expected significance  to the $h\rightarrow\, Z\gamma $ channel is hoped to be $4.9~\sigma$ with $3000~\mathrm{fb}^{-1}$.

In theoretical side,  the loop-induced decays of the SM-like Higgs boson mentioned above are important for searching as well as constraining new physics predicted by recent SM extensions, constructed to explain various current experimental data beyond the SM predictions. In the SM, leading contributions to the amplitudes of both decays $h\rightarrow\gamma\gamma,Z\gamma$  are at the one-loop level and relate  with $W$ and fermion mediation.  On the other hand, SM extensions usually contain  new charged particles including scalar, fermions, and gauge bosons spin 1. If any of them couple with the SM-like Higgs boson, they  will contribute to the decay amplitude $h\rightarrow\gamma\gamma$  from the one-loop level. Normally, these particles  also couple with the SM gauge boson $Z$,  hence give one-loop contributions  to the decay amplitude $h\rightarrow\,Z\gamma$ too.  It seems that the Br of the two  decays $h\rightarrow\gamma\gamma, Z\gamma$ have certain relations so that the recent experimental constraint of $\mu_{\gamma\gamma}$ may result in  a respective constraint on $\mu_{Z\gamma}$.  

The theoretical studies of loop effects caused by new particles  on the SM-like Higgs decays including $h\rightarrow \gamma\gamma,Z\gamma$ have been done recently in many SM extensions such as 2HDM \cite{Fontes:2014xva,Kanemura:2018yai,Bhattacharyya:2013rya,Bhattacharyya:2014oka}, where a  thorough investigation in Ref.~\cite{Fontes:2014xva} concerned  strong correlations between two signal  strengths $\mu_{\gamma\gamma, Z\gamma}$. Hence, the experimental data of  $\mu_{\gamma\gamma}$ can be used as an efficient way to predict theoretically constraints on the $\mu_{Z\gamma}$. In  left-right models, the decay channel $h\rightarrow\gamma\gamma$ can be used  to constrain new heavy charged gauge boson masses~\cite{Bandyopadhyay:2019jzq}. On the other hand, it seems that the  result one-loop contribution to  the decay amplitude  $h\rightarrow Z\gamma$~\cite{Martinez:1989kr,Maiezza:2016bzp} has not been discussed for further studying this decay properties  using the latest experimental data of the SM-like Higgs boson such as  the mass and  decay $h\rightarrow \gamma\gamma$. In a recent  scotogenic model, new singly and doubly charged Higgs bosons contribute to both  loop-induced decay amplitudes $h\rightarrow\gamma\gamma,Z\gamma$~\cite{Chen:2019okl}.   But in this framework, the recent experimental data of the decay  $h\rightarrow\gamma\gamma$ predicts a very small a very small deviation from the SM $|\mu_{Z\gamma}-1|<4\%$. In  Higgs triplet models \cite{Blunier:2016peh},  the situation is the same where it was pointed out that  Br$(h\rightarrow Z\gamma)$ is usually smaller than Br$(h\rightarrow\gamma\gamma)$. Tiny values of  $|\mu_{Z\gamma}-1|$ have been shown recently in other  Higgs extensions of the SM~\cite{Kanemura:2018esc}.

In this work, we will focus on  another class of the SM extensions, called the 3-3-1 models, which are constructed from the gauge group $\mathbf{SU(3)_C}\times SU(3)_L\times U(1)_X$~\cite{Singer:1980sw, Valle:1983dk,Pisano:1991ee,Frampton:1992wt, Foot:1992rh,Foot:1994ym}.   These models have many interesting features which cannot be explained in the SM framework, for example they can give explanations  of the existence of three fermion families, the electric charge quantization \cite{deSousaPires:1998jc}, the sources of  CP violations \cite{Montero:1998yw,Montero:2005yb}, the strong $CP$-problem \cite{Pal:1994ba,Dias:2002gg,Dias:2003zt,Dias:2003iq}. In general, one of the most important parameters to distinguish different 3-3-1 models is denoted as $\beta$, which defines electric charges of new particles through the following electric charge operator,
\bea
Q = T_3 + \beta T_8 + X,
\label{eq:charge_Q}
\eea
where    $T_3$ and  $T_8$ are two diagonal generators of the  $SU(3)$ group, $X$ is the $U(1)$ charge. Apart from the popular 3-3-1 models with values of $\beta=\pm \frac{1}{\sqrt{3}},\pm\sqrt{3}$, other models with $\beta=0,\pm \frac{2}{\sqrt{3}},\;\frac{1}{3\sqrt{3}}$ have been discussed phenomenologically~\cite{Hue:2015mna,Buras:2016dxz,RamirezBarreto:2019bpx}.  Different phenomenological  aspects in  models with arbitrary $\beta$ were also discussed~\cite{CarcamoHernandez:2005ka,Cao:2016uur,Yue:2013qba,Buras:2014yna,Buras:2013dea,Martinez:2014lta,Hue:2017lak,Long:2018fud}. As we will see, the model contains nine electroweak gauge bosons, four of them are identified as the SM-like particles. The remaining include one heavy neutral gauge boson $Z'$ and the two pairs of heavy gauge bosons with electric charges depending on Eq.~\eqref{eq:charge_Q}, see a detailed pedagogical calculation  in Ref.~\cite{Buras:2012dp}. As usual, all particles get masses from three Higgs $SU(3)_L$ triplets, including a neutral $CP$-even Higgs component with a large expectation vacuum value (vev) $v_3$ that generates  masses for heavy $SU(3)_L$ particles. The three  Higgs triplets also contain new charged Higgs bosons that may contribute to the amplitudes of the  loop-induced decays of neutral Higgs bosons, including the SM-like one.  Correlations among these  Higgs and gauge contributions  will predict the allowed regions of the parameter space satisfying the current experimental data of $h\rightarrow\gamma\gamma$. It is interesting to estimate how large of the allowed values of  $\mu_{Z\gamma}$ can be. 

 The decay $h\rightarrow \gamma\gamma$ was mentioned in some particular 3-3-1 models for constraining the parameter space~\cite{Caetano:2013nya,CarcamoHernandez:2019vih}. Both $h\rightarrow \gamma\gamma, Z\gamma$ were also mentioned previously in the 3-3-1 models~\cite{Cao:2016uur,Yue:2013qba}, but some nontrivial contributions to the amplitude of the decay $h\rightarrow \,Z\gamma$ were not included. In this work, we will study effects of heavy particles predicted by the 3-3-1 models on  the  two signal strengths of the two decays of the SM-like Higgs bosons $h\rightarrow\gamma\gamma,Z\gamma$,  using  more general analytic formulas of one-loop contributions to the decay amplitude  $h\rightarrow\,Z\gamma$ introduced recently \cite{Degrande:2017naf,Hue:2017cph}.  For simplicity in calculating the physical states of  the neutral $CP$-even Higgs bosons, the Higgs potential of the 3-3-1 models will be considered as an effective 2HDM after the first breaking step $SU(3)_L\times U(1)_X\rightarrow SU(2)_L\times U(1)_Y$. This form of the Higgs potential was mentioned in detail in Ref.~\cite{Okada:2016whh} for studying a 3-3-1 model with $\beta=-1/\sqrt{3}$. This property of the  3-3-1 models was mentioned  previously~\cite{Martinez:2014lta}.  The Higgs potential in this limit can be applied to a general 3-3-1 model keeping $\beta$ as a free parameter ($331\beta$).  This can be seen by the fact that the model contains two $SU(3)_L$ Higgs triplets having components the same as those appear in the 2HDM.  The physical states of neutral Higgs bosons then can be determined exactly at the tree level.  Recent theoretical constraints  on the Higgs sector of 2HDM~\cite{Haisch:2018djm} can be used to  constrain the allowed regions of the parameter space relating with those included in the $331\beta$.  

 On the other hand,  the 331$\beta$ contains another heavy  neutral Higgs boson $h^0_3$ that does not couple with  SM particles, except the SM-like Higgs boson. Hence, if it is the lightest among those beyond the SM particles, its main decay channels are the tree level decay into a pair of SM-like Higgs boson and loop-induced decays to pairs of gluons and gauge bosons $\gamma\gamma,Z\gamma$.  An investigation to determine which decay channels can be used to distinguish different 3-3-1 models will also be presented. 

Our work is arranged as follows. Section~\ref{models} summarizes contents of the 3-3-1 models investigated in this work. All couplings and analytic formulas needed for calculating the Brs and signal strengths of the $h,h^0_3\rightarrow\gamma\gamma,Z\gamma$  are presented in Section~\ref{coupling}. Numerical results are shown in Section~\ref{numerical}. Important remarks and inclusions are pointed out in Section~\ref{conclusion}.  Finally, there are three appendices listing more detailed calculations on couplings, particular analytic formulas for one-loop contributions to the decay amplitudes $h,h^0_3\rightarrow \gamma\gamma,\,Z\gamma$, and interesting numerical illustrations.  

\section{3-3-1 model with arbitrary $\beta$}
\label{models}
\subsection{The model review}
In this section, we summarize the particle content of the model $331\beta$. Left-  and right-handed leptons are assigned to $SU(3)_L$ antitriplets and 
singlets, respectively: 
\bea && L'_{aL}=\left(
\begin{array}{c}
	e'_a \\
	-\nu'_{a} \\
	E'_a \\
\end{array}
\right)_L \sim \left(1~, 3^*~, -\frac{1}{2}+\frac{\beta}{2\sqrt{3}}\right), \hs a=1,2,3,\crn
&& e'_{aR}\sim   \left(1~,1~, -1\right)  , \hs \nu'_{aR}\sim  \left(1~,1~, 0\right) ,\hs E'_{aR} \sim   \left(1~,1~, -\frac{1}{2}+\frac{\sqrt{3}\beta}{2}\right),  \label{lep}\eea
where in the parentheses present the representations   and the hypercharge $X$   of the gauge groups   $SU(3)_C$, $SU(3)_L$ and    $U(1)_X$, respectively.  The model includes three right handed (RH) neutrinos $\nu'_{aR}$ and heavy exotic leptons $E'^a_{L,R}$.    

The quark sector is arranged to guarantee anomaly cancellation, namely  
\begin{align}
 Q'_{iL}&=\left(
\begin{array}{c}
u'_i \\
d'_{i} \\
J'_i \\
\end{array}
\right)_L \sim \left(3, ~3~, \frac{1}{6}-\frac{\beta}{2\sqrt{3}}\right), \quad  Q'_{3L}=\left(
\begin{array}{c}
d'_3 \\
-u'_{3} \\
J'_3 \\
\end{array}
\right)_L \sim \left(3,~ 3^*~, \frac{1}{6}+\frac{\beta}{2\sqrt{3}}\right),  \label{eq_qfamilyi}\\
u'_{aR}&\sim  \left(3,~ 1~, \frac{2}{3}\right), \; d'_{aR}\sim   \left(3,~1~, \frac{-1}{3}\right) ,\crn 
  J'_{iR} &\sim   \left(3,~1~, \frac{1}{6}-\frac{\sqrt{3}\beta}{2}\right), \; J'_{3R}\sim   \left(3,~1~, \frac{1}{6}+\frac{\sqrt{3}\beta}{2}\right), \label{eq_qR}
\end{align}
where  $ i=1,2$, $a=1,2,3$, and  $J_{aL,R}$ are exotic quarks predicted by the 331$\beta$ model.  There is another arrangement that the model contains three left-handed lepton $SU(3)_L$ triplets, one quark $SU(3)_L$ triplet and  two other quark $SU(3)_L$ antitriplets. But, it was shown that the two arrangements are equivalent in the sense that they predict the same physics~\cite{Descotes-Genon:2017ptp,Hue:2018dqf}.

To generate masses for gauge bosons and fermions, three scalar triplets are introduced as follows  
\begin{align} \label{higgsc}
\chi &=\left(
\begin{array}{c}
\chi^{+A} \\
\chi^{+B} \\
\chi^0 \\
\end{array}
\right)\sim \left(1,~3~, \frac{\beta}{\sqrt{3}}\right), 
\quad  \rho=\left(
\begin{array}{c}
\rho^+ \\
\rho^0 \\
\rho^{-B} \\
\end{array}
\right)\sim \left( 1,~3~, \frac{1}{2}-\frac{\beta}{2\sqrt{3}}\right),\crn  
\eta&=\left(
\begin{array}{c}
\eta^0 \\
\eta^- \\
\eta^{-A} \\
\end{array}
\right)\sim \left(1,~3~, -\frac{1}{2}-\frac{\beta}{2\sqrt{3}}\right),
\end{align}
where $A,B$ denote electric charges defined in \eq{eq:charge_Q}: $A= \frac{1 +\beta \sqrt{2}}{2} $ and $B=\frac{-1 +\beta \sqrt{2}}{2}$. 
These Higgs bosons develop vevs defined as $\langle  \chi^0\rangle =\frac{v_3}{\sqrt{2}} ,\,    \langle  \rho^0\rangle =\frac{v_2}{\sqrt{2}}, \, \langle  \eta^0\rangle =\frac{v_1}{\sqrt{2}} $, leading to 
\begin{align}\label{vevhigg1}  
 \chi^0&=\frac{v_3 + r_3 +i a_3}{\sqrt{2}} ,\quad   \langle  \rho^0\rangle =\frac{v_2 + r_2 +i a_2}{\sqrt{2}}, \quad \langle  \eta^0\rangle =\frac{v_1+ r_1 +i a_1}{\sqrt{2}}. 
\end{align}
The symmetry breaking happens in two steps:  
$SU(3)_L\otimes U(1)_X\xrightarrow{v_3} SU(2)_L\otimes U(1)_Y\xrightarrow{v_1,v_2} U(1)_Q$.  
It is therefore reasonable to assume that $v_3 > v_1,v_2$.  At the second breaking step, $\rho$ and $\eta$ play roles of the two $SU(2)_L$ doublets similar to 2HDM, except differences in coupling with fermions. Masses and physical states of all particles are summarized as follows. 
\subsection{Fermions}

Masses and physical states of the fermions  are derived from the following Yukawa  Lagrangian  
\begin{align}
\mathcal{L}^Y_{\mathrm{lepton}}&=-Y^e_{ab} \overline{L'}_{aL} \eta^*e'_{bR}- Y^\nu_{ab} \overline{L'}_{aL} \rho^*\nu'_{bR} - Y^E_{ab} \overline{L'}_{aL} \chi^*E'_{bR}+\mathrm{h.c.}, \label{ylepton1}\\
\mathcal{L}^Y_{\mathrm{quark}}&= -Y^d_{ia}\overline{Q'}_{iL}\rho d'_{aR} -Y^d_{3a}\overline{Q'}_{3L}\eta^* d'_{aR} -Y^u_{ia}\overline{Q'}_{iL}\eta u_{aR} -Y^u_{3a}\overline{Q'}_{3L}\rho^* u'_{aR} \crn
&-Y^J_{ij}\overline{Q'}_{iL}\chi J'_{jR} -Y^J_{33}\overline{Q'}_{3L}\chi^* J'_{3R} +\mathrm{h.c.} , \label{quark}
\end{align}
We note that depending on particular values of $\beta$, additional Yukawa terms may appear but  a $Z_2$ symmetry can be imposed to exclude them, see an example given in Ref.~\cite{Okada:2016whh}. 

As mentioned above, the SM-like fermions get masses from their couplings to two Higgs triplets  $\eta$ and $\rho$, similarly to the 2HDM. On the other hand, the up (down) quarks couple to both Higgs triplets, leading to a different feature from four popular  types of 2HDM, where all up (down) quarks couple to the same Higgs doublet in order to avoid tree level flavor  changing neutral currents (FCNCs),  see for example in Ref.~\cite{Craig:2012vn}. As a result, many interesting properties  relating with the SM-like fermion couplings   were pointed out  to distinguish 3-3-1 models and 2HDMs~\cite{Okada:2016whh}.  

The exotic fermions couple to only the Higgs triplet $\chi$. Accordingly,  the neutral Higgs sector in Ref.~\cite{Okada:2016whh} has a property that the $\chi^0$ does not contribute to the SM-like Higgs boson,  it therefore decouples with all exotic fermions.   Hence, they do not contribute to the one-loop decay amplitudes  $h\rightarrow\gamma\gamma,Z\gamma$.

The SM-like  fermion masses are determined based on discussions in  refs.~\cite{Okada:2016whh, Buras:2012dp,Hue:2017lak}, where  the mixing between quarks are safely ignored in this work. Then  all fermion mass matrices are  diagonal. Correspondingly,  the original fermion states are physical, hence  they will be denoted by $e_{aL,R},u_{aL,R}$ and $d_{aL,R}$. The fermion masses are given as follows:
\begin{align} \label{eq_Fermionmass}
m_{e_a}&=\frac{Y^e_{aa}v_1}{\sqrt{2}}, \; m_{u_i}=\frac{Y^u_{ii}v_1}{\sqrt{2}}, \; m_{d_i}=\frac{Y^d_{ii}v_2}{\sqrt{2}}, \; m_{u_3}= -\frac{Y^u_{33}v_2}{\sqrt{2}}, \;  m_{d_3}= \frac{Y^d_{33}v_1}{\sqrt{2}}, \;
m_{F_a}= \frac{Y^F_{aa}v_3}{\sqrt{2}}, 
\end{align}
 where  $Y^{f}_{ab}=0~\forall a\neq b$,  $f=e,u,d,F$ and $F=J,E$.    The relations  \eqref{eq_Fermionmass} will be used to determine Feynman rules of Yukawa couplings in Lagrangians~\eqref{ylepton1} and~\eqref{quark}. 

\subsection{Gauge bosons}

The model contains nine  electroweak (EW) gauge bosons corresponding to the 9 generators of the EW gauge group $SU(3)_L\otimes U(1)_X$.  The  covariant derivative is defined as \footnote{This definition is different from Ref. \cite{Okada:2016whh} by $T^9$.}  \cite{Buras:2012dp,Hue:2017lak,Cao:2016uur},
\bea D_{\mu}\equiv \partial_{\mu}-i g T^a W^a_{\mu}-i g_X X T^9X_{\mu},  \label{coderivative1}\eea
where $T^9=1/\sqrt{6}$, $g$ and $g_X$ are coupling constants of the two groups $SU(3)_L$ and $U(1)_X$, respectively. 
The matrix $W^aT^a$, where $T^a =\lambda_a/2$ corresponding to a triplet representation, is 
\bea W^a_{\mu}T^a=\frac{1}{2}\left(
\begin{array}{ccc}
	W^3_{\mu}+\frac{1}{\sqrt{3}} W^8_{\mu}& \sqrt{2}W^+_{\mu} &  \sqrt{2}Y^{+A}_{\mu} \\
	\sqrt{2}W^-_{\mu} &  -W^3_{\mu}+\frac{1}{\sqrt{3}} W^8_{\mu} & \sqrt{2}V^{+B}_{\mu} \\
	\sqrt{2}Y^{-A}_{\mu}& \sqrt{2}V^{-B}_{\mu} &-\frac{2}{\sqrt{3}} W^8_{\mu}\\
\end{array}
\right),
\label{wata1}\eea 
where we have defined the mass eigenstates of the nondiagonal gauge bosons as
\bea W^{\pm}_{\mu}=\frac{1}{\sqrt{2}}\left( W^1_{\mu}\mp i W^2_{\mu}\right), \; 
Y^{\pm A}_{\mu}=\frac{1}{\sqrt{2}}\left( W^4_{\mu}\mp i W^5_{\mu}\right),\; 
V^{\pm B}_{\mu}=\frac{1}{\sqrt{2}}\left( W^6_{\mu}\mp i W^7_{\mu}\right),
\label{gbos}\eea 
and  $A,B$ are electric charges of the corresponding gauge bosons  calculated based on \eq{eq:charge_Q},
\begin{equation}
A=\fr{1}{2}+\beta\fr{\sqrt{3}}{2}, \quad 
B=-\fr{1}{2}+\beta\fr{\sqrt{3}}{2}\label{charge_AB}.
\end{equation}
We note that $B$ is also the electric charge of the new leptons $E_a$.

The symmetry breaking happens in two steps:  
$SU(3)_L\otimes U(1)_X\xrightarrow{v_3} SU(2)_L\otimes U(1)_Y\xrightarrow{v_1,v_2} U(1)_Q$, corresponding to the following transformation of  the neutral gauge bosons form the original basis to the final physical one:  $ X_{\mu},\, W^3_{\mu}, \, W^8_{\mu} \xrightarrow{\theta_{331}}\, B_{\mu},\, W^3_{\mu}, \, Z'_{\mu}  \xrightarrow{\theta_W}  A_{\mu},\, Z_{\mu}, \, Z'_{\mu} \xrightarrow{\theta}   A_{\mu},\, Z_{1\mu}, \, Z_{2\mu}$.  
After the first step, five gauge bosons will be massive and the remaining four massless gauge bosons 
can be identified with the before-symmetry-breaking SM gauge bosons.  The two physical states $Z_{1,2}$ are mixed from the SM and heavy gauge bosons $Z_{\mu}$ and $Z'_{\mu}$.  

 It is well-known that 
\bea
g_2 = g,\quad g_1 = g_X\fr{g}{\sqrt{6g^2 + \beta^2 g_X^2}},
\label{matching_coupl} 
\eea
where $g_2$ and $g_1$ are the two couplings  of  the of the SM gauge groups $SU(2)_L$ and $U(1)_Y$, respectively.  Using  the weak mixing angle defined as $t_W = \tan\theta_W = g_1/g_2$ and denoting  $s_W = \sin\theta_W$ and $c_W=\cos\theta_W$,  it is derived that   
\bea
\fr{g_X^2}{g^2} = \fr{6s_W^2}{1-(1+\beta^2)s_W^2}= \fr{6s_W^2}{c^2_W(1 -\beta^2 t^2_W)},
\label{eq_beta_coupl}
\eea
which gives a constraint $|\beta| \le \sqrt{3}$ used in the numerical analysis. 

The masses of the  gauge bosons given in~\eqref{gbos} are
\bea 
m^2_Y\equiv m^2_{Y^{\pm A}} = \frac{g^2}{4}(v_3^2+v_1^2),\; m^2_V \equiv\
m^2_{V^{\pm B}}=\frac{g^2}{4}(v_3^2+v_2^2), \; 
m^2_W\equiv m^2_{W^\pm} = \fr{g^2}{4}(v_1^2 + v^2_2).
\label{masga}\eea 
The matching condition with the SM gives $v^2\equiv v_1^2+v_2^2\simeq 246\, [\mathrm{GeV}^2]$. Based on Refs.  \cite{Buras:2012dp,Buras:2014yna},  
the ratios between  vevs are used to define three mixing parameters  as follows 
\begin{align}
 s_{ij}\equiv &  \frac{v_i}{\sqrt{v_i^2+v_j^2}}, \;
c_{ij}\equiv \sqrt{1-s^2_{ij}}, \quad t_{ij} \equiv\tan\beta_{ij}=\frac{s_{ij}}{c_{ij}}, 
 \label{vevang} 
\end{align}
where $i<j$ and $i,j=1,2,3$. 

The model predicts three neutral gauge bosons including the massless photon.    Defining \cite{Buras:2012dp} 
\begin{align} \label{sc331}
s_{331}\equiv \sin\theta_{331}&=\frac{\sqrt{6}g}{\sqrt{6g^2+\beta^2g_X^2}}=\sqrt{1-\beta^2t_W^2}, \; 
c_{331}\equiv \cos\theta_{331} =\beta t_W,
\end{align}
the relation between the original and physical base  of the  neutral gauge bosons are
\begin{align}
\begin{pmatrix}
X_\mu \\
W^3_\mu\\
W^8_\mu
\end{pmatrix} &= \begin{pmatrix}
s_{331} &0 & c_{331}  \\
0 & 1 & 0 \\
c_{331} & 0 & -s_{331}
\end{pmatrix}
\begin{pmatrix}
c_W & -s_W & 0 \\
s_W & c_W& 0 \\
0 & 0 & 1
\end{pmatrix}
\begin{pmatrix}
1 &   0 & 0\\
0 & c_\theta & -s_\theta \\
0 & s_\theta & c_\theta
\end{pmatrix}\begin{pmatrix}
A_\mu \\
Z_{1\mu} \\
Z_{2\mu}
\end{pmatrix}=C\begin{pmatrix}
A_\mu \\
Z_{1\mu} \\
Z_{2\mu}
\end{pmatrix},\crn 
C&=\begin{pmatrix}
s_{331}c_W, &   \left(- s_{331}s_Wc_\theta+c_{331}s_\theta\right),  & \left( s_{331}s_Ws_\theta +c_{331}c_\theta\right) \\
s_W ,& c_Wc_\theta ,& -s_\theta c_w \\
c_{331}c_W, &   -\left( c_{331}s_Wc_\theta+s_{331}s_\theta\right),  &\left(  c_{331}s_Ws_\theta -s_{331}c_\theta\right) 
\end{pmatrix},
 \label{neutralgaugebosonmix}
\end{align}
where in the limit $v^2\ll v_3^2$, the mixing angle $\theta$ is determined as   \cite{Buras:2014yna} 
\begin{align} \label{eq_stheta}
s_{\theta}\equiv \sin \theta &= \left(3\beta \frac{s_W^2}{c_W^2}+\frac{\sqrt{3}(t^2_{21}-1)}{t^2_{21} +1}\right)  \frac{\sqrt{1-\beta^2 t_W^2}v^2}{ 4 c_Wv_3^3},   
\end{align}
and  $M^2_{Z'}=g^2v^2_3/(3s^2_{331}) +\mathcal{O}(v^2)$. 

To continue, the  neutral gauge bosons will be identified as $Z_1\equiv Z$ and $Z_2\equiv Z'$, where $Z$ is the  one found experimentally.

\subsection{Higgs bosons}

The scalar potential is
\bea V_{h}&=&\mu_1^2 \eta^{\dagger}\eta+\mu_2^2\rho^{\dagger}\rho+\mu_3^2\chi^{\dagger}\chi
+\lambda_1 \left(\eta^{\dagger}\eta\right)^2
+\lambda_2\left(\rho^{\dagger}\rho\right)^2
+\lambda_3\left(\chi^{\dagger}\chi\right)^2\crn
&+& \lambda_{12}(\eta^{\dagger}\eta)(\rho^{\dagger}\rho)
+\lambda_{13}(\eta^{\dagger}\eta)(\chi^{\dagger}\chi)
+\lambda_{23}(\rho^{\dagger}\rho)(\chi^{\dagger}\chi)\crn
&+&\tilde{\lambda}_{12} (\eta^{\dagger}\rho)(\rho^{\dagger}\eta) 
+\tilde{\lambda}_{13} (\eta^{\dagger}\chi)(\chi^{\dagger}\eta)
+\tilde{\lambda}_{23} (\rho^{\dagger}\chi)(\chi^{\dagger}\rho)  -\sqrt{2} f\left(\epsilon_{ijk}\eta^i\rho^j\chi^k +\mathrm{h.c.} \right). \label{Vh1}\eea
The minimum conditions of the Higgs potential can be figured out easily \cite{Buras:2012dp,Hue:2017lak}.
After that, we can take $\mu_i^2$ as functions of other independent parameters. These functions  are  inserted into the Higgs potential \eqref{Vh1}, which is used to determine the masses and physical states of all Higgs bosons.

The relations between original and mass eigenstates of charged Higgs bosons are \cite{Buras:2012dp,Hue:2017lak}: 
\begin{align}
\left(
\begin{array}{c}
\phi^{\pm}_W \\
H^{\pm} \\
\end{array}
\right)&=R(\beta_{12})
\left(
\begin{array}{c}
\rho^{\pm} \\
\eta^{\pm} \\
\end{array}
\right), \; m_{\phi_W}=0,\; m^2_{H^{\pm}}= \frac{\tilde{\lambda}_{12} v^2}{2}  + \frac{fv_3}{2s_{12}c_{12}},
\label{scHigg}\\
\left(
\begin{array}{c}
\phi^{ \pm A}_Y \\
H^{\pm A} \\
\end{array}
\right)&= R(\beta_{13})\left(
\begin{array}{c}
\chi^{\pm A} \\
\eta^{\pm A} \\
\end{array}
\right),  \; m_{\phi_Y}=0,\; m^2_{H^{\pm A}}= \left(\frac{\tilde{\lambda}_{13}}{2}  + \frac{ f}{t_{12}v_3} \right) \left( v_1^2  +v_3^2\right), 
\label{qacHigg}\\
\left(
\begin{array}{c}
\phi^{\pm B}_V \\
H^{\pm B} \\
\end{array}
\right)&= R(\beta_{23})\left(
\begin{array}{c}
\chi^{\pm B} \\
\rho^{\pm B} \\
\end{array}
\right),  \; m_{\phi_V}=0,\; m^2_{H^{\pm B}}= \left(\frac{\tilde{\lambda}_{23}}{2}  + \frac{ t_{12}f}{v_3} \right) \left( v_2^2  +v_3^2\right),
\label{qbcHigg}   
\end{align}
where we have define a rotation $R(x)$ as 
\begin{equation}\label{eq_Rx}
R(x)\equiv\left(
\begin{array}{cc}
c_{x} & -s_{x} \\
s_{x} & c_{x} \\
\end{array}
\right).
\end{equation}
The massless states $\phi^{\pm}_W$,   $\phi^{\pm A}_Y$, and  $\phi^{\pm\,B}_V$ are Goldstone bosons absorbed by the physical gauge bosons.

For neutral Higgs bosons, to avoid the tree level contribution of SM-like Higgs bosons to the  flavor  changing neutral currents (FCNC) in the quark sector,   we follow the aligned limit introduced in Ref.~\cite{Okada:2016whh}, namely
\begin{equation}\label{eq_alignH0}
f=\lambda_{13}t_{12}v_3 =\frac{\lambda_{23}v_3}{t_{12}}.
\end{equation}
From this, we will choose $f$ and $\lambda_{23}$ as functions of the remaining,  leading  to the following form of the squared mass matrix corresponding to the basis $(r_1,r_2,r_3)$:
\begin{align}\label{eq_m2r}
M^2_r= \begin{pmatrix}
2\lambda_1 s^2_{12}v^2 +\lambda_{13}v_3^2&  t_{12} \left( \lambda_{12} c_{12}^2v^2 -\lambda_{13}v_3^2 \right)&0  \\ 
 t_{12} \left( \lambda_{12} c_{12}^2v^2 -\lambda_{13}v_3^2 \right) &  2 c^2_{12} \lambda_2 v^2 +t^2_{12} \lambda_{13}v_3^2& 0 \\ 
0&0  &  s_{12}^2 \lambda_{13}  v^2 +2\lambda_3 v_3^2 
\end{pmatrix}.
\end{align}
As a result, $r_3\equiv h^0_3$ is a physical $CP$-even neutral Higgs boson with mass $m^2_{h^0_3}= \lambda_{13}s^2_{12}v^2 +2\lambda_3 v_3^2 $.  The submatrix $2\times 2$ in Eq.~\eqref{eq_m2r} is denoted as $M'^2_r$.  It is diagonalized as follows~\cite{Okada:2016whh},
\begin{align}\label{eq-mixingh0}
R(\alpha)M'^2_rR^T(\alpha)= \mathrm{diag}(m^2_{h^0_1},  m^2_{h^0_2}),
\end{align}
where  
\begin{align}
\alpha&\equiv \beta_{12}  -\frac{\pi}{2} + \delta, \label{eq_alpha} \\
\tan2\delta&= \frac{2M^2_{12}}{M^2_{22} -M^2_{11}} \sim\mathcal{ O}\left(\frac{v^2}{v_3^2}\right), \label{eq_mixingh0}\\
m^2_{h^0_1}&= M^2_{11}\cos^2\delta +M^2_{22}\sin^2\delta - M^2_{12}\sin2\delta,\label{eq_mh01}\\
m^2_{h^0_2}&=  M^2_{11}\sin^2\delta +M^2_{22}\cos^2\delta + M^2_{12}\sin2\delta, \label{eq_mh02}\\
M^2_{11}&= 2\left( s_{12}^4 \lambda_{1} +c_{12}^4 \lambda_{2} +s_{12}^2 c_{12}^2 \lambda_{12}\right) v^2=\mathcal{O}(v^2),  \crn 
M^2_{12}&= \left[- \lambda_1 s^2_{12} + \lambda_2 c^2_{12} +\lambda_{12} (s^2_{12} -c^2_{12})\right]s_{12}c_{12} v^2 =\mathcal{O}(v^2),\crn
M^2_{22}&= 2s^2_{12}c^2_{12}\left[ \lambda_1 +\lambda_2 -\lambda_{12}\right] v^2 +\frac{\lambda_{13}v^2_3}{c^2_{12}}. \nn
\end{align}
We also have 
\begin{align}
\left(
\begin{array}{c}
r_1 \\
r_2\\
\end{array}
\right)&= R^T(\alpha)\left(
\begin{array}{c}
h^0_1 \\
h^0_2 \\
\end{array}
\right). 
\end{align}
To determine the SM-like Higgs boson, we first look at the Eq.~\eqref{eq_mixingh0}, which give $\delta=\mathcal{O}(\frac{v^2}{v_3^2}) \simeq0$ when $v^2\ll v_3^2$. In this limit, $m^2_{h}= M^2_{11} + v^2\times \mathcal{O}(\frac{v^2}{v_3^2})\sim  M^2_{11}$ while $m^2_{h^0_2}= M^2_{22} + v^2 \times  \mathcal{O}(\frac{v^2}{v_3^2}) \simeq M^2_{22}$. Hence, $h^0_1\equiv\, h$ is identified with the SM-like Higgs boson found at LHC. Furthermore, in the following calculation  we will see more explicitly that the couplings of this Higgs boson are the same as those given in the SM in the limit $\delta\rightarrow0$. 

Because the two  mass matrices $M'^2_r$ given in Eqs.~\eqref{eq_m2r} and the one given in Eq.~\eqref{eq-mixingh0} differ from each other by the  unitary transformation $R(\alpha)$,  their  traces  are equal,  namely  Tr$[M'^2_r]=\mathrm{Tr}[R(\alpha)M'^2_rR^T(\alpha)]=m^2_{h^0_1} +m^2_{h^0_2}$.  Accordingly,  $\lambda_{13}$ can be written as
\begin{equation}\label{eq_la13} 
\lambda_{13}=\frac{c^2_{12}}{v_3^2} \left[ m^2_{h^0_1} +m^2_{h^0_2} -2v^2\left( s_{12}^2\lambda_1 +c_{12}^2\lambda_2  \right)\right]. 
\end{equation}
 We will choose $\delta$, $m_{h^0_1}\equiv m_h$ and $m_{h^0_2}$ as input parameters.  The $\lambda_{13}$, $\lambda_{12}$ and $\lambda_{2}$ are dependent parameters, namely 
 \begin{align}
 \label{eq_la122}
 \lambda_{2}&=t^4_{12}\lambda_1 +\frac{ -\left[c^2_{\delta} (t^2_{12}-1) +t_{12} s_{2\delta}\right] m^2_{h} +\left[s^2_{\delta}(1-t^2_{12}) +s_{2\delta} t_{12}\right]m^2_{h^0_2}}{2 c^2_{12} v^2} , \crn 
 \lambda_{12}&=-2t^2_{12} \lambda_1  + \frac{\left( s_{2\delta} +2t_{12}c^2_{\delta} \right) m^2_{h} +\left( -s_{2\delta} +2t_{12}s^2_{\delta} \right)m^2_{h^0_2}}{2 s_{12}c_{12} v^2},
 \end{align}
 and $\lambda_{13}$ was given in Eq.~\eqref{eq_la13}.
 
The  Higgs self-couplings should satisfy all constraints discussed recently to guarantee the vacuum stability of the Higgs potential~\cite{Sanchez-Vega:2018qje}, the perturbative limits, and the positive squared masses of all Higgs bosons. We note that  in the case of absence the relations in Eq.~\eqref{eq_alignH0}, the mixing between SM-like Higgs bosons with other heavy neutral Higgs still suppressed due to large $v_3>5$ TeV enoungh to cancel the FCNCs  in 3-3-1 models~\cite{Huitu:2019kbm}. 

\section{Couplings and analytic formulas  involved with loop-induced Higgs decays}
\label{coupling}
\subsection{Couplings}
From the above discussion on the Higgs potential, we can derive all  Higgs self-couplings of the SM-like Higgs boson relating to the decays $h \rightarrow Z\gamma$ and $h\rightarrow \gamma\gamma$, using the interacting Lagrangian $\mathcal{L}_{hHH}=-V_h$. The Feynman rules are given in Table~\ref{table_h0coupling}, where each factor $-i\lambda_{hss}$ corresponds to a vertex $hss$, where $s=H^{\pm},H^{\pm A}, H^{\pm B}$.
\begin{table}[ht]
	\centering 
	\begin{tabular}{|c|c|}
		\hline 
		Vertex		& Coupling: $-i\lambda_{hss}$ \\ 
		\hline 
		$-i\lambda_{h H^+ H^-}$	& $iv\left[ 2s_{12}c_{12} \left(-\lambda_1 c_{12}\,c_{\alpha} + \lambda_2 s_{12}\,s_{\alpha}\right) + \left(s_{\alpha}c^3_{12} - c_{\alpha}s^3_{12}\right)\lambda_{12} -c_{\delta} \tilde{\lambda}_{12} \right]   $ \\ 
		\hline 
		$-i\lambda_{h H^{A} H^{-A}}$	&  $i c^2_{13}\left\{   v\left[ s_{\alpha}c_{12}\left( \lambda_{12} +t^2_{13}\lambda_{23} \right) -c_{\alpha}s_{12} \left(2 \lambda_1 +t^2_{13}(\lambda_{13} +\tilde{\lambda}_{13}) \right)  \right] + v_3  t_{13} \left( \frac{2fs_{\alpha}}{v_3} -c_{\alpha} \tilde{\lambda}_{13}\right) \right\}$\\
		
		\hline 
		$-i\lambda_{h H^{B} H^{-B}}$	& $i c^2_{23}\left\{ v\left[s_{\alpha}c_{12}\left(2\lambda_2  +t^2_{23}(\lambda_{23} +\tilde{\lambda}_{23})\right) -c_{\alpha}s_{12}\left( \lambda_{12} +t^2_{23} \lambda_{13} \right)  \right] + v_3 t_{23}\left( s_{\alpha} \tilde{\lambda}_{23} -\frac{2fc_{\alpha}}{v_3} \right)\right\} $\\
			\hline 
		\end{tabular} 
	\caption{Feynman rules for  the SM-like  Higgs boson couplings with charged Higgs bosons} \label{table_h0coupling}
\end{table}

Based on the Yukawa Lagrangians \eqref{ylepton1} and \eqref{quark}, the couplings of the SM-like Higgs boson with SM fermions can be determined, see also in table~\ref{table_h0coupling}, where we have used the relation~\eqref{eq_alpha}.  The notation of the Feynman rule is $-i\left(Y_{h\bar{f}fL}P_L + Y_{h\bar{f}fR}P_R\right)$ for each vertex $h\bar{f}f$. For simplicity, the Yukawa couplings of the SM-like fermions in this case were identified with those in the SM, as discussed before. Then, we  have $Y_{\bar{f}fL}=Y_{\bar{f}fR}$, which are given in table~\ref{table_h01ff}. 
\begin{table}[ht]
	\centering 
	\begin{tabular}{cccccc}
		\hline 
	&$-iY_{h\overline{e_a}e_aL,R}$ &   $-iY_{h\overline{u_i}u_iL,R}$ &   $-iY_{h\overline{u_3}u_3L,R}$ & $ -iY_{h\overline{d_i}d_iL,R}$ &  $ -iY_{h\overline{d_3}d_3L,R}$ \\
	\hline 
	& $ -i \frac{m_{e_a}}{v} \left( c_{\delta}  - \frac{s_{\delta}}{t_{12}}\right) $ & $ -i \frac{m_{u_i}}{v} \left( c_{\delta}  - \frac{s_{\delta}}{t_{12}}\right) $ & $ -i \frac{m_{u_3}}{v} \left(c_{\delta} +  t_{12}s_{\delta}\right) $ & $ -i \frac{m_{d_i}}{v} \left(c_{\delta} +  t_{12}s_{\delta}\right) $ & $ -i \frac{m_{d_3}}{v} \left( c_{\delta}  - \frac{s_{\delta}}{t_{12}}\right) $\\
	\hline 
	\end{tabular}
	\caption{Yukawa couplings of the SM-like Higgs boson} \label{table_h01ff}
\end{table}
Both neutral $CP$-even Higgs bosons  $h$  and $h^0_2$ do not couple to exotic fermion in the aligned limit \eqref{eq_alignH0}.  In contrast, $h^0_3$ couples  only to the exotic fermions, while it  does not couple to the SM  ones.  

The couplings of Higgs and gauge bosons are contained in the covariant kinetic terms of the Higgs bosons
\begin{align}\label{eq_lkHiggs}
\mathcal{L}^{H}_{\mathrm{kin}}&=\left(D_{\mu}\chi \right)^{\dagger} \left(D^{\mu}\chi \right) + \left(D_{\mu}\rho \right)^{\dagger} \left(D^{\mu}\rho \right) + \left(D_{\mu}\eta \right)^{\dagger} \left(D^{\mu}\eta \right)\crn
= & \sum_{v}g_{hvv} g_{\mu\nu}hv^{-Q\mu} v^{Q\nu} \crn 
+&\sum_{s,v}\left[ -ig^*_{hsv}v^{-Q\mu}\left( s^{+Q}\partial_{\mu}h -h\partial_{\mu}s^{+Q} \right) + ig_{hsv}v^{Q\mu}\left( s^{-Q}\partial_{\mu}h -h\partial_{\mu}s^{-Q} \right) \right]\crn 
+&\sum_{s}ig_{Zss}Z^{\mu} \left( s^{-Q}\partial_{\mu}s^{Q} -s^{Q}\partial_{\mu}s^{-Q} \right) +\sum_{s,v}\left[ ig_{Zvs}Z^{\mu} v^{Q\nu}s^{-Q} g_{\mu\nu} + ig^*_{Zvs}Z^{\mu} v^{-Q\nu}s^{Q} g_{\mu\nu}\right] \crn
+ &\sum_{s} ie Q A^{\mu}\left( s^{-Q}\partial_{\mu}s^{Q} -s^{Q}\partial_{\mu}s^{-Q} \right) +..., 
\end{align}
 where sums are taken over  $s=H^{\pm},H^{\pm A},H^{\pm B}$ and $v=W,Y,V$. In addition, we only list the relevant terms contributing to the decays $h\rightarrow Z\gamma,\gamma\gamma$  and ignore the remaining terms. The  Feynman  rules for particular couplings in \eqref{eq_lkHiggs} are shown in Table~\ref{table_HGcoupling}, where $\partial_{\mu}h\rightarrow -ip_{0\mu}h$ and $\partial_{\mu}s^{\pm Q}\rightarrow -ip_{\pm\mu}s^{\pm Q}$ and the relation ~\eqref{eq_alpha} was used. The notations $p_0$, $p_{\pm}$  are incoming momenta.
\begin{table}[ht]
	\centering 
	\begin{tabular}{|c|c|c|c|}
		\hline 
	Vertex	& Coupling: & Vertex&Coupling\\ 
	\hline 
$g_{h W^+W^-}$	&$g \,m_W \,c_{\delta}$& $g_{hY^{+A}Y^{-A}}$	& $ g \,m_W \,c_{\alpha}s_{12}$\\
\hline 
 $g_{h V^{+B}V^{-B}}$& $- g \,m_W \,s_{\alpha}c_{12}$ & $g_{h H^-W^+}$ & $ \frac{g\,s_{\delta}}{2} $\\
\hline 
  $g_{h H^{-A}Y^{A}}$&  $-\frac{g\,c_{13}c_{\alpha}}{2} $ & $g_{hH^{-B}V^{B}}$& $ \frac{g\,c_{23}s_{\alpha}}{2} $\\
\hline
	\end{tabular}
	\caption{Feynman rules for couplings of the SM-like Higgs boson to Higgs and  gauge bosons.  }\label{table_HGcoupling}
\end{table}

Similarity to the SM-like Higgs boson case,the Feynman rules for the couplings of  $Z$ to  charged Higgs and gauge bosons  in \eqref{eq_lkHiggs} are given in table~\ref{table_Z1A}.
\begin{table}[ht]
	\centering 
	\begin{tabular}{|c|c|}
		\hline
	Vertex& Coupling\\
	\hline
	$g_{ZH^+H^-}$	& $\frac{g}{2c_W}\left(c_{\theta}\,c_{2W} +\frac{s_{\theta}\left[\sqrt{3}c^2_W(1 -2s^2_{12}) +3\beta s^2_W\right]}{3c_W\sqrt{1-\beta^2t^2_W}} \right) $\\
	\hline 	
	$g_{ZH^{A}H^{-A}}$	& $\frac{g}{2c_W}\left( c_{\theta} \left[ s^2_{13} -(1 +\sqrt{3}\beta)s_W^2 \right] +\frac{s_{\theta}\left[\sqrt{3}c^2_W(s^2_{13}-2) +3\beta (\sqrt{3}\beta +c^2_{13})  s^2_W\right]}{3c_W\sqrt{1-\beta^2t^2_W}} \right) $\\
	\hline 	
	$g_{ZH^{B}H^{-B}}$	& $\frac{ig}{2c_W}\left( -c_{\theta} \left[ s^2_{23} +(\sqrt{3}\beta-1)s_W^2 \right] +\frac{s_{\theta}\left[\sqrt{3}c^2_W(s^2_{23}-2) +3\beta (\sqrt{3}\beta -c^2_{23})  s^2_W\right]}{3c_W\sqrt{1-\beta^2t^2_W}} \right) $\\
	\hline 	
	$g_{ZW^{+}H^-}$& $ -\frac{g\,m_W (2s_{12}c_{12}s_{\theta})}{\sqrt{3(1-\beta^2t^2_W)}}$\\
	\hline 
		$g_{ZY^{A}H^{-A}}$, & $ \frac{g^2c_{13}}{4}\left\{\frac{}{} c_{\theta}c_W \left[ s_{12}\left( 1 +(2 +\sqrt{3}\beta)t^2_W \right)v +t_{13}(1 -\sqrt{3}\beta t^2_W) v_3 \right]  \right.$\\
		$g_{ZY^{-A}H^{A}}$	& $\left. +\frac{s_{\theta}}{3\sqrt{1-\beta^2t^2_W}} \left[s_{12} \left(\sqrt{3} -3\beta(2 +\sqrt{3}\beta)t^2_W\right) v +\sqrt{3}t_{13} \left(1 +3 \beta^2t_W^2 \right)v_3 \right]\right\}$\\
	\hline 	
		$g_{ZV^{B}H^{-B}}$, &  $ \frac{g^2c_{23}}{4}\left\{\frac{}{} c_{\theta}c_W \left[ c_{12} \left( -1 +(-2 +\sqrt{3}\beta)t^2_W \right)v  -t_{23} (1 +\sqrt{3}\beta t^2_W) v_3 \right]  \right.$\\
			$g_{ZV^{-B}H^{B}}$ & $\left. +\frac{s_{\theta}}{3\sqrt{1-\beta^2t^2_W}} \left[ c_{12} \left(\sqrt{3} -3\beta( -2 +\sqrt{3}\beta)t^2_W\right) v +\sqrt{3}t_{23} \left(1 +3 \beta^2t_W^2 \right)v_3 \right]\right\}$\\
\hline 
	\end{tabular}
\caption{Feynman rules of couplings with $Z$ to charged Higgs and gauge bosons. }\label{table_Z1A}
\end{table}

The couplings of $Z$ and photon $A_{\mu}$ with fermions arise from the covariant kinetic of fermions:
\begin{align}\label{eq_Lkinf}
\mathcal{L}^{f}_{\mathrm{kin}} &=\sum_{a=1}^3\left( \overline{L_{aL}}\gamma^{\mu}D_{\mu}L_{aL} + \overline{\nu_{aR}}\gamma^{\mu}\partial_{\mu}\nu_{aR} + \overline{e_{aR}}\gamma^{\mu} D_{\mu}e_{aR} + \overline{E_{aR}}\gamma^{\mu} D_{\mu}E_{aR}\right) \crn
& +\sum_{a=1}^3\left( \overline{Q_{aL}}\gamma^{\mu}D_{\mu}Q_{aL} + \overline{u_{aR}}\gamma^{\mu}D_{\mu}u_{aR} + \overline{d_{aR}}\gamma^{\mu} D_{\mu}d_{aR} + \overline{J_{aR}}\gamma^{\mu} D_{\mu}J_{aR}\right) \crn
&\supset \sum_f \left[ \frac{g\,c_{\theta}}{c_W}  \overline{f}\gamma^{\mu} \left( g^f_L P_L +g^f_R P_R\right)f Z_{\mu} + e Q_f \overline{f}\gamma^{\mu}f A_{\mu}\right],
\end{align}
where $f$ runs over all fermions  in the 331$\beta$ model, $Q_f$ is the electric charge of the fermion $f$. Values of $g^f_{L,R}$ are  shown in table~\ref{table_Z1ff}.
\begin{table}[ht]
	\centering
	\begin{tabular}{|c|c|c|}
		\hline
		$f$ & $g^f_L$&  $g^f_R$\\
		\hline 
			$e_a$ &  $  -\frac{1}{2} +s^2_W + \frac{t_\theta\, c_W(1 -\sqrt{3}\beta t^2_W)}{2\sqrt{3(1-\beta^2 t^2_W)}} $ & $s^2_W \left(1 - \frac{t_\theta\, \beta}{c_W\sqrt{1 -\beta^2 t^2_W}} \right) $  \\
			\hline 
			$u_i$ & $\frac{1}{2} -\frac{2}{3}s^2_W + \frac{t_\theta\, c_W (\beta t^2_W -\sqrt{3})}{6\sqrt{1-\beta^2 t^2_W}} $ &  $ -\frac{2}{3}s^2_W  \left(1 -  \frac{t_\theta\, \beta}{c_W\sqrt{1 -\beta^2 t^2_W}} \right) $ \\
				\hline 
			$u_3$ & $ \frac{1}{2} -\frac{2}{3}s^2_W + \frac{t_\theta\, c_W (\beta t^2_W +\sqrt{3})}{6\sqrt{1-\beta^2 t^2_W}} $ &  $ -\frac{2}{3}s^2_W \left(1 -  \frac{t_\theta\, \beta}{c_W\sqrt{1 -\beta^2 t^2_W}} \right) $ \\
			\hline 
			$d_i$ &  $-\frac{1}{2} +\frac{1}{3}s^2_W + \frac{t_\theta c_W(\beta t^2_W-\sqrt{3})}{6\sqrt{1-\beta^2 t^2_W}}$ & $ \frac{1}{3}s^2_W  \left(1 -  \frac{t_\theta\, \beta}{c_W\sqrt{1 -\beta^2 t^2_W}} \right) $ \\
		\hline 
		$d_3$ &  $-\frac{1}{2} +\frac{1}{3}s^2_W + \frac{t_\theta c_W(\beta t^2_W +\sqrt{3})}{6\sqrt{1-\beta^2 t^2_W}}$ & $ \frac{1}{3}s^2_W  \left(1 -  \frac{t_\theta\, \beta}{c_W\sqrt{1 -\beta^2 t^2_W}} \right) $ \\
	\hline 
	\end{tabular}
	\caption{ Couplings of $Z$ with fermions}\label{table_Z1ff}
\end{table}

The triple couplings of three gauge bosons arise from the covariant kinetic Lagrangian of the non-Abelian gauge bosons:
\begin{equation}\label{eq_LDg}
\mathcal{L}^g_D= -\frac{1}{4}\sum_{a=1}^8F^a_{\mu\nu}F^{a\mu\nu},
\end{equation} 
where 
\begin{equation}\label{eq_Famunu}
F^a_{\mu\nu}=\partial_{\mu}W^a_{\nu} -\partial_{\nu}W^a_{\mu} + g \sum_{b,c=1}^8f^{abc}W^b_{\mu} W^c_{\nu},
\end{equation}
$f^{abc}$ $(a,b,c=1,2,...,8)$ are structure constants of the $SU(3)$ group. They are defined as 
\begin{align}\label{eq_ZAgg}
\mathcal{L}^g_D \rightarrow& -g_{Zvv}Z^{\mu}(p_0)v^{+Q\nu}(p_+)v^{-Q\lambda}(p_-)\times \Gamma_{\mu\nu\lambda}(p_0,p_+,p_-), \crn 
&-eQA^{\mu}(p_0)v^{+Q\nu}(p_+)v^{-Q\lambda}(p_-)\times \Gamma_{\mu\nu\lambda}(p_0,p_+,p_-),
\end{align}
where $\Gamma_{\mu\nu\lambda}(p_0,p_+,p_-) \equiv g_{\mu\nu}(p_0-p_+)_{\lambda} +g_{\nu\lambda}(p_+ -p_-)_{\mu} +g_{\lambda\mu}(p_--p_0)_{\lambda}$, and $v=W,V,Y$. 
The involved  couplings of $Z$ are given in table~\ref{table_3gaugcoupling}.
\begin{table}[ht]
	\centering \begin{tabular}{|c|c|}
		\hline
		Vertex & Coupling\\
\hline 
$-ig_{ZW^{+\nu}W^{-\lambda}}$& $-igc_Wc_{\theta}
 $\\
\hline 
$-ig_{ZY^{A}Y^{-A}}$& $ \frac{ i g}{2} \left[  c_{\theta} \left(-c_W +\sqrt{3} \beta  s_W t_W\right) +s_{\theta} \sqrt{3 -3 \beta^2  t_W^2}\right] 
 $\\
\hline 
$-ig_{ZV^{B}Y^{-B}}$& $ \frac{ i g}{2} \left[  c_{\theta} \left(c_W +\sqrt{3} \beta  s_W t_W\right) +s_{\theta} \sqrt{3 -3 \beta^2  t_W^2}\right] 
$\\
\hline 
\end{tabular}
	\caption{Feynman rules for  triple gauge couplings  relating with the decay $h\rightarrow Z\gamma,\gamma\gamma$.
	} \label{table_3gaugcoupling}
\end{table}
  These triple couplings were also given in Ref.~\cite{Diaz:2004fs,Cao:2016uur} in the limit $\theta=0$.

\subsection{Partial decay widths and signal strengths of the decays $h\rightarrow Z\gamma, \gamma\gamma$}
\begin{figure}[ht]
\includegraphics[width=12cm]{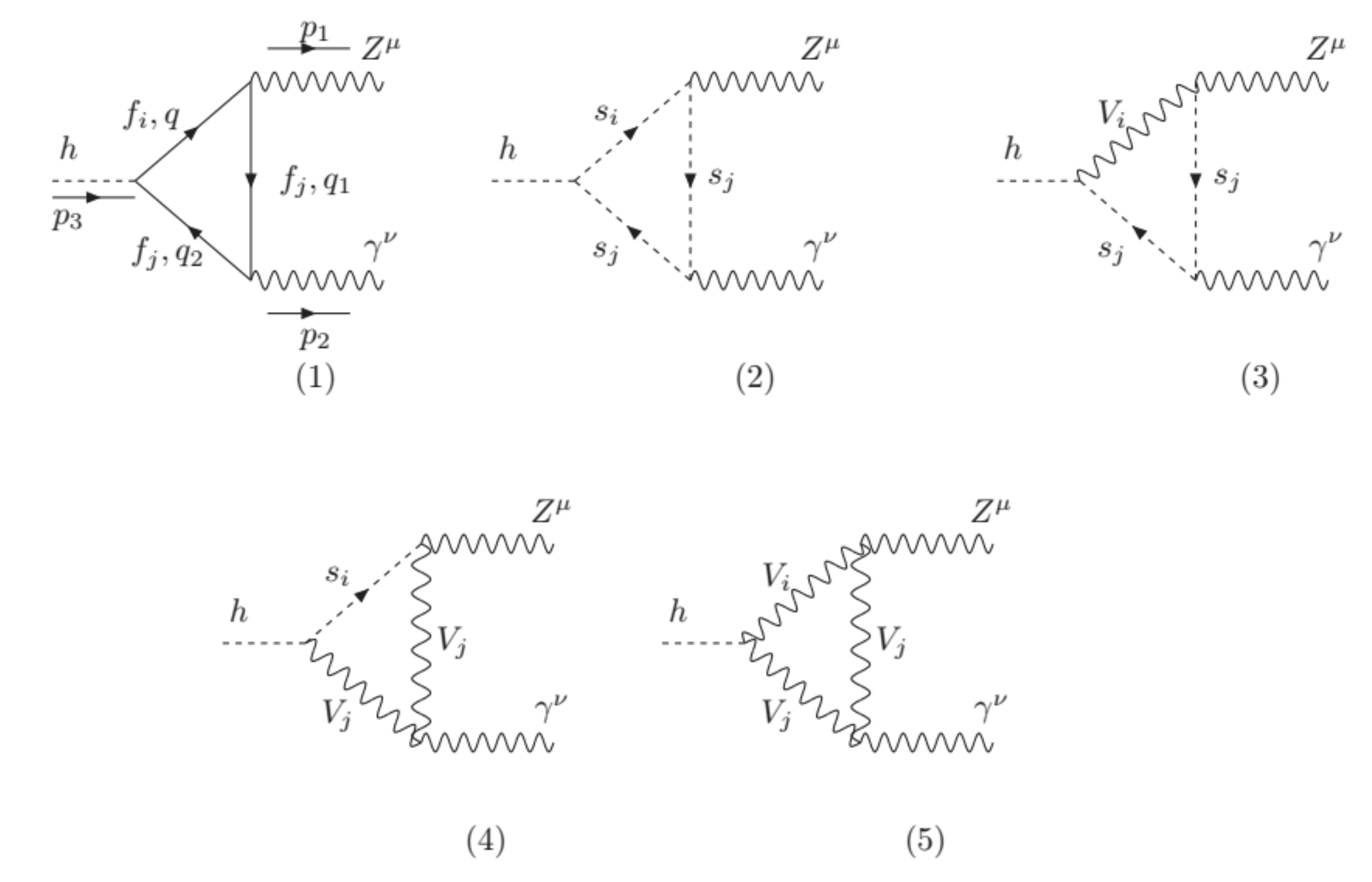}
	\caption{ One-loop three-point Feynman diagrams contributing to the decay $h\rightarrow\,Z\gamma$ in the unitary gauge, where $f_{i,j}$ are the SM leptons,  $s_{i,j}=H^{\pm},H^{\pm A},H^{\pm B}$, $v_{i,j}=W^{\pm}, Y^{\pm A}, V^{\pm B}$.}\label{fig_hzgaDiagram}
\end{figure}

In the unitary gauge, the above couplings generate  one-loop three point Feynman diagrams  which contribute to the decay amplitude of the SM-like Higgs boson $h\rightarrow\, Z\gamma$, as  given in Fig.~\ref{fig_hzgaDiagram}.  

 The partial decay width is~\cite{Gunion:1989we, Degrande:2017naf}
\be
\Gamma(h\rightarrow  Z\gamma)=\frac{m_h^3}{32\pi}
\times \left(1-\frac{m_Z^2}{m_h^2}\right)^3 |F_{21}|^2,
\label{GaHZga1}
\ee
where the scalar factor  $F_{21}$  is determined from one-loop contributions. More general formulas were given in Ref.~\cite{Hue:2017cph}, leading the following expression
\begin{align}
F^{331}_{21}&= \sum_{f}F^{331}_{21,f} + \sum_{s}F^{331}_{21,s} +\sum_{v}F^{331}_{21,v}+  \sum_{\{s,v\}}\left(F^{331}_{21,vss} + F^{331}_{21,svv} \right) .
\label{eq_F21331be}
\end{align} 
We note that $F^{331}_{21,vss}$ and $F^{331}_{21,svv}$ were not included in previous works~\cite{Cao:2016uur,Yue:2013qba}. 

The detailed analytic formulas of particular  notations in \eqref{eq_F21331be}  are given in appendix~\ref{app_hzga1loop}.
The partial decay width of the decay $h\rightarrow \gamma\gamma$ can be calculated as \cite{Degrande:2017naf, Hue:2017cph}
\begin{equation}\label{eq_Gah2gamma}
\Gamma(h\rightarrow  \gamma\gamma)=\frac{m_h^3}{64\pi}
\times |F^{331}_{\gamma \gamma}|^2 ,
\end{equation}
where 
\begin{align}\label{eq_F2gamma}
F^{331}_{\gamma\gamma}&=\sum_{f}F^{331}_{\gamma \gamma,f} + \sum_{s}F^{331}_{\gamma \gamma,s} +\sum_{v}F^{331}_{\gamma \gamma,v},
\end{align}
see  detailed analytic formulas in appendix~\ref{app_hzga1loop}. 
To determine the Br of a SM-like Higgs decay, we need to know the total decay width. In the SM, this quantity is~\cite{Heinemeyer:2013tqa, deFlorian:2016spz}
\begin{align}\label{eq_SMGa}
\Gamma_h^{\mathrm{SM}}&= \sum_{q\neq t}\Gamma^{\mathrm{SM}}(h\rightarrow \bar{q}q) + \sum_{\ell=e,\mu\tau}\Gamma^{\mathrm{SM}}(h\rightarrow \ell^+\ell^-) + \Gamma^{\mathrm{SM}}(h\rightarrow WW^*) + \Gamma^{\mathrm{SM}}(h\rightarrow ZZ^*)    \crn
& +\Gamma^{\mathrm{SM}}(h\rightarrow \gamma\gamma)  +\Gamma^{\mathrm{SM}}(h\rightarrow Z\gamma) +\Gamma^{\mathrm{SM}}(h\rightarrow gg), 
\end{align}
where the partial decay widths are well-known with Higgs boson mass of 125.09 GeV found experimentally~\cite{Tanabashi:2018oca}. The Br of a particular decay channel $h\rightarrow X$, $X=gg,\gamma\gamma, Z\gamma$, is:
\begin{equation}\label{eq_brSM}
\mathrm{Br}^{\mathrm{SM}}(h\rightarrow X)\equiv \frac{\Gamma^{\mathrm{SM}}(h\rightarrow  X)}{\Gamma_h^{\mathrm{SM}}}.
\end{equation}
The numerical values are given in table~\ref{table_GaSMhtoXY}~\cite{Heinemeyer:2013tqa, deFlorian:2016spz}, where the diphoton decay is consistent with that used in Ref.~\cite{Aaboud:2018xdt}, Br$(h\rightarrow \gamma\gamma)=(2.27\pm 0.07)\times 10^{-3}$.
\begin{table}[ht]
	\begin{tabular}{|c|c|c|c|c|c|c|c|c|c|}
		\hline 
		$b\bar{b}\;$& $\tau^+\tau^-$ & $\mu^+\mu^-$ & $ c\bar{c}\;$ & $ gg$ & $\gamma \gamma$ & $Z \gamma$ & $WW$ & $ZZ$ & $\Gamma^{\mathrm{SM}}_h$ (GeV)\\
		\hline 	
		0.5809	& 0.06256 & $2.171\times 10^{-4}$ &  $0.02884$ & $0.0818$ &  $0.00227$ &  $0.001541$ &  $0.2152$ & $0.02641$ &$4.10\times 10^{-3}$\\
		\hline 	
	\end{tabular}
	\caption{ Branching ratios of the SM Higgs boson decays with mass of 125.09 GeV.} \label{table_GaSMhtoXY}
\end{table}
The recent global signal strength found experimentally by ATLAS is $\mu_{\gamma\gamma}=0.99\pm0.14$~\cite{Aaboud:2018xdt}\footnote{This value gives the same numerical discussion with that reported in~\cite{Khachatryan:2014ira,Aad:2014eha}. }.

The  total decay width of the SM-like Higgs boson predicted by the $331\beta$ is computed based on the  deviations of the Higgs couplings with fermions and gauge bosons between  the two models SM and $331\beta$, as  given in tables~\ref{table_h0coupling} and~\ref{table_HGcoupling}. The result is
\begin{align}\label{eq_SMlikeGa}
\Gamma_h^{331}
&=  0.6725\left( c_{\delta} -\frac{s_{\delta}}{t_{12}}\right) ^2\Gamma_h^{\mathrm{SM}} \crn
&+c^2_{\delta} \left[ 0.2152+ \left(1 -\frac{2c_{\theta}s_{\theta}c_W}{\sqrt{1-\beta^2t_W^2}} \left( \beta t_W^2 + \frac{s_{12}c_{\alpha} +c_{12}s_{\alpha}}{\sqrt{3}c_{\delta}}\right)\right) ^2 0.02641\right] \Gamma_h^{\mathrm{SM}} \crn 
&+\Gamma^{331}(h\rightarrow \gamma \gamma) +\Gamma^{331}(h\rightarrow Z\gamma) +\Gamma^{331}(h\rightarrow gg).
\end{align}

There are three loop-induced decays $h\rightarrow \gamma\gamma, Z\gamma,gg$. The SM-like Higgs boson does not couple with the exotic quarks in the $331\beta$,  we can consider only the top quark contribution to the loop contributing to the  decay $h\rightarrow gg$. This results in
\begin{equation}\label{eq_htogg}
\Gamma^{331}(h\rightarrow gg)=\left( c_{\delta} +t_{12}s_{\delta}\right)^2 \Gamma^{\mathrm{SM}}(h\rightarrow gg),
\end{equation}
where the deviation comes from the $ht\bar{t}$ coupling listed in table~\ref{table_h0coupling}. This is consistent with recent investigation for $h\rightarrow\gamma\gamma$ in a 3-3-1 model \cite{CarcamoHernandez:2019vih}.

In the 331$\beta$ framework, the branching ratio  of the decay $h\rightarrow\,X$ with $X=\gamma \gamma,Z\gamma$  is  
\begin{equation}\label{eq_br1}
\mathrm{Br}^{331}(h\rightarrow X)\equiv \frac{\Gamma^{331}(h\rightarrow  X)}{\Gamma_h^{331}}.
\end{equation}

Many experimental measurements relating to the SM-like Higgs boson were reported in Ref.~\cite{Khachatryan:2016vau}. We consider the SM-like Higgs production through the gluon fusion process $ggF$ at LHC. The respective signal strength predicted by $331\beta$ is defined as:
\begin{equation}\label{eq_331muggF}
\mu^{331}_{ggF}\equiv \frac{\sigma^{331}(gg\rightarrow h)}{\sigma^{\mathrm{SM}}(gg\rightarrow h)} \simeq \left( c_{\delta} +t_{12}s_{\delta}\right)^2,
\end{equation}
where the last value comes from our assumption that  only the main  contribution from top quark in the loop is considered. 
The signal strength of an individual 	loop-induced decay channel is 
\begin{equation}\label{eq_331mubr}
\mu^{331}_{X}\equiv   \left( c_{\delta} +t_{12}s_{\delta}\right)^2 \times  \frac{\mathrm{Br}^{331}(h\rightarrow X)}{ \mathrm{Br}^{\mathrm{SM}}(h\rightarrow X)}. 
\end{equation}
The recent signal strengths of the two loop-induced decay $h\rightarrow Z\gamma$ is   $\mu_{Z\gamma}<6.6(5.2)$~\cite{Aaboud:2017uhw, Tanabashi:2018oca}.

\subsection{Decays of the neutral Higgs boson $h^0_3$}
In the above discussion we derived only couplings that contribute to the one-loop amplitudes of the two SM-like Higgs decay channels $h\rightarrow \gamma\gamma, Z\,\gamma$.  Other interesting couplings  are listed in the appendix~\ref{app_coupling}. Here we  stress a very interesting property of the heavy neutral Higgs boson $h^0_3$ that  it has only one non-zero coupling with two SM particles, namely only $\lambda_{h^2h^0_3}\neq 0$.  We have  $m_{h^0_3}> 2 m_{h}$ then  if  $h^0_3$ is lighter than all other exotic particles predicted by the 331$\beta$ model, only the tree level decay  $h^0_3\rightarrow hh$ appears.  Loop-induced decays such as $h^0_3\rightarrow\,gg,\gamma\gamma,Z\gamma$ also appear, as we will present below.  Hence, the total decay width of $h^0_3$  cannot satisfy the stable condition of a dark matter, $\Gamma_{h^0_3}<1.3\times 2\pi\times 10^{-42}$ GeV~\cite{Banerjee:2016vrp, DeLopeAmigo:2009dc, Eiteneuer:2017hoh,Belyaev:2016lok}.  Anyway, DM candidates as scalar 3-3-1 Higgs bosons were pointed out previously \cite{Filippi:2005mt, Cogollo:2014jia, deS.Pires:2007gi}.  
  
The couplings  of neutral heavy Higgs bosons   $h^0_{2,3}$ to fermions are 
\begin{align}\label{eq_Yh23ff}
Y_{h^0_2ffL,R}&=\begin{cases}
\frac{m_{f}}{v} \left(  \frac{c_{\delta}}{t_{12}} + s_{\delta}  \right), ~&f= e_a,u_i,d_3\\
\frac{m_{f}}{v} \left( - c_{\delta}t_{12} + s_{\delta}  \right), ~&f= u_3,d_i\\
0,&~f=E_a,J_a.
\end{cases},  \quad 
Y_{h^0_3ffL,R} =\begin{cases}
0& f=e_a,u_a,d_a\\
\frac{m_{f}}{v_3}& f=E_a,J_a
\end{cases}.
\end{align}
One interesting point is that $h^0_3$ couples to only exotic fermions,  similar to the heavy neutral Higgs appeared in a $SU(2)_1\times SU(2)_2\times U(1)_Y$
model~\cite{Boucenna:2016qad}, where  the partial decay width $h^0_3\rightarrow\,gg$ is ~\cite{Gunion:1989we,Spira:1997dg},  
\begin{equation}\label{eq_h03toggexact}
\Gamma(h^0_3\rightarrow\,gg)\simeq \frac{\alpha_s^2m^3_{h^0_3}}{32\pi^3v_3^2}\left|\sum_{a=1}^3 t_a\left[1 +(1-t_a) f(t_a)\right]\right|^2 ,
\end{equation}
where $t_a\equiv 4m^2_{J_a}/m^2_{h^0_3}$,
\begin{align}\label{eq_fta}
f(x)&=\begin{cases}
\mathrm{arcsin}^2\frac{1}{\sqrt{x}},& x\ge 1\\
-\frac{1}{4}\left[\ln\frac{1+\sqrt{1-x}}{1-\sqrt{1-x}} -i\pi\right]^2,& x<1.\\
\end{cases} 
\end{align}
In the limit $t_a\gg1~\forall a=1,2,3$, Eq.~\eqref{eq_h03toggexact}  can be estimated as~\cite{Boucenna:2016qad} 
\begin{equation}\label{eq_h03togg}
\Gamma(h^0_3\rightarrow\,gg)\simeq \frac{\alpha_s^2m^3_{h^0_3}}{8\pi^3v_3^2}.
\end{equation}
Furthermore, the production cross section of this Higgs boson through the gluon-gluon fusion can be estimated from the two gluon decay channel~\cite{Boucenna:2016qad}. 

The partial width of the  tree level decay $h^0_3\rightarrow hh$ when $m_{h^0_3}> 2 m_{h}$ is \cite{Okada:2016whh}
\begin{align}
\label{eq_h03to2h}
\Gamma(h^0_3\rightarrow hh)&=\frac{|\lambda_{h^0_3hh}|^2}{8\pi\,m_{h^0_3}}\sqrt{1-\frac{4m^2_h}{m^2_{h^0_3}}}= \frac{\lambda_{13}^2s^4_{\delta}v_3^2}{8\pi\,c^4_{12}\,m_{h^0_3}}\sqrt{1-\frac{4m^2_h}{m^2_{h^0_3}}},
\end{align}
where $\lambda_{13}$ was given in Eq.~\eqref{eq_la13}.

The total decay width of the $h^0_3$ is then 
\begin{align}
\Gamma_{h^0_3}= \Gamma(h^0_3\rightarrow\, hh) + \Gamma(h^0_3\rightarrow gg) + \Gamma(h^0_3\rightarrow\,\gamma\gamma) + \Gamma(h^0_3\rightarrow\,Z\gamma).
\end{align}
The last two decays are calculated as follows,
\begin{align}
\Gamma(h^0_3\rightarrow  Z\gamma)&=\frac{m_{h^0_3}^3}{32\pi}
\left(1-\frac{m_Z^2}{m_{h^0_3}^2}\right)^3|F_{21}(h^0_3\rightarrow\,Z\gamma)|^2,
\label{Gah03Zga} \\
\Gamma(h^0_3\rightarrow  \gamma\gamma)&=\frac{m_{h^0_3}^3}{64\pi}
\times |F^{331}_{\gamma \gamma}(h^0_3\rightarrow  \gamma\gamma)|^2 \label{Gah032ga},
\end{align}
where 
\begin{align}
F^{331}_{21}(h^0_3\rightarrow\, Z\gamma)&= \sum_{F=E_a,J_a}F^{331}_{21,F}(h^0_3\rightarrow\, Z\gamma) +  \sum_{s}F^{331}_{21,s}(h^0_3\rightarrow\, Z\gamma) + \sum_{v=Y,V}F^{331}_{21,v} (h^0_3\rightarrow\, Z\gamma)\crn 
&+ \sum_{\{s,v\}}\left[ F^{331}_{21,vss}(h^0_3\rightarrow\, Z\gamma) + F^{331}_{21,svv}(h^0_3\rightarrow\, Z\gamma)\right] ,  \label{eq_Fh03Zga}\\ 
 F^{331}_{\gamma\gamma}(h^0_3\rightarrow\, \gamma\gamma)&= \sum_{F=E_a,J_a}F^{331}_{\gamma\gamma,F}(h^0_3\rightarrow\, Z\gamma) +  \sum_{s}F^{331}_{\gamma\gamma,s}(h^0_3\rightarrow\, Z\gamma) + \sum_{v=Y,V}F^{331}_{\gamma\gamma,v} (h^0_3\rightarrow\, Z\gamma),\nn 
\end{align}
where $s=H^{\pm},H^{\pm,A},H^{\pm,B}$, $v=Y^{\pm,A},V^{\pm,B}$, and $\{s,v \}=\{H^{\pm,A},Y^{\pm,A}\}, \{H^{\pm,B},V^{\pm,B}\}$. The explicit forms in \eqref{eq_Fh03Zga} were shown in appendix~\ref{app_hzga1loop}.

\section{\label{numerical} Numerical discussions}
\subsection{\label{setuppa} Signal  of the decay $h\rightarrow Z\gamma$ under   recent constraints of parameters and the decay $h\rightarrow \gamma\gamma$}

In this section, to express quantitative  deviations between predictions of  the two models 331$\beta$ and the SM for  decays $h\rightarrow X$ ($X=\gamma\gamma,Z\gamma$), we define a quantity $\delta\mu_X$   as follows 
\begin{align}\label{eq_demu}
\delta{\mu}_{X}\equiv \left(\mu^{331}_{X}-1\right)\times 100\%. 
\end{align}
We also introduce a new quantity $R_{Z\gamma/\gamma\gamma}\equiv |\delta\mu_{Z\gamma}/\delta\mu_{\gamma\gamma}|$ to investigate the relative difference between the two signal  strengths, which have many similar properties. The recent allowed values relating with the two photon decay is $-15\% \le \delta\mu_{\gamma\gamma}\le 13\%$, corresponding to the recent experimental constraint  $\mu_{\gamma\gamma}=0.99\pm0.14$~\cite{Aaboud:2018xdt}. The future sensitivities obtained by experiments we accept here are $\mu_{\gamma\gamma}=1\pm 0.04$ and $\mu_{Z\gamma}=1\pm0.23$~\cite{Cepeda:2019klc}, i.e. $|\delta\mu_{\gamma\gamma}|\leq4\%$ and $|\delta\mu_{Z\gamma}|\leq23\%$, respectively.

Many well-known quantities used in this section are fixed from experiments~\cite{Tanabashi:2018oca}, namely the SM-like Higgs mass  $m_h=125.09$ GeV;  the gauge boson  masses $m_W$, $m_Z$;  well-known charged fermion  masses; the vev $v\simeq 246$ GeV; and the $SU(2)_L$ couplings $g\simeq 0.651$,  $\alpha_{\mathrm{em}}=1/137$, $e=\sqrt{4\pi\alpha_{\mathrm{em}}}$, $s^2_W=0.231$. 

The unknown independent parameters used as inputs are  $\beta$, $t_{12}$, $SU(3)_L$ scale $v_3$, the neutral Higgs mixing $s_{\delta}$, the heavy neutral Higgs boson  masses  $m_{h^0_2}$, $m_{h^0_3}$, the  triple Higgs self couplings including $\lambda_1$, $\tilde{\lambda}_{12}$,  $\tilde{\lambda}_{13}$, $\tilde{\lambda}_{23}$, and the exotic fermion masses $m_{E_a}$, $m_{J_a}$.

The exotic fermion masses $m_{E_a}$, $m_{J_a}$ affect only the loop-induced decays of $h^0_3$. We can put $m_{E_a}=m_{J_a}=m_F$ for simplicity. There is a more general case that the mixing between different exotic leptons appear, then the loop with two distinguished fermions will contribute to the $h^0_3\rightarrow\,Z\gamma$ decay amplitude only.

The $SU(3)_L$ scale depends  strongly on the heavy neutral gauge boson mass  $m_{Z'}$, which the lower bound is constrained from experimental  searches for decays to pairs of SM leptons  $Z'\rightarrow \ell \bar{\ell}$,    for 3-3-1 models see~\cite{Coutinho:2013lta}, where decays into exotic lepton pairs were  included.  Accordingly, at LHC@14TeV, $m_{Z'}<4$ TeV is excluded at the integrated luminosity of $23~\mathrm{fb}^{-1}$ for $\beta=-1/\sqrt{3}$.  Recent works have used the  $m_{Z'}\ge 4$ TeV for models with $\beta=-1/\sqrt{3}$~\cite{Ferreira:2019qpf,Long:2018dun}, based on the latest LHC search~\cite{Aaboud:2017sjh,Sirunyan:2018exx,Aad:2019fac}.   Because  $v_3\sim \mathcal{O}(1)$ TeV, the  $m_{Z'}$ is approximately calculated from   $m_{Z'}^2=\frac{g^2v_3^2c^2_W}{3[1 -(1+\beta^2)s^2_W]}$. From this, the lower bound of $m_{Z'} >4$ TeV corresponds to lower bounds of $v_3\ge 10.6,\,10.1,\,8.2,\,3.3$ TeV with  respective values  of $\beta=0,\pm1/\sqrt{3},\pm2/\sqrt{3},\pm\sqrt{3}$.  Recent discussion on  3-3-1 models with heavy right-handed neutrinos where $\beta=-1/\sqrt{3}$ and $m_{Z'}=3$ TeV is allowed~\cite{Freitas:2018vnt,Arcadi:2019uif} because the decay of $Z'$ into  a pair of light exotic neutrinos is included. The respective lower bound of the  $SU(3)_L$ scale is $v_3\ge   7.6$ TeV, which is still the same mentioned bounds. 
  On the other hand, a model with $\beta=\sqrt{3}$ still allows rather low $SU(3)_L$ scale, for example   $m_{Z'}\simeq  3.25$ TeV, corresponding to $v_3\simeq 2.7$ TeV~\cite{Coriano:2018coq}. Because the numerical results do not change significantly in the range $7.6\;\mathrm{TeV}<v_3<14$ TeV, we will fixed $v=14$ TeV for $|\beta|<\sqrt{3}$ and $v=3$ TeV for $|\beta|=\sqrt{3}$.

The perturbative limits require that the absolute values of  all Yukawa and Higgs self-couplings   should be less than $\sqrt{4\pi}$ and $4\pi$, respectively.  This leads to an upper bound of  $t_{12}$ derived from the Yukawa coupling of the top quark in Eq.~\eqref{eq_Fermionmass}, namely $t_{12}<\sqrt{2\pi}v/m_t\simeq 3.5$. Other studies on the 2HDM suggest  that $t_{12}>1/60$~\cite{Cepeda:2019klc}. We will limit that $0.1\le\,t_{12}\le3$, which is consistent with Ref.~\cite{Okada:2016whh} and allows large $|s_{\theta}|\ge 5\times 10^{-3}$.

Considering $m_{h^0_2}, m_A,m_{H^{\pm}}$, $t_{12}$ and $s_{\delta}$ as  parameters of a 2HDM model mentioned in Ref.~\cite{Kanemura:2018yai}, an important constraint can be found as $c_{\delta}>0.99$ for all 2HDMs, leading to rather large range of  $|s_{\delta}|<0.14$.  But large $s_{\delta}$ prefers that  $t_{12}$ is  around 1~\cite{Aad:2015pla}.  The recent global fit for 2HDM gives the same result~\cite{Haller:2018nnx}. Lower masses of  heavy Higgs bosons are around 1 TeV.   As we will show, the recent signal strength of the SM-like Higgs decay $h\rightarrow\gamma\gamma$ gives more strict constraint on $s_{\theta}$, hence we focus on the interesting range $|s_{\theta}|\leq 0.05$.  The  parameters $\lambda_2$ and $\lambda_{12}$  relating with 2HDM  affect strongly on  $m_{h^0_2}$. Large $|s_{\theta}|$ results in small allowed values of $m_{h^0_2}$ in order to keep $\lambda_{12}$ satisfying the perturbative limit.   In contrast, all other quantities  relating with the $SU(3)_L$ symmetry are well allowed. The regions of parameter space chosen here are consistent with the recent works on 2HDM~\cite{Ferreira:2019iqb,Kainulainen:2019kyp,Babu:2018uik}.   The recent experimental  searches for Higgs bosons predicted by  2HDM have been paid much attentions \cite{Aaboud:2017gsl}.  The value of 300 GeV for  lower bounds of charged and $CP$-even neutral Higgs bosons are accepted in recent studies  on 2HDM~\cite{Babu:2018uik}. Following that, values of $m_{h^0_2}$ and $m_{H^\pm}$  will be  chosen to  satisfy  $m_{h^0_2},m_{H^{\pm}}\ge300$ GeV.

We will also consider the case of  light charged Higgs masses, which loop contributions to the decay $h\rightarrow \gamma\gamma,Z\gamma$ may be large. Accordingly, the Higgs self-couplings $\tilde{\lambda}_{ij}$ relating with charged Higgs masses in Eqs.~\eqref{scHigg},~\eqref{qacHigg} and~\eqref{qbcHigg}, should be negative. Our investigation suggests that $|\tilde{\lambda}_{13,23}|\le \mathcal{O}(10^{-3})$ while $|\tilde{\lambda}_{12}|$ can be 
reach order 1.  We will consider more detailed in particular numerical investigations.

Strict constraints of the Higgs self-couplings for a 3-3-1 model with right handed neutrino  were discussed in Ref.~\cite{Sanchez-Vega:2018qje}, where the Higgs potential is forced to satisfy the vacuum stability condition. Accordingly, interesting results can be applied to the 3-3-1 models with arbitrary $\beta$, namely  
\begin{align}\label{eq_laconstraint}
\lambda_i>0,\quad f_{ij} \equiv \lambda_{ij}+ 2\sqrt{\lambda_i \lambda_j}&>0,\quad  \tilde{f}_{ij}\equiv \lambda_{ij}+ \tilde{\lambda}_{ij} + 2\sqrt{\lambda_i \lambda_j}>0, 
\end{align}
with $i,j=1,2,3$ and $i<j$.
Note that the constraints on the Higgs self-couplings  $\lambda_{1,2,12}$ correspond to  the particular cases  of the 2HDM \cite{Tanabashi:2018oca, Kanemura:2018yai,Kanemura:1999xf}. Because   $\left( t_{12} +t^{-1}_{12} + c_{12}(v/v_3)^2\right)fv_3\sim m_A^2$ being the squared mass of the $CP$-odd neutral Higgs boson,  the requirement $m^2_A>0$ shows $f>0$ \cite{Okada:2016whh, Dias:2009au}. The other conditions guarantee that all squared Higgs masses must be positive and SM-like Higgs mass is identified with the experimental value.   It can be seen in Eqs.~\eqref{scHigg}-\eqref{qbcHigg} that  all  charged Higgs squared masses are always positive if all $\tilde{\lambda}_{ij}>0$, but their values seem very large. More interesting cases correspond to the existence of  light charged Higgs bosons, which may contribute significant contributions to loop-induced decays of the SM-like Higgs bosons. Based on eq.~\eqref{eq_alignH0}, a first estimation  suggests that $f$ has the same order with $SU(3)_L$ scale $v_3$, leading to the requirement that   $\tilde{\lambda}_{12,13,23}<0$ for the existence of light charged Higgs bosons.  Furthermore, the relation~\eqref{eq_la13} results in a consequence that $\lambda_{13}$ will be small for the case of our interest with large $v_3\ge 3$ TeV and small $m_{h^0_2}$ around 1 TeV.   In this case $f$ is also small, as we realize in the numerical investigation as well as it has been shown recently~\cite{Palcu:2019nld}. Taking this into account to the charged Higgs masses in Eqs.\eqref{qacHigg} and~\eqref{qbcHigg} we derive that the absolute values of negative values of $\tilde{\lambda}_{13,23}$ seems very small.  In contrast, the appearance of a light charged Higgs $H^\pm$  allows negative $\tilde{\lambda}_{12}$ and  rather large $|\tilde{\lambda}_{12}|$  that  satisfy the inequality $\tilde{f}_{12}>0$ given in~\eqref{eq_laconstraint}. We will consider the two separate cases:  $\tilde{\lambda}_{ij}\ge 0$ with all $i>j$, $i,j=1,2,3$;  and  $\tilde{\lambda}_{12}< 0$. The values   of $\lambda_{13,23}$  are always chosen to get large absolute values of $F^{331}_{21,s}$, and/or  $F^{331}_{21,sv}\equiv F^{331}_{21,svv} +F^{331}_{21,vss}$.

The above discussion allows  us to choose the default values of unknown independent parameters as follows:  $\beta=1/\sqrt{3}$,  $s_{\delta}=0.01$, $\lambda_{1}=1$, $t_{12}=0.8$, $\tilde{\lambda}_{12}=  \tilde{\lambda}_{13}= \tilde{\lambda}_{23}=0.1$, $m_{h^0_2}=1.2$ TeV, $m_{h^0_3}=1$ TeV,  $v_3=14$ TeV, $m_{E_a}=m_{J_a}=1.5$ TeV. We choose the perturbative limit of Higgs self-couplings is $10$, which is a bit more strict than $4\pi$ \footnote{ We thank the referee for reminding us this point.}.  In addition, depending on the particular discussions, changing any numerical values will be noted.

\subsubsection{Case 1: $\tilde{\lambda}_{12}\ge 0$}
\begin{figure}[ht]
	\centering
	\begin{tabular}{cc}
		\includegraphics[width=7.5cm]{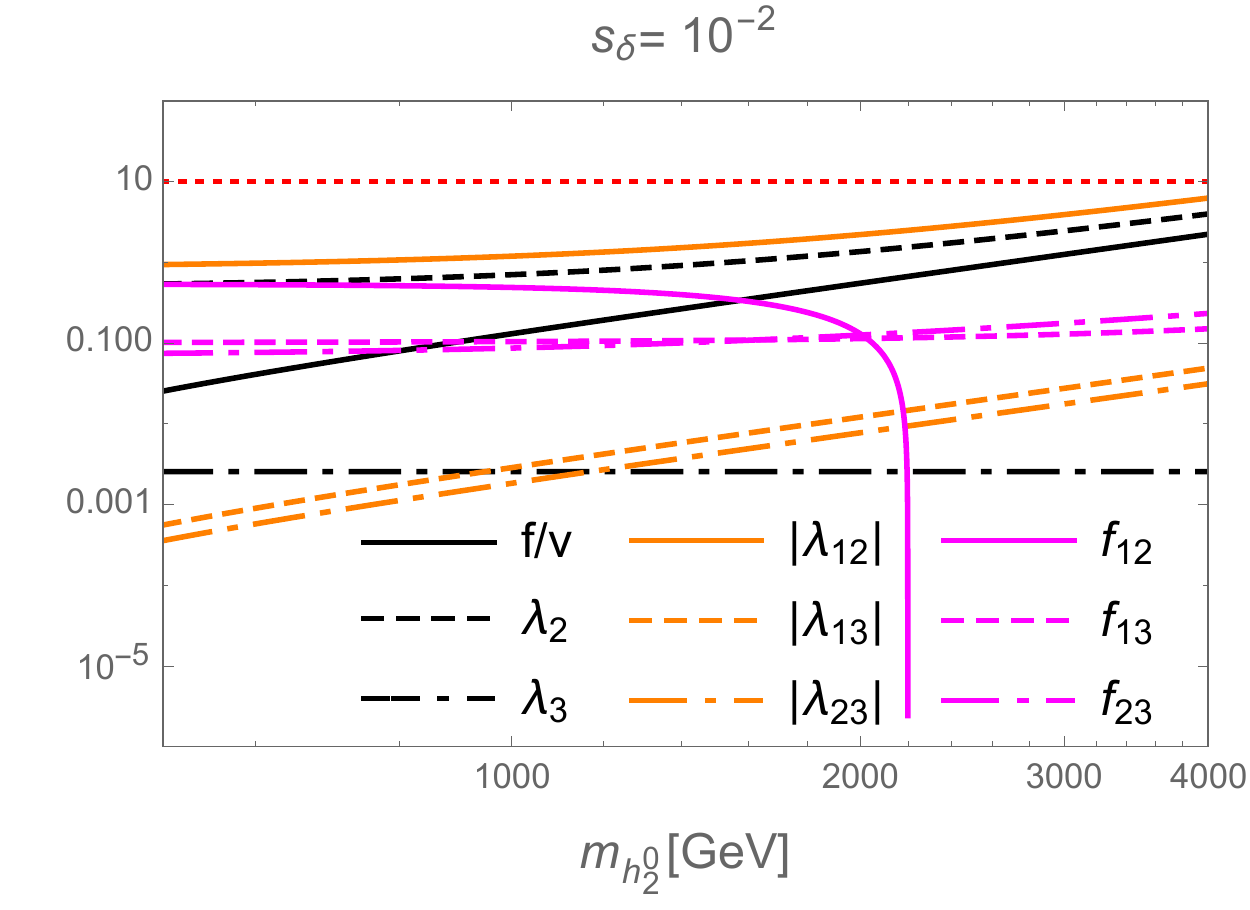} &
		\includegraphics[width=7.5cm]{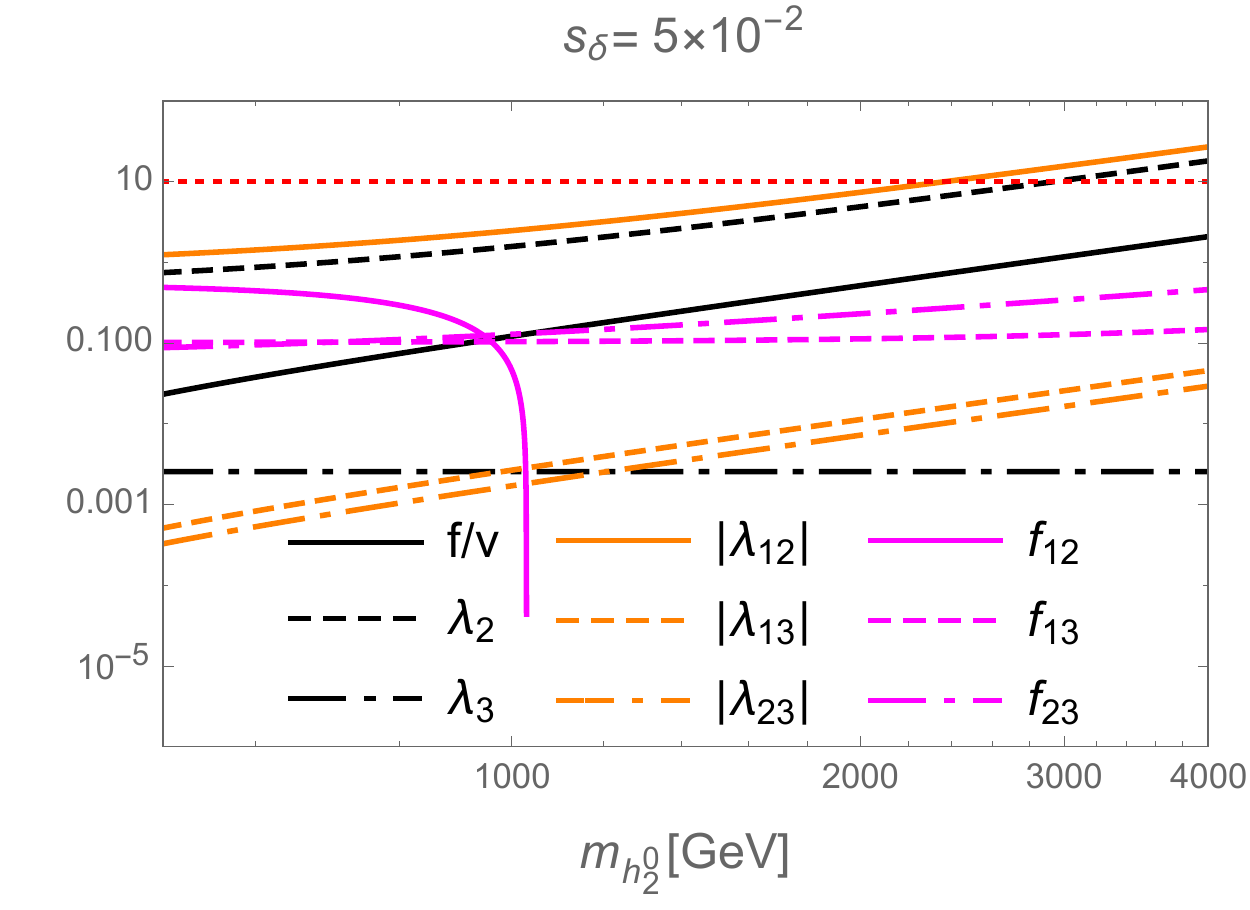} \\
		\end{tabular}%
	\caption{ $f_{ij}$ and Higgs self-couplings as functions of $m_{h^{0}_2}$ with $s_{\delta} >0$ and  $t_{12}=0.8$. The horizontal lines at the value of $10$ correspond to the perturbative limit of the Higgs self-couplings.}
	\label{fig_lacontrainsdp}
\end{figure}

\begin{figure}[ht]
	\centering
	\begin{tabular}{cc}
		\includegraphics[width=7.5cm]{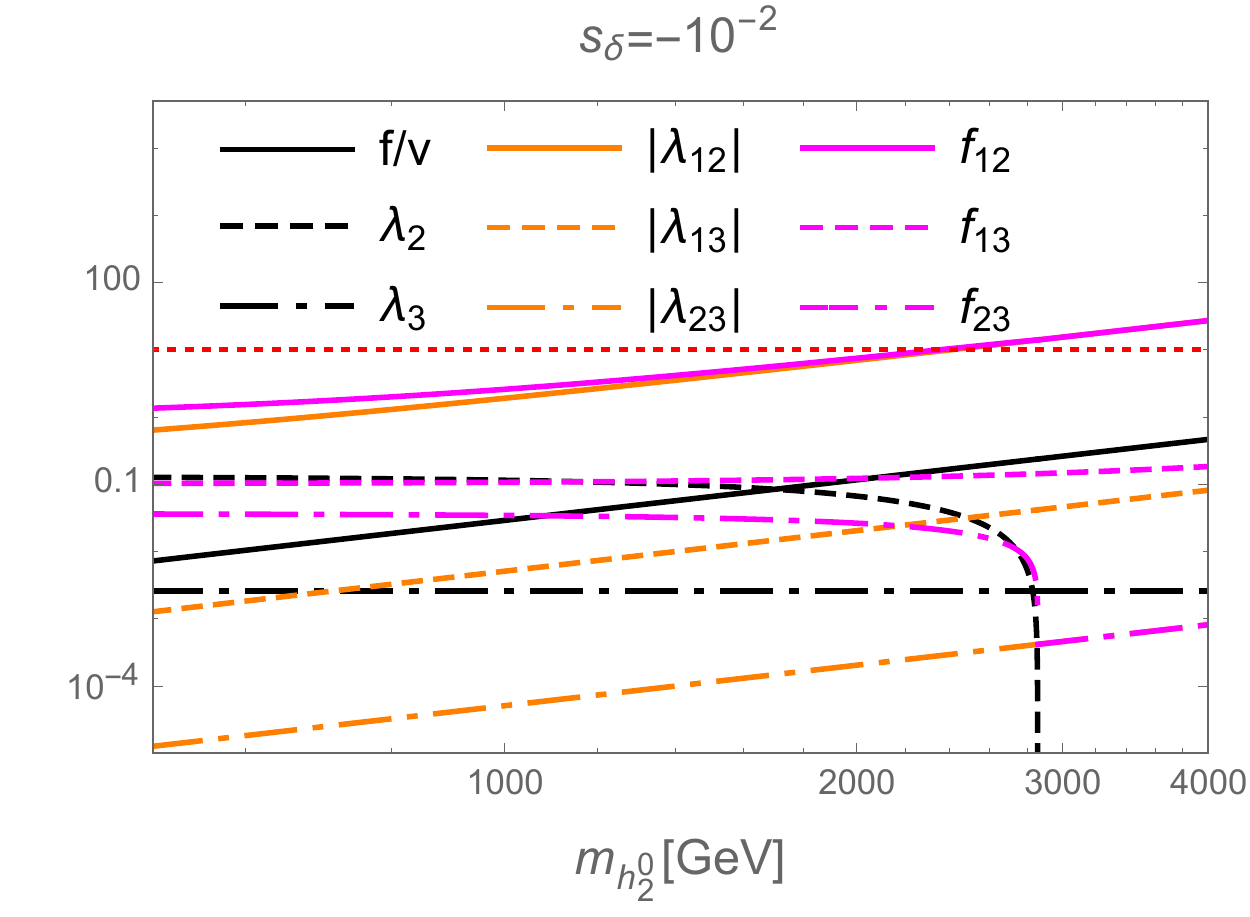} &
		\includegraphics[width=7.5cm]{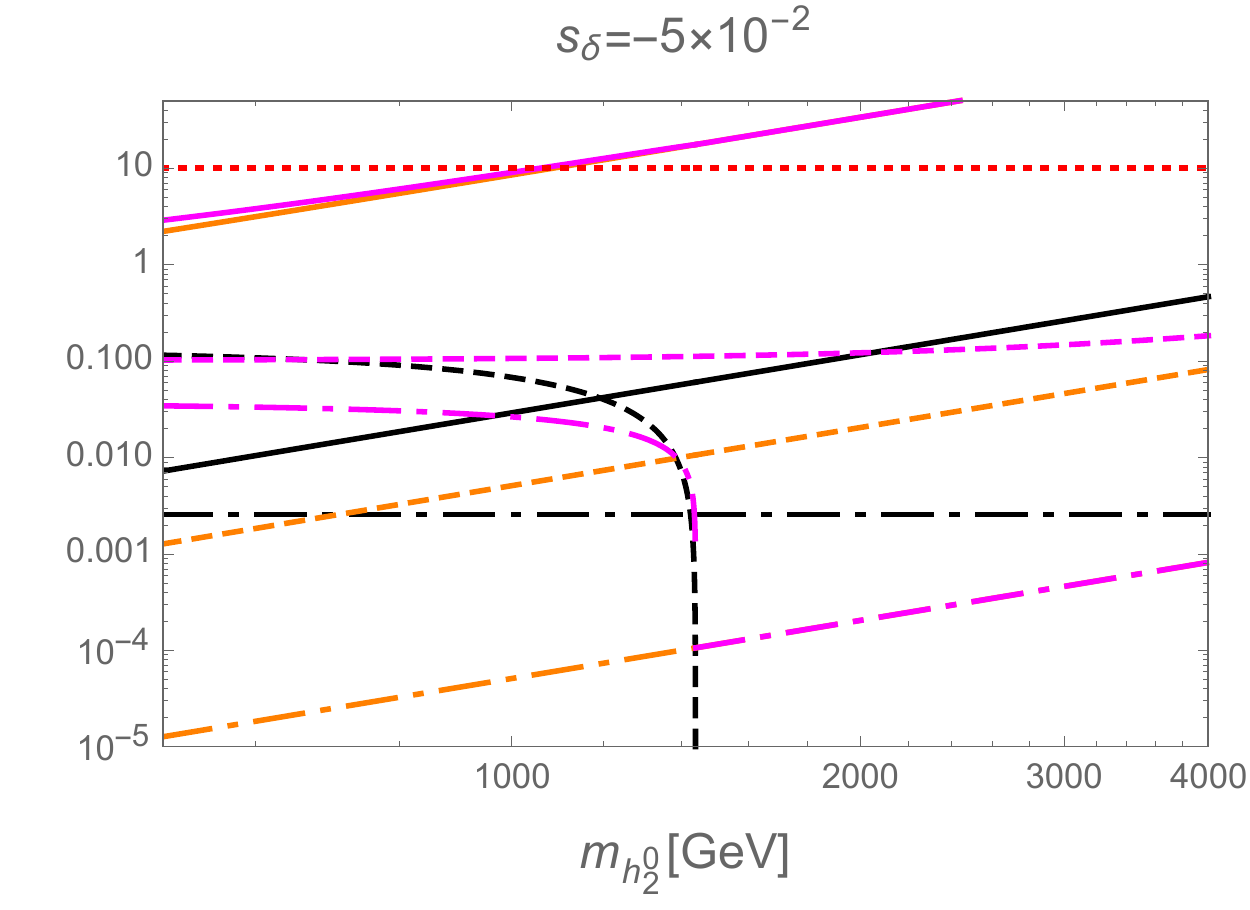} \\
	\end{tabular}%
	\caption{$f_{ij}$ and Higgs self-couplings as functions of $m_{h^{0}_2}$ with $s_{\delta} <0$ and $t_{12}=0.1$.}
	\label{fig_lacontrainsdn}
\end{figure}

First, we focus on the 2HDM parameters. Fig.~\ref{fig_lacontrainsdp} and~\ref{fig_lacontrainsdn} illustrate numerically Higgs self-couplings  and $f_{ij}$ as functions of $m_{h^0_2}$, and other independent  parameters are fixed as  $t_{12}=0.8$ and changing  $s_{\delta}=\pm 10^{-2},\pm 5\times 10^{-2}$, which are significantly large. 
 For $s_{\delta}>0$,  the $t_{12}$  is chosen large enough to satisfy  $f_{12}>0$ and $m_{h^0_2}>1$ TeV.  The parameters  $\tilde{\lambda}_{12}\ge0$ and negative $\tilde{\lambda}_{13,23}\rightarrow0$ do not affect the quantities investigated in this figure. We conclude that the vacuum stability requirement $f_{12}>0$ gives strong upper bound on $m_{h^0_2}$, where larger $s_{\delta}$ gives smaller allowed $m_{h^0_2}$.

Fig.~ \ref{fig_lacontrainsdn} illustrates allowed regions for  $s_{\delta}<0$, where we choose  $t_{12}=0.1$, enough small to allow  $\lambda_2>0$ and $m_{h^0_2}>1$ TeV.
Again we derive that larger $|s_{\delta}|$ gives smaller upper bound of $m_{h^0_2}$. 

\begin{figure}[ht]
	\centering
	\begin{tabular}{cc}
		\includegraphics[width=7.5cm]{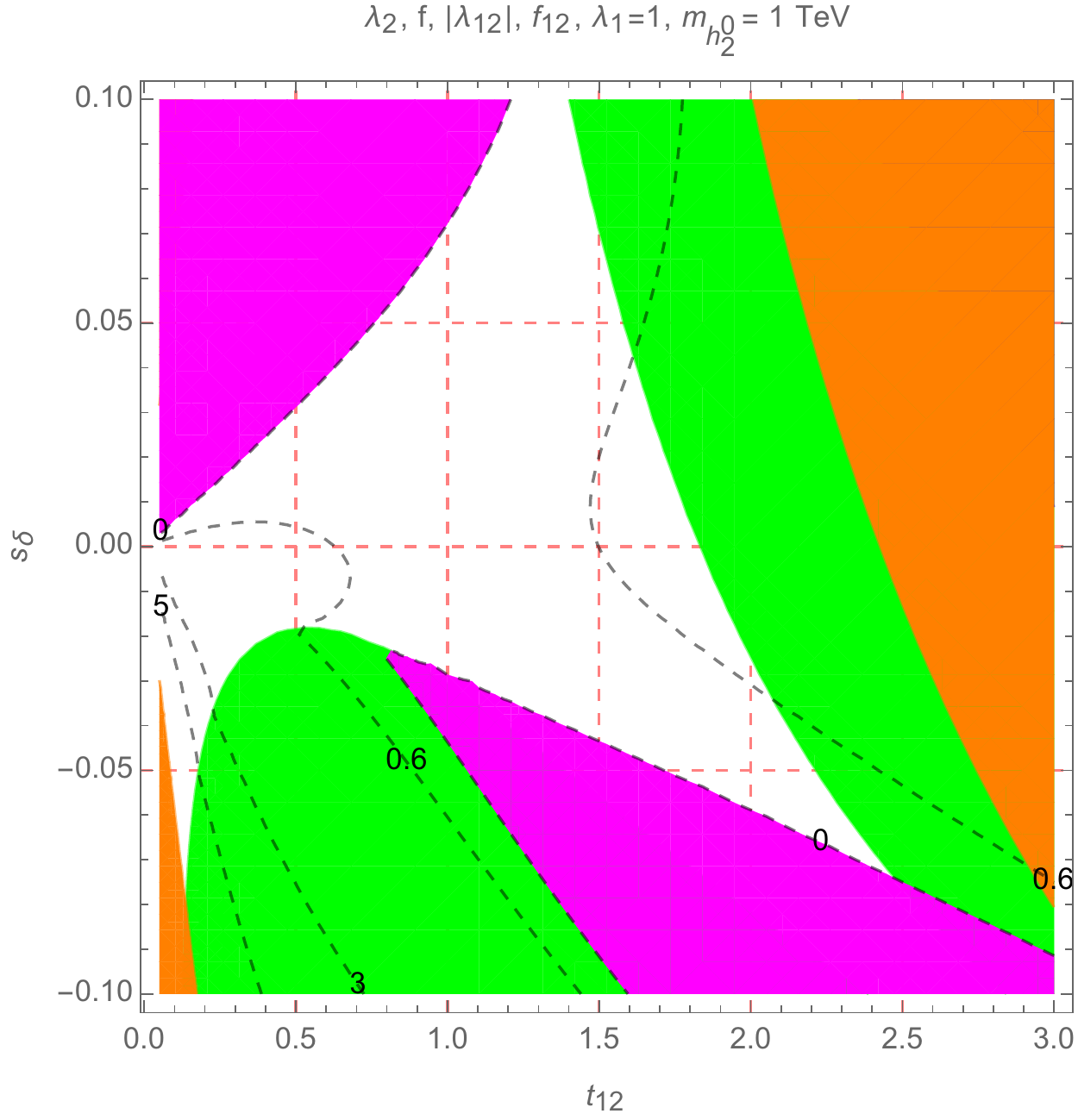} &
		\includegraphics[width=7.5cm]{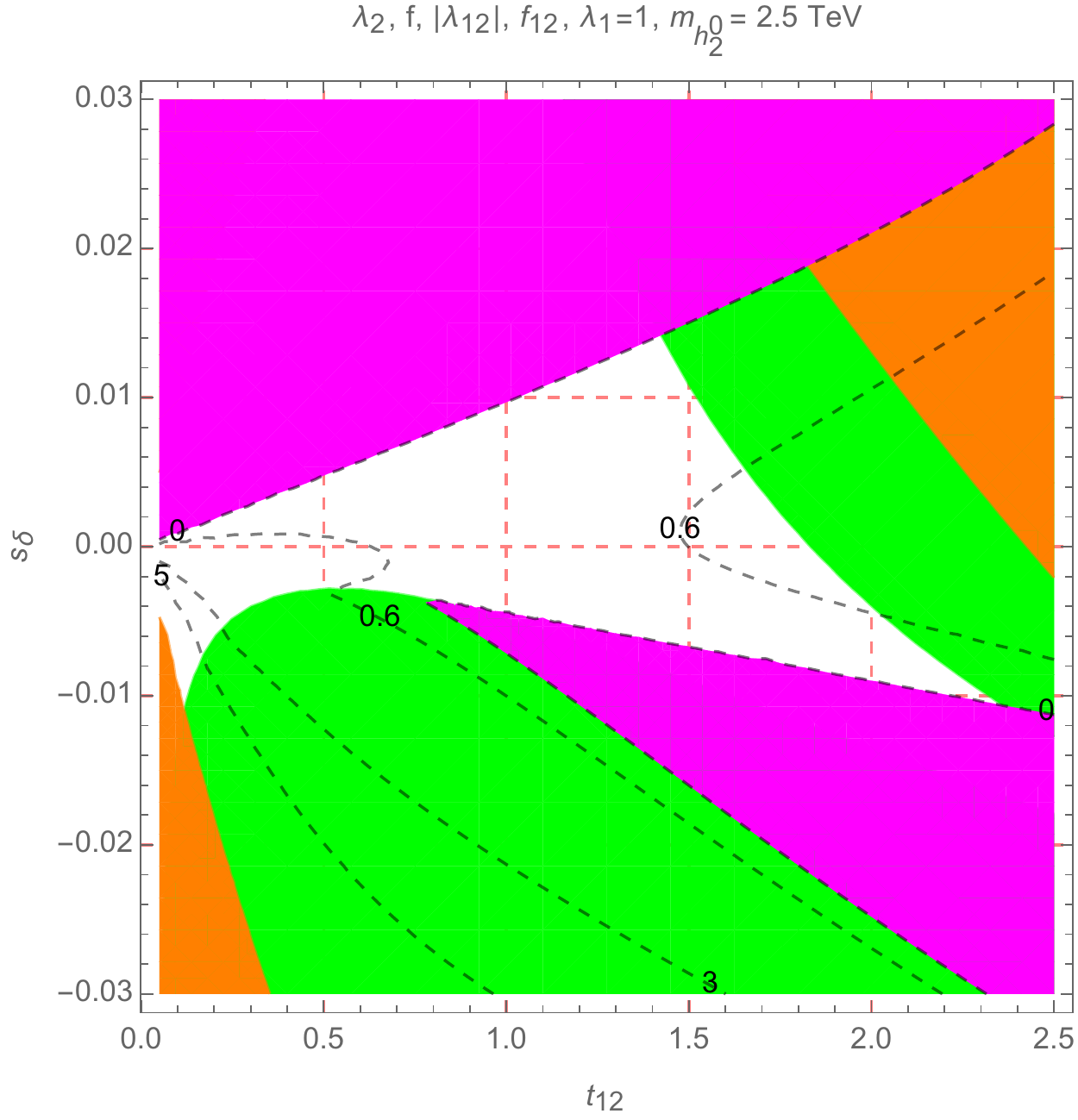} \\
	\end{tabular}%
	\caption{ Contour plots of $\lambda_{2}$, $f$, $|\lambda_{12}|$ and $f_{12}$ as functions of $s_{\theta}$ and $t_{12}$. The green, blue, orange, magenta regions are excluded by requirements that  $0<\lambda_2<10$, $f>0$, $|\lambda_{12}|<10$, and $f_{12}>0$, respectively. Dashed-black curves present constant values of $f_{12}$.}
	\label{fig_lacontour}
\end{figure}

 In general, our scan  shows that allowed $t_{12}$ and $s_{\theta}$ are affected the most strongly by  $m_{h^0_2}$.   As illustration,  the Fig.~\ref{fig_lacontour} presents allowed regions of $t_{12}$ and $s_{\theta}$ with two fixed $m_{h^0_2}=1$ TeV and $2.5$ TeV. 
It can be seen that larger $m_{h^0_2}$ results in smaller allowed  $|s_{\theta}|$.  The  dashed black curves presenting constant values of $f_{12}$ will be helpful for the discussion on the case of $\tilde{\lambda}_{12}<0$. This is because the constraint from $\tilde{f}_{12}>0$ will be more strict than that from $f_{12}>0$ when $\tilde{\lambda}_{12}<0$, namely it will be equivalent to $f_{12}>|\tilde{\lambda}_{12}|$. Hence $f_{12}$ plays role as the upper bound of  $|\tilde{\lambda}_{12}|$.

\begin{figure}[ht]
	\centering
	\begin{tabular}{cc}
	\includegraphics[width=7.5cm]{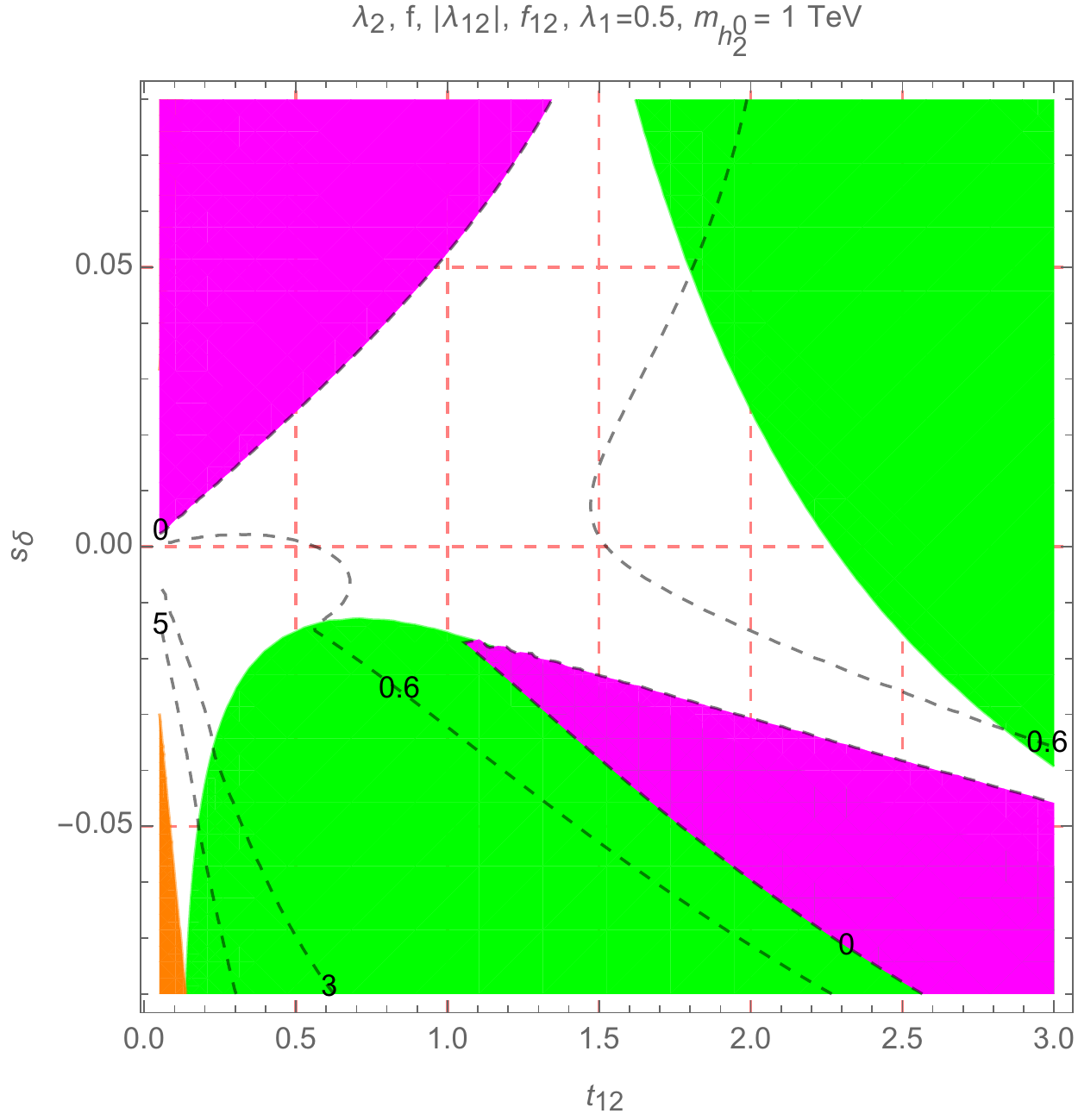} &
		\includegraphics[width=7.5cm]{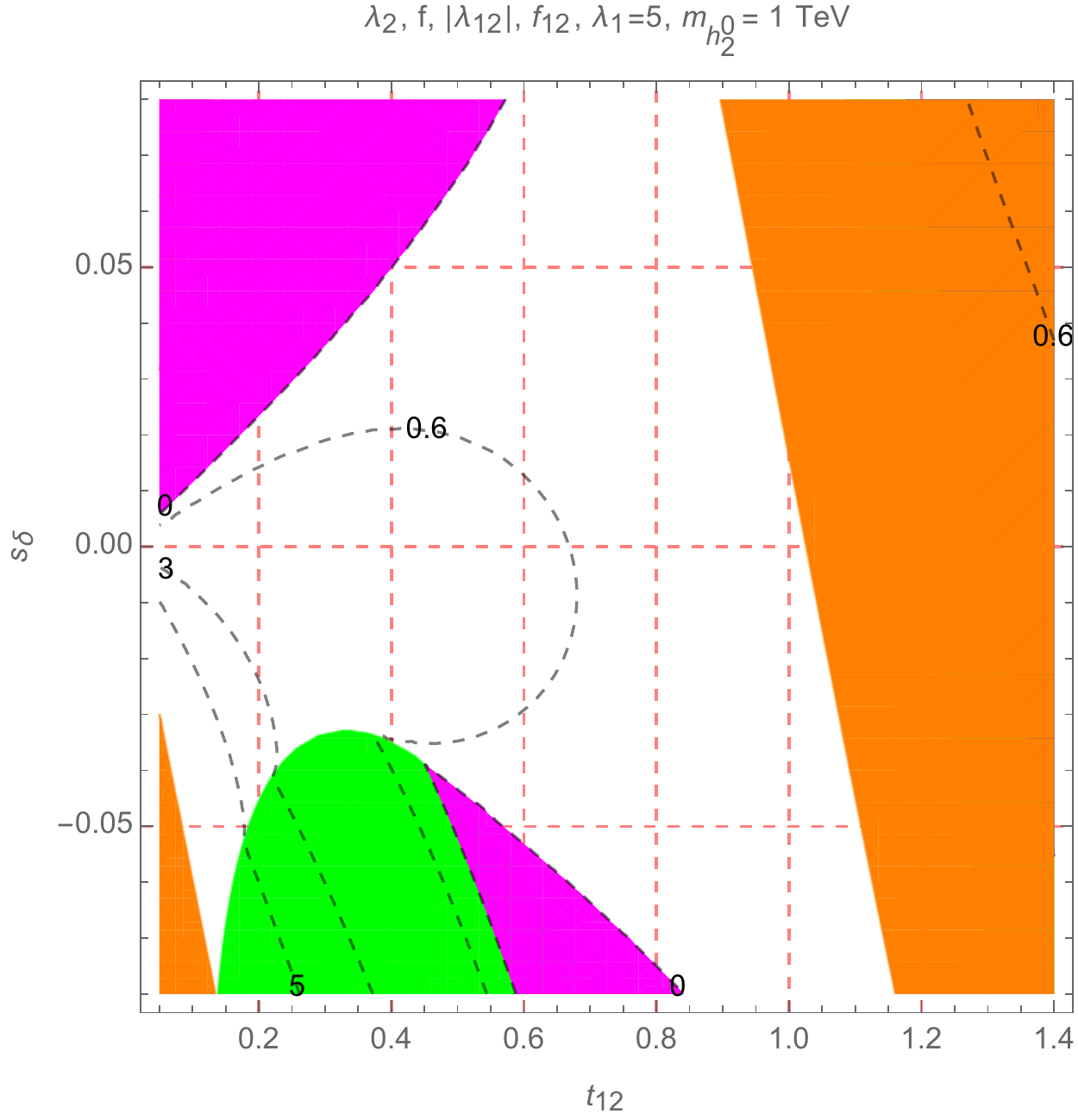} \\
	\end{tabular}%
	\caption{Contour plots of $\lambda_{2}$,  $|\lambda_{12}|$ and $f_{12}$ as functions of $s_{\delta}$ and $t_{12}$. The green, blue, orange, magenta regions are excluded by requirements that $0<\lambda_2<10$, $f>0$, $|\lambda_{12}|<10$, and $f_{12}>0$, respectively. Dashed-black curves presents constant values of $f_{12}$.}
	\label{fig_la1contour}
\end{figure}

The allowed regions also depend on $\lambda_1$, see  contour plots in Fig.~\ref{fig_la1contour} corresponding to $\lambda_1=0.5,5$.
It can be seen that $\lambda_1$ should be large enough to allow large $|s_{\theta}|$, see illustrations in Fig.~\ref{fig_la1contour1} for $\lambda_1=0.1, 10$ in appendix~\ref{app_numerical}.

\begin{figure}[ht]
	\centering
	\begin{tabular}{cc}
		\includegraphics[width=7.5cm]{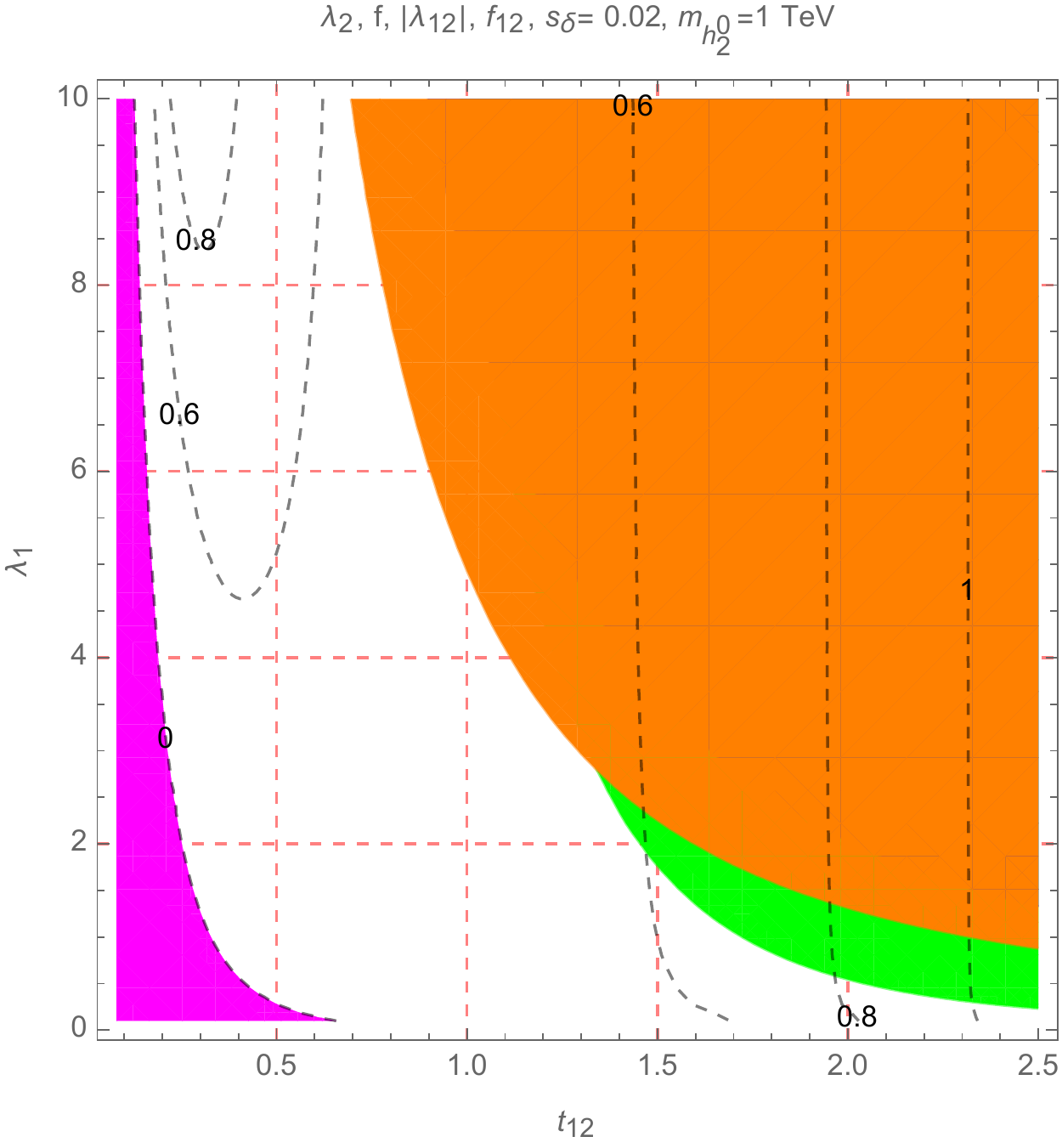} &
		\includegraphics[width=7.5cm]{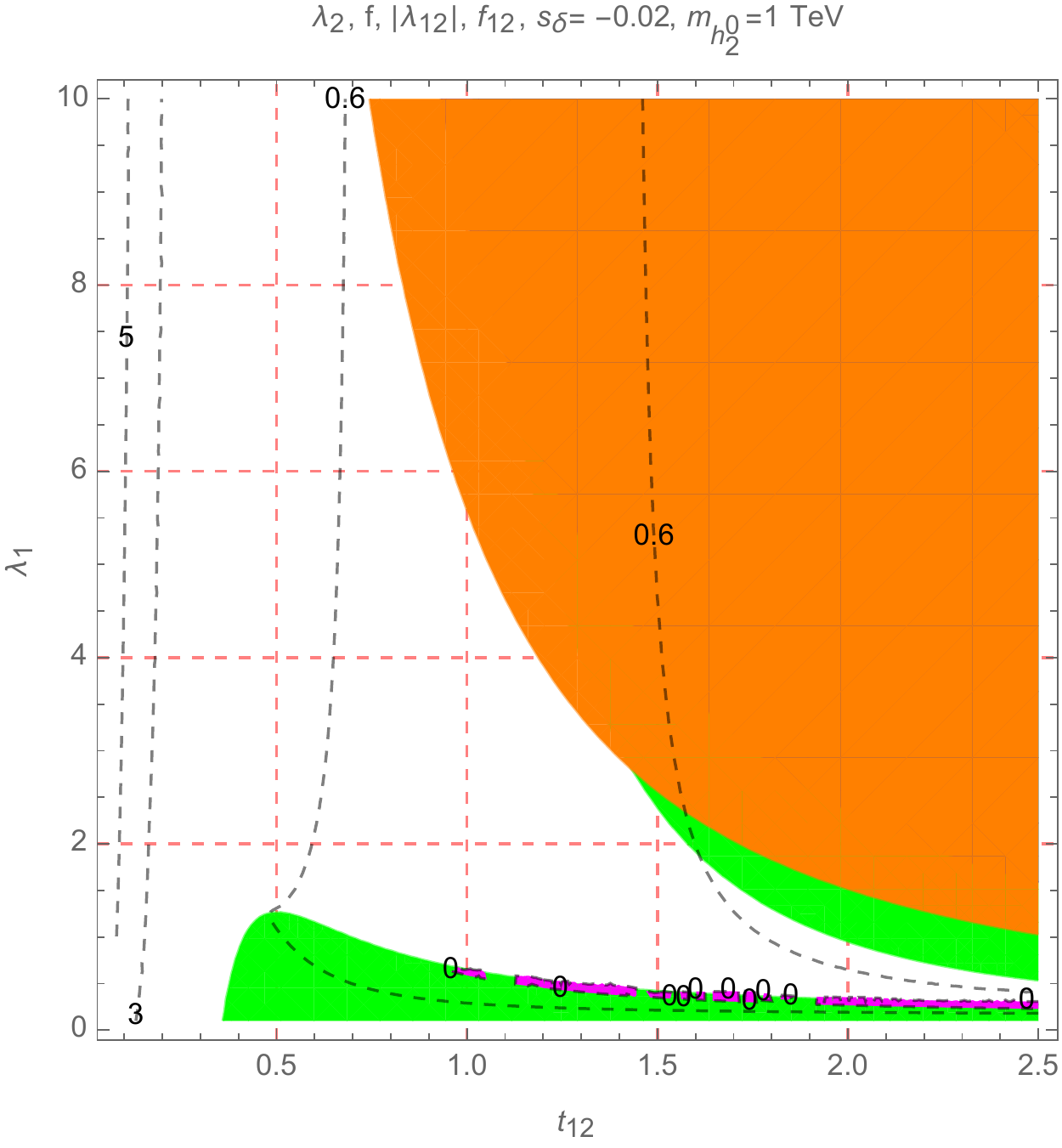} \\
	\end{tabular}%
	\caption{Contour plots of $\lambda_{2}$,  $|\lambda_{12}|$ and $f_{12}$ as functions of $\lambda_1$ and $t_{12}$ with some fixed $m_{h^0_2}$ . The green, blue, orange, magenta regions are excluded by requirements that $0<\lambda_2<10$, $f>0$, $|\lambda_{12}|<10$, and $f_{12}>0$, respectively. Dashed-black curves present constant values of $f_{12}$.}
	\label{fig_contourft12la1}
\end{figure}

In the case of large $|s_{\theta}|=0.02$, the allowed values  $\lambda_1$ and $t_{12}$ are shown in Fig.~\ref{fig_contourft12la1}.
It can be seen that only negative $s_{\theta}$ allows large $f_{12}$. The case of larger $|s_{\theta}|=0.05$ is shown in Fig.~\ref{fig_contourft12la1p} of the appendix~\ref{app_numerical}. We can choose $m_{h^0_2}=1.2$ TeV so that $|s_{\theta}|=0.05$ is still allowed.  Both large $|s_{\delta}|$ and $m_{h^0_2}$ give narrow allowed regions of $t_{12}$ and $\lambda_1$, and small $f_{12}$.  For small $|s_{\delta}|<10^{-2}$,  the allowed values of  $m_{h^0_2}$ and $t_{12}$ will relax. But it will not result in  much deviation from the SM prediction.

\begin{figure}[ht]
	\centering
	\begin{tabular}{cc}
			\includegraphics[width=7.8cm]{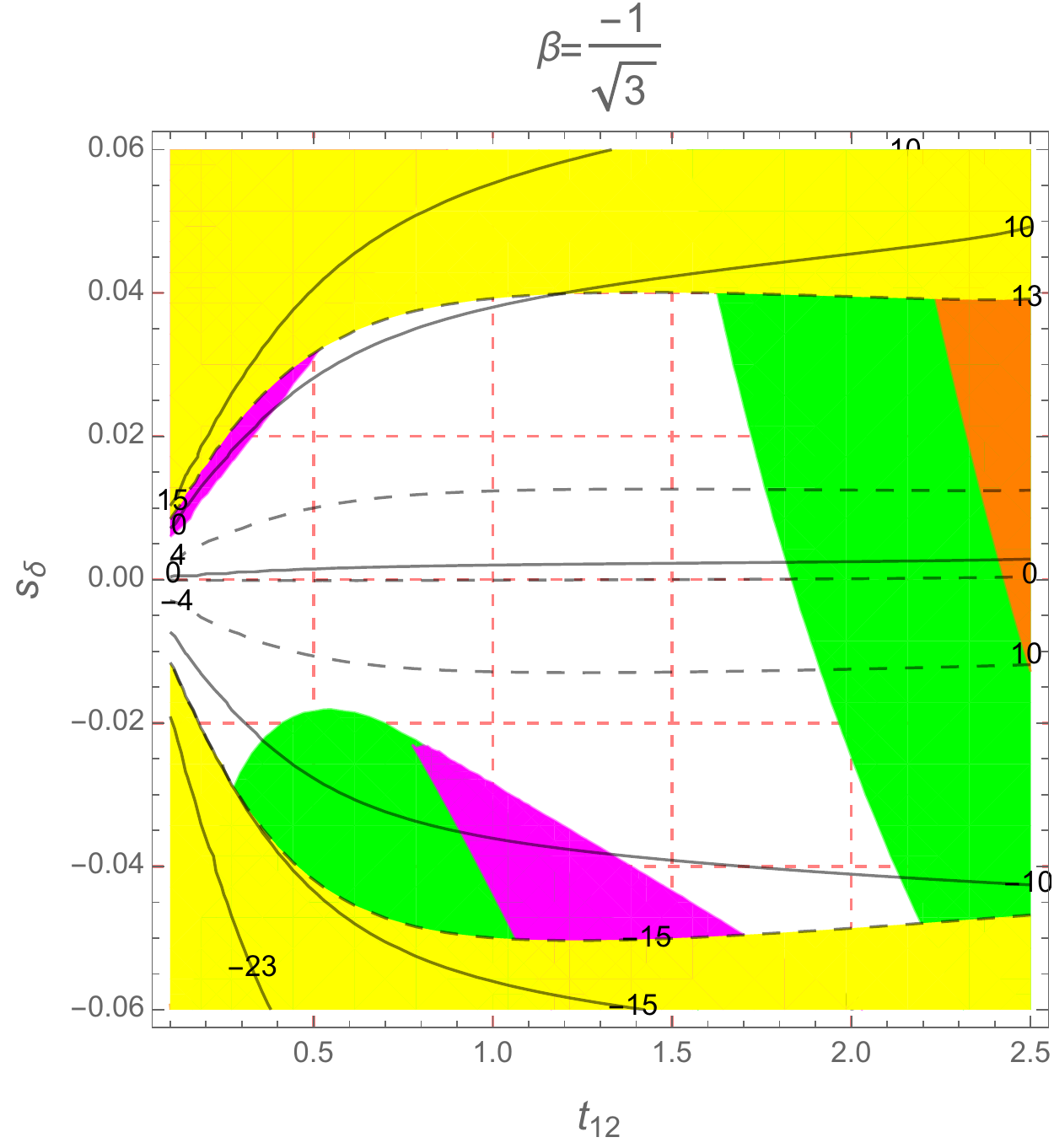} &
		\includegraphics[width=7.8cm]{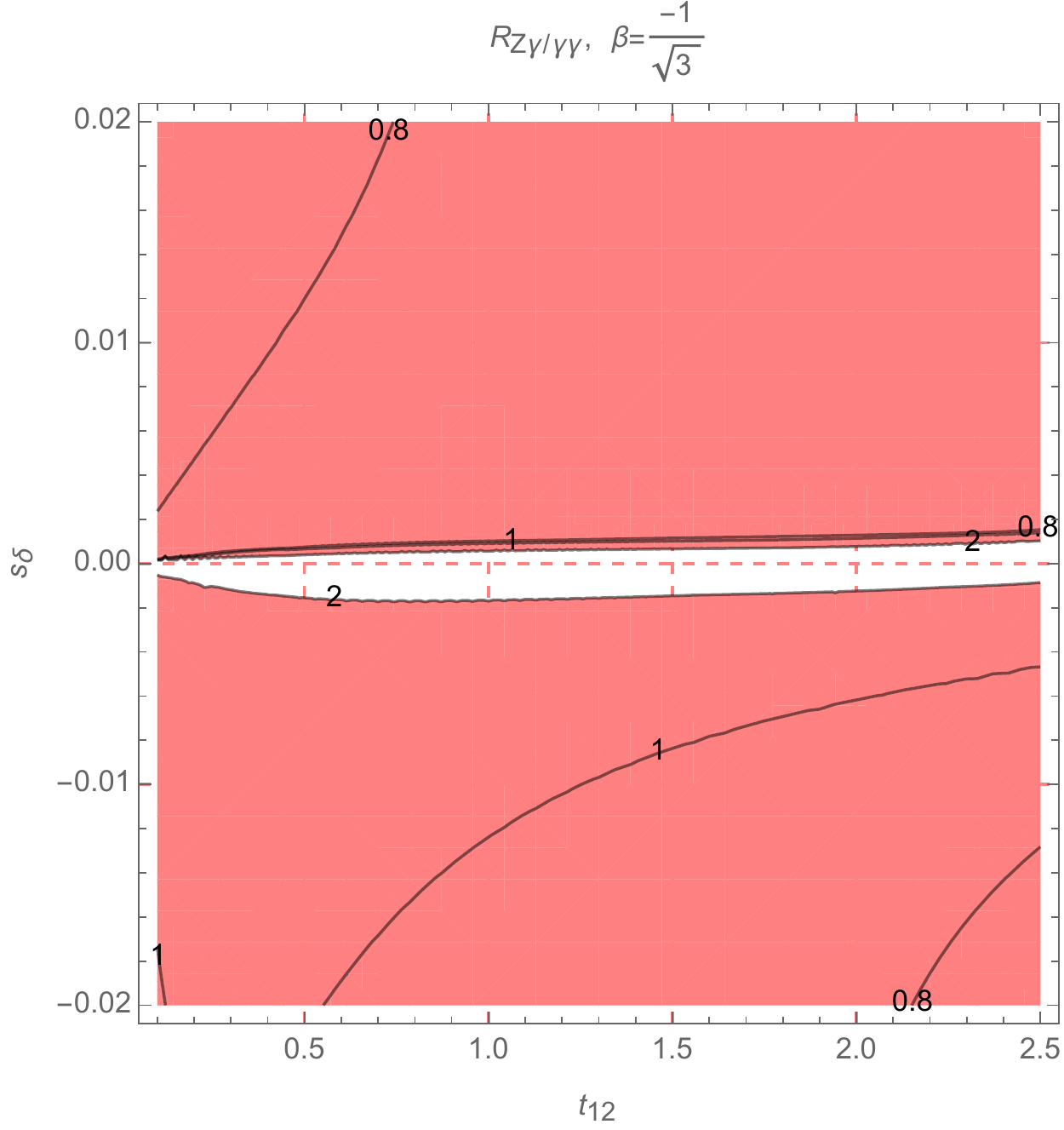} 
\\
	\end{tabular}%
	\caption{Contour plots showing allowed regions of  $s_{\delta}$ and $t_{12}$ (left) and $R_{Z\gamma/\gamma\gamma}$ as a function of $s_{\delta}$ and $t_{12}$. The green, blue, orange, magenta and yellow regions are excluded by valid requirements of $\lambda_2,f,\lambda_{12},f_{12}$, and $\delta\mu_{\gamma\gamma}$, respectively.  The black and dotted black curves show constant values of $\delta\mu_{Z\gamma}$ and $\delta\mu_{\gamma\gamma}$, respectively. The non color region in the right panel corresponds to $R_{Z\gamma/\gamma\gamma} \ge2$.}
	\label{fig_conhto2ga}
\end{figure}

The left panel of Fig.~\ref{fig_conhto2ga} illustrates the contour plots with fixed $\beta=-1/\sqrt{3}$ for allowed values of $\delta\mu_{Z\gamma}$ corresponding to the noncolored✿ regions that satisfy the constraints of parameters and the recent experimental bound on $\delta\mu_{\gamma\gamma}$. 
The right panel of Fig.~\ref{fig_conhto2ga} shows the contour plots of $R_{Z\gamma/\gamma\gamma}$, where the noncolored  region satisfies  $R_{Z\gamma/\gamma\gamma}\ge2$. In this region, we can see that $|s_{\delta}|\sim \mathcal{O}(10^{-3})$ and negative. In addition, $\delta\mu_{\gamma\gamma}<0.04$. Hence, the current constraints $\mu_{\gamma\gamma}=0.99\pm0.14$ predicts $|\delta\mu_{Z\gamma}|<0.15$ which is still smaller than the future sensitivity $\delta\mu_{Z\gamma}=\pm0.23$ mentioned in Ref.~\cite{Cepeda:2019klc}. In addition, most of the allowed regions satisfy  $0.8\le\; R_{Z\gamma/\gamma\gamma} \le\;2$, hence the approximation  Br$(h\rightarrow\gamma\gamma)\simeq \mathrm{Br}(h\rightarrow\;Z\gamma)$ is accepted for simplicity in previous works.

 For large $v_3=14$ TeV and recent uncertainty of the $\delta\mu_{\gamma\gamma}$, our investigation shows generally that the above discussions on the allowed regions as well as $R_{Z\gamma/\gamma\gamma}$ illustrated in Fig.~\ref{fig_conhto2ga} depend weakly  on  $\beta$. The results are also unchanged for lower bound of $v_3=8$ TeV which is allowed for $\beta=\pm2/\sqrt{3}$.   This property can be explained by the fact that, large $v_3\simeq 10$ TeV results in heavy charged gauge bosons $m_{Y},m_V$ having masses around  4 TeV, and the charged Higgs masses being not less than 1 TeV.   As a by-product, one loop contributions from $SU(3)_L$ particles to $F^{331}_{21}$ and $F^{331}_{\gamma\gamma}$ are  at least four orders smaller than the corresponding SM amplitudes $F^\mathrm{SM}_{21,\gamma\gamma}$, illustrations are given  in table~\ref{table_F331numerical}.
 \begin{table}[ht] 
 	\centering 
 	\begin{tabular}{cccccccccc}
 		\hline 
 		$\beta$ & $s_{\delta}$ &$t_{12}$& $\frac{F^{331}_{21,s}}{\mathrm{Re}[F^{\mathrm{SM}}_{21}]}$& $\frac{F^{331}_{21,v}}{\mathrm{Re}[F^{\mathrm{SM}}_{21}]}$& $\frac{F^{331}_{21,sv}}{\mathrm{Re}[F^{\mathrm{SM}}_{21}]}$& $\frac{F^{331}_{\gamma\gamma,s}}{\mathrm{Re}[F^{\mathrm{SM}}_{\gamma\gamma}]}$& $\frac{F^{331}_{\gamma\gamma,v}}{\mathrm{Re}[F^{\mathrm{SM}}_{\gamma\gamma}]}$ & $\delta\mu_{Z\gamma}$& $\delta\mu_{\gamma\gamma}$\\
 		\hline 
 		$\frac{2}{\sqrt{3}}$& $2\times 10^{-2}$& $1.5$& $-3.3\times 10^{-4}$ &$3\times 10^{-5}$& $-1.6 \times 10^{-4}$& $-6\times 10^{-4}$& $5.5\times 10^{-4}$ & $4.4$& $6.5$\\
 		$\frac{2}{\sqrt{3}}$& $-2\times 10^{-2}$& $1.5$& $ \sim  10^{-6}$ &$3\times 10^{-5}$& $-1.5 \times 10^{-4}$& $\sim 10^{-6}$& $5.3\times 10^{-4}$ & $-5.4$& $-6$\\
 		$\frac{2}{\sqrt{3}}$& $2\times 10^{-2}$& $0.5$& $ 1.3\times 10^{-4}$ &$-9\times 10^{-5}$& $-5 \times 10^{-5}$& $2.3\times 10^{-4}$& $2.2\times 10^{-4}$ & $6.8$& $8.1$\\
 		$\frac{2}{\sqrt{3}}$& $-2\times 10^{-2}$& $0.5$& $ -4.2\times 10^{-4}$ &$-9\times 10^{-5}$& $-4 \times 10^{-5}$& $-7.5\times 10^{-4}$& $2.1\times 10^{-4}$ & $-7.5$& $-7.4$\\
 		$\frac{2}{\sqrt{3}}$& $ -10^{-3}$& $1.5$& $ -1.6\times 10^{-4}$ &$3\times 10^{-5}$& $-1.6 \times 10^{-4}$& $-2.9\times 10^{-4}$& $5.4\times 10^{-4}$ & $-0.8$& $-0.2$\\
 		\hline 
 	\end{tabular}
 	\caption{Numerical contributions of $SU(3)_L$ particles to $F^{331}_{21}$ and $F^{331}_{\gamma\gamma}$, see Eqs.~\eqref{eq_F21331be} and ~\eqref{eq_F2gamma}, where $F^{331}_{21,sv}\equiv F^{331}_{21,svv} +F^{331}_{21,vss}$.}\label{table_F331numerical}
 \end{table} 
 Here we use the SM amplitudes predicted by the SM, namely  $\mathrm{Re}[F^{\mathrm{SM}}_{21}]=-5.6\times 10^{-5} \, [\mathrm{GeV}^{-1}]$ and $\mathrm{Re}[F^{\mathrm{SM}}_{\gamma\gamma}]= -3.09 \times 10^{-5}\, [\mathrm{GeV}^{-1}]$, and ignore the tiny imaginary parts. We can see that both $\delta\mu_{Z\gamma}$ and $\delta\mu_{\gamma\gamma}$ depend strongly on $s_{\delta}$ and $t_{12}$. In contrast, the one-loop contributions from new particles are suppressed, as shown in the last line  in table~\ref{table_F331numerical}: suppressed $s_{\delta}$ results in  $|\delta\mu_{Z\gamma}|\simeq 4 |\delta\mu_{\gamma\gamma}|=0.8\%\ll4\%$, which is even much smaller than the expected sensitivity of $\delta\mu_{\gamma\gamma}=4\%$. Anyway, it can be noted that $F^{331}_{21,sv}$ may significantly larger than $F^{331}_{21,v}$, hence  both of them should be included simultaneously into the decay amplitude $h\rightarrow\;Z\gamma$ in general. Suppressed contributions of new particles to $\delta\mu_{Z\gamma,}$ are shown explicitly in the left panel of Fig.~\ref{fig_conhto2ga}, where three constant curves  $s_{\delta}=\delta\mu_{Z\gamma} =\delta\mu_{\gamma\gamma}=0$ are very close together. 
 
 For large and positive $\tilde{\lambda}_{12}$ and small $m_{h^0_2}$,   one loop contributions from $H^{\pm}$ to $F^{331}_{21}$ and $F^{331}_{\gamma\gamma}$ are dominant but still not large enough to give significant deviations to $\delta\mu_{Z\gamma}$, see an illustration with suppressed $s_{\delta}=10^{-3}$ in the first line of  table~\ref{table_F331numerical2}.
 \begin{table}[ht] 
	\centering 
	\begin{tabular}{cccccccccc}
		\hline 
		$\beta$ & $s_{\delta}$ &$t_{12}$& $\frac{F^{331}_{21,s}}{\mathrm{Re}[F^{\mathrm{SM}}_{21}]}$& $\frac{F^{331}_{21,v}}{\mathrm{Re}[F^{\mathrm{SM}}_{21}]}$& $\frac{F^{331}_{21,sv}}{\mathrm{Re}[F^{\mathrm{SM}}_{21}]}$& $\frac{F^{331}_{\gamma\gamma,s}}{\mathrm{Re}[F^{\mathrm{SM}}_{\gamma\gamma}]}$& $\frac{F^{331}_{\gamma\gamma,v}}{\mathrm{Re}[F^{\mathrm{SM}}_{\gamma\gamma}]}$ & $\delta\mu_{Z\gamma}$& $\delta\mu_{\gamma\gamma}$\\
		\hline 
		$\frac{2}{\sqrt{3}}$& $10^{-3}$& $1.7$& $-1.46\times 10^{-2}$ &$4\times 10^{-5}$ & $-1.7 \times 10^{-4}$& $-2.64\times 10^{-2}$& $5.7\times 10^{-4}$ & $-3.1$& $-4.7$\\
		$\frac{2}{\sqrt{3}}$& $-10^{-3}$& $1.7$& $-1.44\times 10^{-2}$ &$4\times 10^{-5}$ & $-1.7 \times 10^{-4}$& $-2.61\times 10^{-2}$& $5.7\times 10^{-4}$ & $-3.6$& $-5.3$\\
		$\frac{2}{\sqrt{3}}$& $3\times 10^{-2}$& $1.5$& $-1.24\times 10^{-2}$ &$3\times 10^{-5}$& $-1.6 \times 10^{-4}$& $-2.23\times 10^{-2}$& $5.5\times 10^{-4}$ & $4.4$& $5.2$\\
		$\frac{2}{\sqrt{3}}$& $-3\times 10^{-2}$& $1.5$& $-9.6\times 10^{-3}$ &$3\times 10^{-5}$& $-1.5 \times 10^{-4}$& $-1.75\times 10^{-3}$& $5.3\times 10^{-4}$ & $-9.6$& $-12.3$\\
			\hline 
	\end{tabular}
	\caption{Numerical contributions of $SU(3)_L$ particles to $F^{331}_{21}$ and $F^{331}_{\gamma\gamma}$ for large $\tilde{\lambda}_{12}=5$ and small $m_{h^0_2}=600$ GeV.}\label{table_F331numerical2}
\end{table} 
Here we always force $|\delta\mu_{\gamma\gamma}|\le4\%$ being the future sensitive of $\mu_{\gamma\gamma}$. On the other hand, large deviations can result from large $|s_{\delta}|$. In this case, all $\delta\mu_{Z\gamma,\gamma\gamma}$ and $s_{\delta}$ have the same signs.

\begin{figure}[ht]
 	\centering
 	\begin{tabular}{cc}
 		\includegraphics[width=7.8cm]{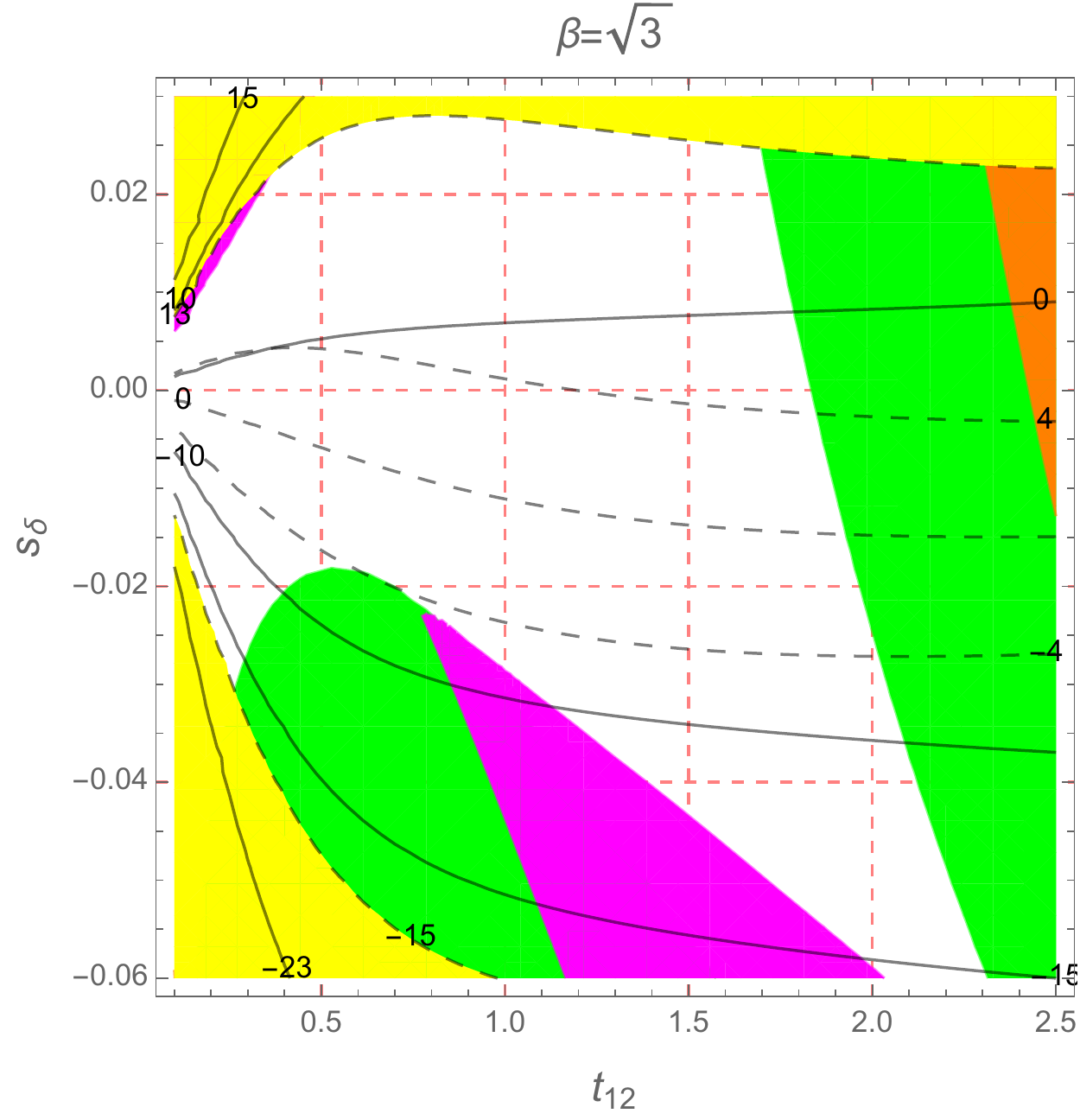}
 		\\
 	\end{tabular}%
 	\caption{Contour plots showing allowed regions of  $s_{\delta}$ and $t_{12}$ with $v_3=3$ TeV following Ref.~\cite{Coriano:2018coq}. The green, blue,  orange, magenta and yellow regions are excluded by valid requirements of $\lambda_2,f,\lambda_{12},f_{12}$, and $\delta\mu_{\gamma\gamma}$, respectively.  The black and dotted black curves show constant values of $\delta\mu_{Z\gamma}$ and $\delta\mu_{\gamma\gamma}$, respectively.}
 	\label{fig_conhto2gabe3}
 \end{figure}

 Regarding $\beta=\sqrt{3}$ corresponding to the model discussed in Ref.~\cite{Coriano:2018coq}, where $v_3=3$ TeV is still accepted, the allowed regions change significantly, as illustrated in Fig.~\ref{fig_conhto2gabe3}.
In particularly, the model gives more strict positive $s_{\delta}<0.03$.  One-loop contributions  from $SU(3)_L$ particles can give deviations up to few percent for both $\delta\mu_{Z\gamma}$, $\delta\mu_{\gamma\gamma}$, as shown in Fig.~\ref{fig_conhto2gabe3} that the two contours $\delta_{Z\gamma}=\delta\mu_{\gamma\gamma}=0$ distinguish with  the line $s_{\delta}=0$.  Interesting numerical values are illustrated in table~\ref{table_F331numericalbe3}.
\begin{table}[ht] 
	\centering 
	\begin{tabular}{cccccccccc}
		\hline 
		$\beta$ & $s_{\delta}$ &$t_{12}$& $\frac{F^{331}_{21,s}}{\mathrm{Re}[F^{\mathrm{SM}}_{21}]}$& $\frac{F^{331}_{21,v}}{\mathrm{Re}[F^{\mathrm{SM}}_{21}]}$& $\frac{F^{331}_{21,sv}}{\mathrm{Re}[F^{\mathrm{SM}}_{21}]}$& $\frac{F^{331}_{\gamma\gamma,s}}{\mathrm{Re}[F^{\mathrm{SM}}_{\gamma\gamma}]}$& $\frac{F^{331}_{\gamma\gamma,v}}{\mathrm{Re}[F^{\mathrm{SM}}_{\gamma\gamma}]}$ & $\delta\mu_{Z\gamma}$& $\delta\mu_{\gamma\gamma}$\\
		\hline 
		$\sqrt{3}$& $ 10^{-3}$& $1.5$& $-1.8\times 10^{-4}$ &$-1.6\times 10^{-3}$& $-4 \times 10^{-3}$& $-3.2\times 10^{-4}$& $2.2\times 10^{-2}$ & $-1.6$& $4.8$\\
		$\sqrt{3}$& $ -10^{-3}$& $1.5$& $-1.6\times 10^{-4}$ &$-1.7\times 10^{-3}$& $-4 \times 10^{-3}$& $-2.9\times 10^{-4}$& $2.2\times 10^{-2}$ & $-2$& $4.2$\\
			\hline 
	\end{tabular}
	\caption{For model in Ref.~\cite{Coriano:2018coq}, numerical contributions of $SU(3)_L$ particles to $F^{331}_{21}$ and $F^{331}_{\gamma\gamma}$. Notations are given from caption of  table~\ref{table_F331numerical}.}\label{table_F331numericalbe3}
\end{table} 
We emphasize two important properties. First, one loop contributions from gauge $SU(3)_L$ bosons are dominant, which can give $\delta\mu_{\gamma\gamma}$ to reach the future sensitivity.  Values of $F^{331}_{21,v}$ and $F^{331}_{21,sv}$ can have the same order of $10^{-3}$ compared with the SM part, but  these contributions are not large enough to result in large deviation of $|\delta\mu_{Z\gamma}|> 23\%$.    

To finish the case of $\lambda_{ij}>0$ we mentioned above, we see that in this case all of the charged Higgs boson masses are order of $\mathrm{O}(1)$TeV and they have small couplings with $h$. For large $\lambda_1$, $\lambda_{12}$ and small $m_{h^0_2}=800$ GeV,  there may give small $\delta\mu_{\gamma\gamma}$ but large $\delta\mu_{Z\gamma}$, see examples in table~\ref{table_F331numericalbe3la}.
\begin{table}[ht] 
	\centering 
	\begin{tabular}{ccccccccccc}
		\hline 
		$\lambda_1$ & $\tilde{\lambda}_{12}$ &$s_{\delta}$ &$t_{12}$& $\frac{F^{331}_{21,s}}{\mathrm{Re}[F^{\mathrm{SM}}_{21}]}$& $\frac{F^{331}_{21,v}}{\mathrm{Re}[F^{\mathrm{SM}}_{21}]}$& $\frac{F^{331}_{21,sv}}{\mathrm{Re}[F^{\mathrm{SM}}_{21}]}$& $\frac{F^{331}_{\gamma\gamma,s}}{\mathrm{Re}[F^{\mathrm{SM}}_{\gamma\gamma}]}$& $\frac{F^{331}_{\gamma\gamma,v}}{\mathrm{Re}[F^{\mathrm{SM}}_{\gamma\gamma}]}$ & $\delta\mu_{Z\gamma}$& $\delta\mu_{\gamma\gamma}$\\
		\hline 
		$1.95$& $8$ & $ 10^{-3}$& $1.5$& $-1.22\times 10^{-2}$ &$-1.7\times 10^{-3}$& $-4.4 \times 10^{-3}$& $-2.21\times 10^{-2}$& $2.2\times 10^{-2}$ & $-4.$& $0.4$\\
		$1.95$& $8$ & $-10^{-3}$& $1.5$& $-1.21\times 10^{-2}$ &$-1.7\times 10^{-3}$& $-4.4 \times 10^{-3}$& $-2.19\times 10^{-2}$& $2.2\times 10^{-2}$ & $-4.5$& $-0.17$\\
		$1.$& $5$ & $-2\times 10^{-2}$& $1.95$& $-1.6\times 10^{-3}$ &$-9.7\times 10^{-4}$& $-5 \times 10^{-3}$& $-1.2\times 10^{-2}$& $2.4\times 10^{-2}$ & $-7.7$& $-4$\\
		\hline 
	\end{tabular}
	\caption{For model in Ref.~\cite{Coriano:2018coq}, numerical contributions of $SU(3)_L$ particles to $F^{331}_{21}$ and $F^{331}_{\gamma\gamma}$ with $m_{h^0_2}=800$ GeV. Notations are given from caption of  table~\ref{table_F331numerical}.}\label{table_F331numericalbe3la}
\end{table} 
We  stress here an interesting point that with the existence of new Higgs and gauge bosons, their contributions  $F^{331}_{\gamma\gamma,s}$ and $F^{331}_{\gamma\gamma,v}$ to the decay amplitude $h\rightarrow \gamma\gamma$ may be destructive and the same order, hence keep the respective signal strength satisfying the small experimental constraint. Simultaneously, all of the contributions to the decay amplitude $h \rightarrow\,Z\gamma$ are constructive so that the deviation can be large. For the model with $\beta= \sqrt{3}$ and $v_3=3$ TeV, we can find this deviation can reach around $-10$, but this values is still far from the expected sensitive $\delta\mu_{Z\gamma}=\pm23\%$ in the  HL-LHC project.   For the models with $v_3\ge 8$ TeV, heavy gauge contributions are suppressed, hence large contribution from charged Higgs bosons is dominant. Then, the constraint from $\delta\mu_{\gamma\gamma}$ will give more strict constraint on $\delta\mu_{Z\gamma}$ than that obtained from the experiments. 

\subsubsection{Case 2:  $\tilde{\lambda}_{12}<0$. }
As we can see in eq.~\eqref{scHigg}, negative $\tilde{\lambda}_{12}$ may result in small charged Higgs mass $m_{H^{\pm}}$. In addition, large $|\tilde{\lambda}_{12}|$ may give large coupling of this Higgs boson with the SM-like one, leading to large $|F^{331}_{21,s}|$ and $|F^{331}_{\gamma\gamma,s}|$. We will focus on this interesting case. 

One of the conditions given in \eqref{eq_laconstraint}, namely  $\tilde{f}_{12}>0$,  will   automatically satisfy  if $f_{12}>0$ and  $\tilde{\lambda}_{12}\ge0$. In the case of $\tilde{\lambda}_{12}<0$, the inequality $\tilde{f}_{12}>0$ is equivalent to the more strict condition  $f_{12}>|\tilde{\lambda}_{12}|>0$ or $-f_{12}<\tilde{\lambda}_{12}<0$. This helps us determine the allowed regions with large $|\tilde{\lambda}_{12}|$, which give large one-loop contributions of charged Higgs boson $H^{\pm}$ to the two decay amplitudes $h\rightarrow\,Z\gamma,\gamma\gamma$. Based on the fact that allowed regions with large positive $f_{12}$ will allow large $\tilde{\lambda}_{12}$, two Figs. \ref{fig_lacontour} and \ref{fig_la1contour} show that large $\tilde{\lambda}_{12}$ corresponds to regions   having negative $s_{\delta}$ and small $t_{12}$. Small $s_{\theta}$ allows small $|\tilde{\lambda}_{12}|$. The Fig.~\ref{fig_contourft12la1} shows that values of $\lambda_1$ seems not affect allowed $\tilde{\lambda}_{12}$ in the regions of negative $s_{\delta}$.  For large $v_3$, large  $|\tilde{\lambda}_{12}|$ in this case  does not affect significantly on both $\delta\mu_{Z\gamma,\gamma\gamma}$ see illustration in table~\ref{table_F331numerical1}. 
\begin{table}[ht] 
	\centering 
	\begin{tabular}{cccccccccc}
		\hline 
		$\beta$ & $m_{h^0_2}$ [TeV] &$s_{\delta}$& $\frac{F^{331}_{21,s}}{F^{\mathrm{SM}}_{21}}$& $\frac{F^{331}_{21,v}}{F^{\mathrm{SM}}_{21}}$& $\frac{F^{331}_{21,sv}}{F^{\mathrm{SM}}_{21}}$& $\frac{F^{331}_{\gamma\gamma,s}}{F^{\mathrm{SM}}_{\gamma\gamma}}$& $\frac{F^{331}_{\gamma\gamma,v}}{F^{\mathrm{SM}}_{\gamma\gamma}}$ & $\delta\mu_{Z\gamma}$& $\delta\mu_{\gamma\gamma}$\\
		$\frac{2}{\sqrt{3}}$& $1$  & $-10^{-3}$ & $3.4\times 10^{-4}$ & $-1.4\times 10^{-4}$& $\simeq0$ & $6.1\times 10^{-3}$ & $9\times 10^{-5}$ & $-1.8$ & $-1.2$\\
		$\frac{2}{\sqrt{3}}$& $0.6$  & $-10^{-3}$ & $1.2\times 10^{-3}$ &  $- 1.4\times 10^{-4}$ &  $\simeq0$ & $-2.1 \times 10^{-3}$ & $9\times 10^{-5}$ & $-1.7$ & $-0.9$ \\
		$\frac{2}{\sqrt{3}}$& $1$  & $-2\times 10^{-2}$ & $-1.4\times 10^{-3}$ & $-1.4\times 10^{-4}$ & $\simeq0$&  $-2.6 \times 10^{-3}$ & $9\times 10^{-5}$& $-23.9$ & $-23.9$ \\
		$\frac{2}{\sqrt{3}}$& $0.6$  & $-2\times 10^{-2}$ & $6.6\times 10^{-4}$ & $-1.2\times 10^{-3}$ & $\simeq0$&  $-1.2 \times 10^{-3}$ & $9\times 10^{-5}$& $-23.7$ & $-23.6$ \\ 		 		
		\hline 
	\end{tabular}
	\caption{Numerical contributions of $SU(3)_L$ particles to $F^{331}_{21}$ and $F^{331}_{\gamma\gamma}$. Numerical  fixed values of unknown parameters are: $\beta=2/\sqrt{3}$, $t_{12}= 0.1$, $\tilde{\lambda}_{12}=-1$.}\label{table_F331numerical1}
\end{table}
Hence, the case of negative $\tilde{\lambda}_{12}$ may results in light charged Higgs boson $H^{\pm}$, but it does not support large one-loop contributions from charged Higgs mediation to decay amplitudes $h\rightarrow\,Z\gamma,\gamma\gamma$.

Regarding the model with $\beta=\sqrt{3}$ discussed in Ref.~\cite{Coriano:2018coq}, the main difference is the small $v_3=3 $ TeV, leading to a significant effect of heavy gauge bosons to the one-loop contributions  $|F^{331}_{21,v}/F^{\mathrm{SM}}_{21}|,\, |F^{331}_{21,sv}/F^{\mathrm{SM}}_{21}|,\, |F^{331}_{21,s}/F^{\mathrm{SM}}_{21}|,\,  |F^{331}_{\gamma\gamma,v}/F^{\mathrm{SM}}_{\gamma\gamma}|,\,   |F^{331}_{\gamma\gamma,s}/F^{\mathrm{SM}}_{\gamma\gamma}|\sim \mathcal{O}(10^{-2})$. But with $\tilde{\lambda}_{12}<0$,  constructive contributions appear in the decay amplitude $h\rightarrow\gamma\gamma$, while destructive contributions appear in the decay amplitude $h\rightarrow\,Z\gamma$. Hence, the constraint from experimental data of the decay $h\rightarrow \gamma\gamma$ predicts smaller deviation of the $\mu_{Z\gamma}$ than that corresponding to $\tilde{\lambda}_{12}>0$. 

To finish, from above discussion we emphasize that in other gauge extensions of the SM such as the $SU(2)_1\otimes SU(2)_2\otimes U(1)_Y$ models, which  still allow low values of new gauge and charged Higgs bosons masses~\cite{Boucenna:2016qad,Hue:2016nya,He:2017bft,Yue:2019kky,Abdullah:2018ets}, the contributions like $F_{12,sv}$ may be as large as usual ones, hence it should be included in the decay amplitude $h\rightarrow\,Z\gamma$. In addition, these models may predict large $\delta\mu_{Z\gamma}$, which also satisfies  $|\delta\mu_{\gamma\gamma}|\leq0.04$. This interesting topic deserves to be paid attention more detailed.

\subsection{$h^0_3$ decays as a signal of the $331\beta$ model}
\begin{figure}[ht]
	\centering
	\begin{tabular}{cc}
		\includegraphics[width=7.8cm]{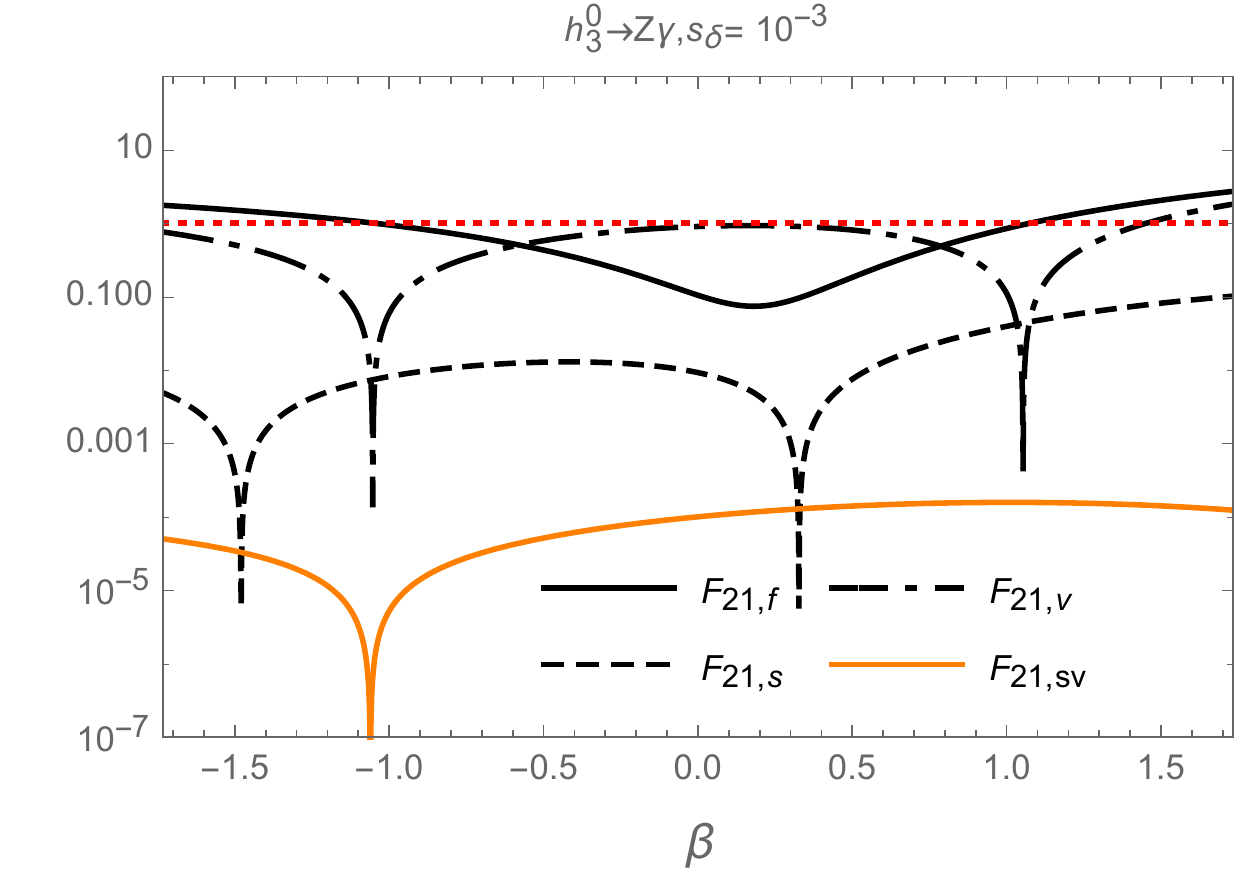} &
		\includegraphics[width=7.8cm]{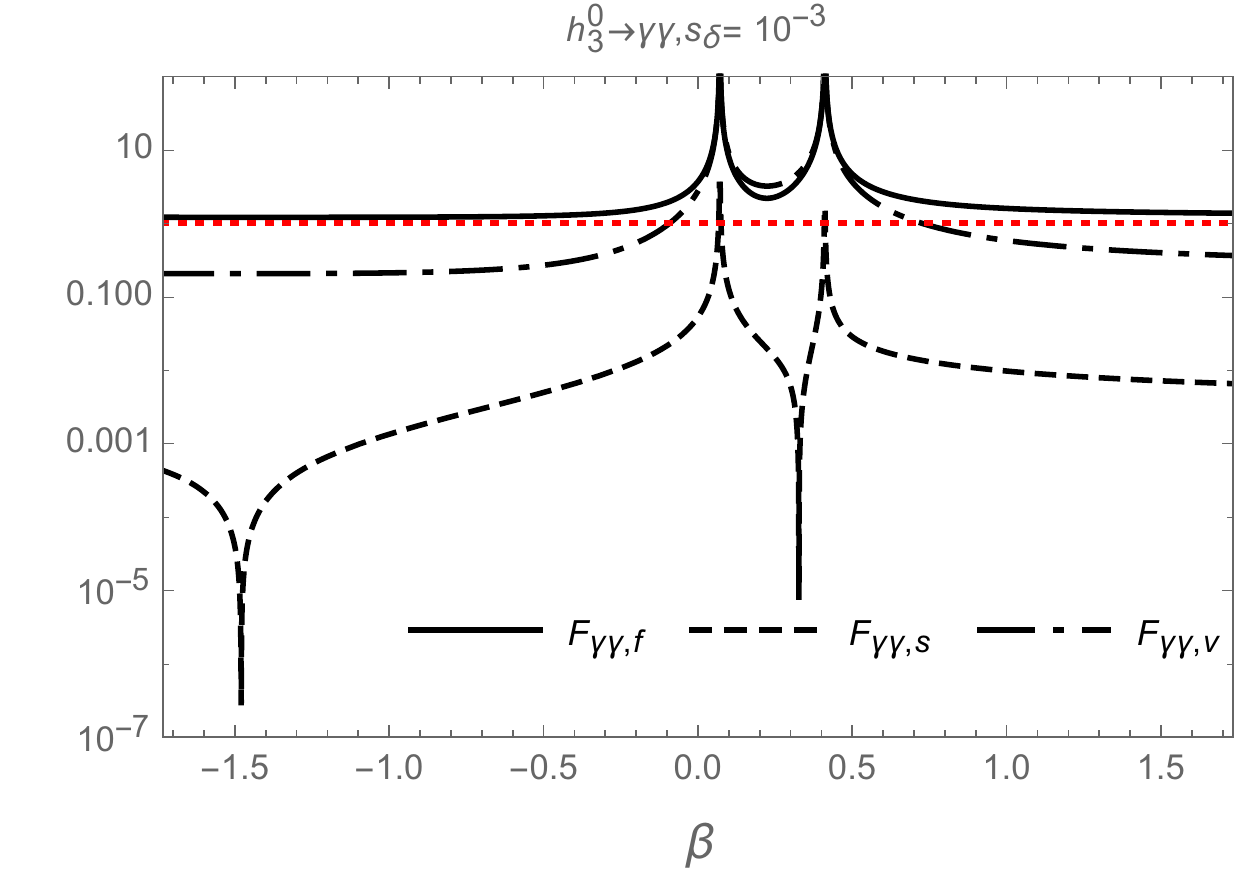} \\
	\end{tabular}%
	\caption{Different contributions to loop-induced decays $h^0_3\rightarrow \gamma\gamma,Z\gamma$ as functions of $\beta$.}
	\label{fig_h03F21ga}
\end{figure}

Different contributions to loop-induced decays $h^0_3\rightarrow \gamma\gamma,Z\gamma$ with small  $s_{\theta}=10^{-3}$, $m_{h^0_3}=700$ GeV, $t_{12}=0.8$ are illustrated in Fig.~\ref{fig_h03F21ga}, where the ratios $|F_{21,x}(h^0_3\rightarrow\,Z\gamma)|/|F_{21}(h^0_3\rightarrow\,Z\gamma)|$ and $|F_{\gamma\gamma,x}(h^0_3\rightarrow\,Z\gamma)|/|F_{\gamma\gamma}(h^0_3\rightarrow\,Z\gamma)|$ are presented, $x=f,s,v,sv$. 
In addition, our scan shows  that the curves in the Fig.~\ref{fig_h03F21ga} do not sensitive with the changes of  $s_{\delta}$.  We can conclude that contributions from heavy exotic fermions are alway dominant  for large $\beta$. While $F_{21,sv}$ is suppressed.  For the decay $h^0_3\rightarrow \gamma\gamma$, the destructive correlation between $F_{\gamma\gamma,v}$ and $F_{\gamma\gamma,f}$ happens with small $|\beta|$. This results in two peaks in the figure, where $|F_{\gamma\gamma}|\ll |F_{\gamma\gamma,f}|, |F_{\gamma\gamma,v}|$.

\begin{figure}[ht]
	\centering
	\begin{tabular}{cc}
		\includegraphics[width=7.8cm]{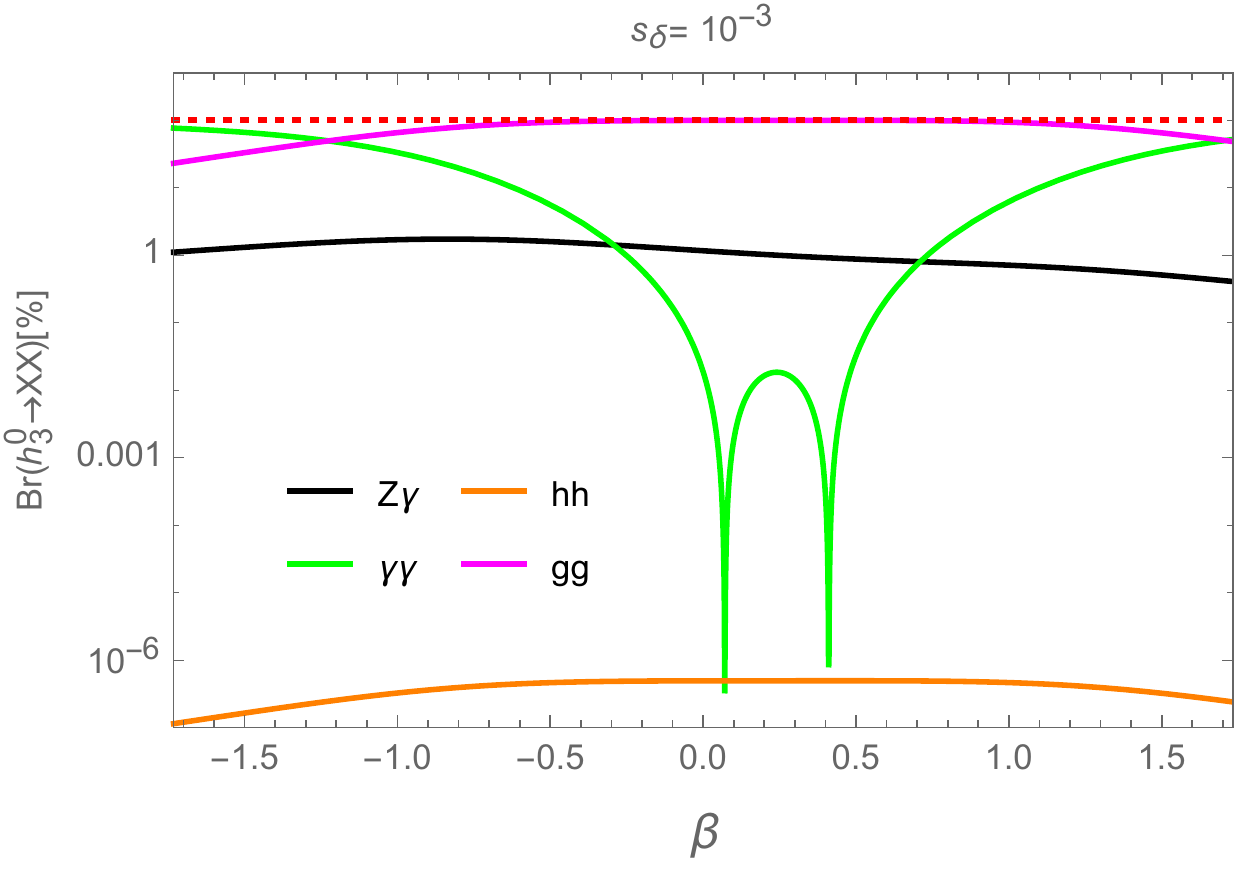} &
		\includegraphics[width=7.8cm]{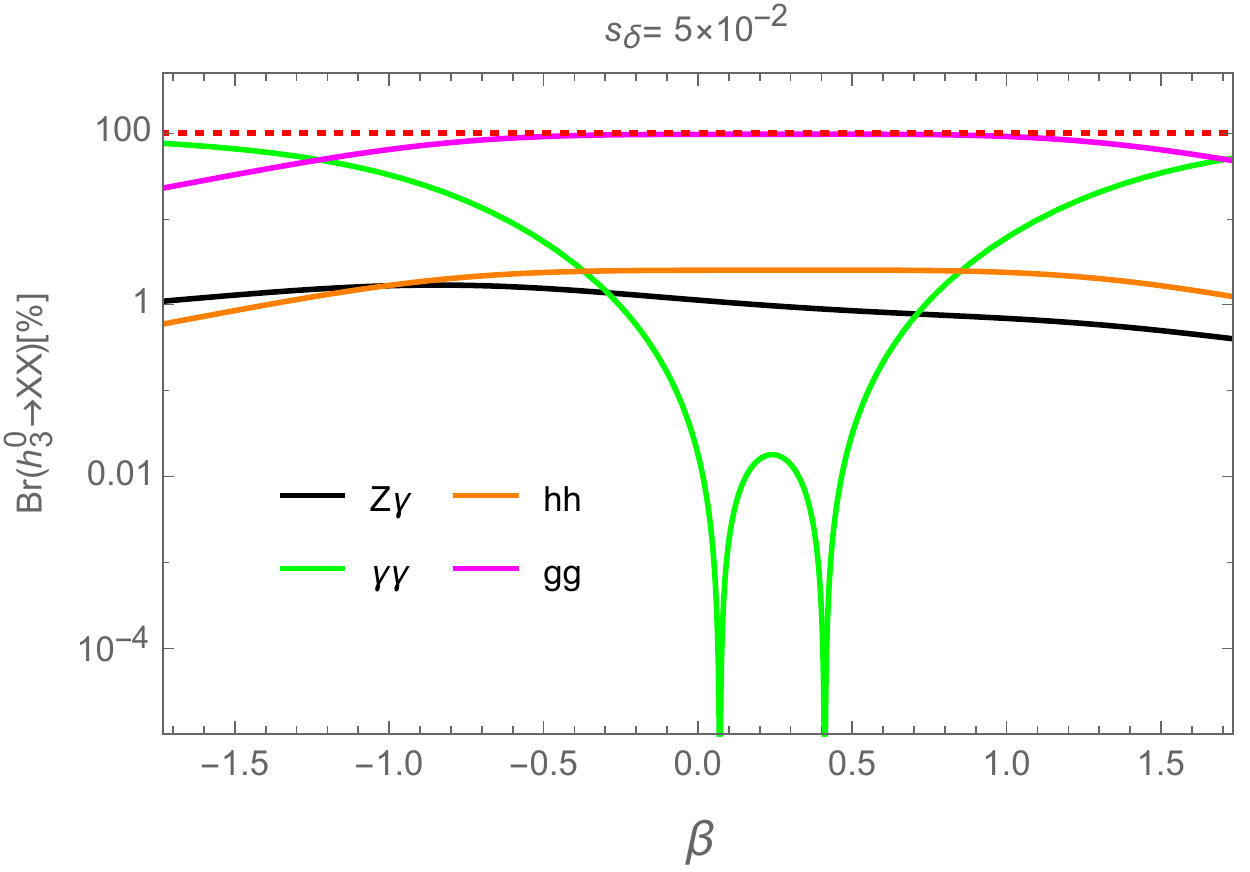} \\
	\end{tabular}%
	\caption{ Branching ratios of  the $h^0_3$ decays as functions of $\beta$.}
	\label{fig_brh03}
\end{figure}

Individual branching ratios of $h^0_3$ are shown in Fig.~\ref{fig_brh03}.
The most interesting property is that, the Br$(h^0_3\rightarrow\gamma\gamma)$ may have large values and it is very sensitive with the change of $\beta$. Hence this decay is a promising channel to fix the $\beta$ value once $h^0_3$ exists. On the other hand, Br$(h^0_3\rightarrow\,hh)$ is sensitive with $s_{\delta}$: it increases significantly with large $s_{\delta}$, but the values is always small Br$(h^0_3\rightarrow\,hh)<1\%$. 

\begin{figure}[ht]
	\centering
	\begin{tabular}{cc}
		\includegraphics[width=7.8cm]{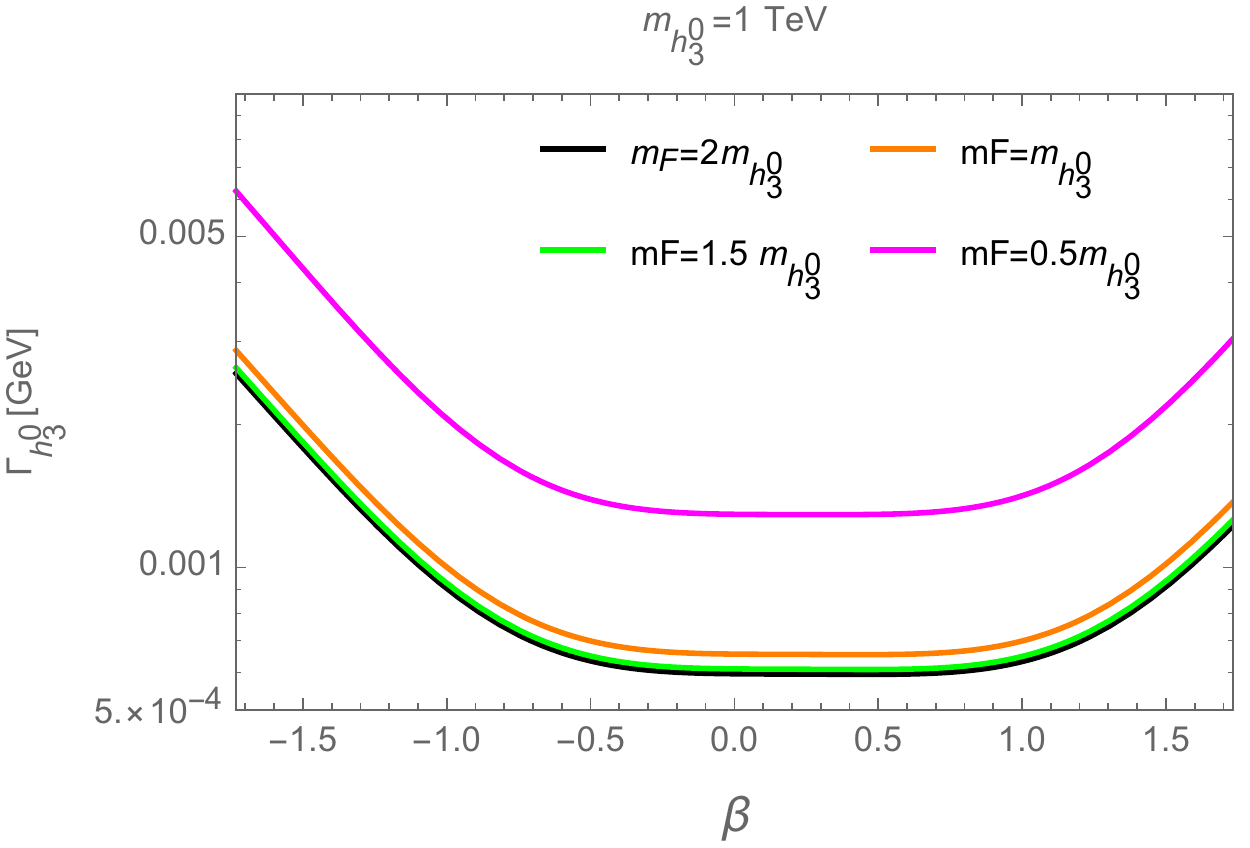} &
		\includegraphics[width=7.8cm]{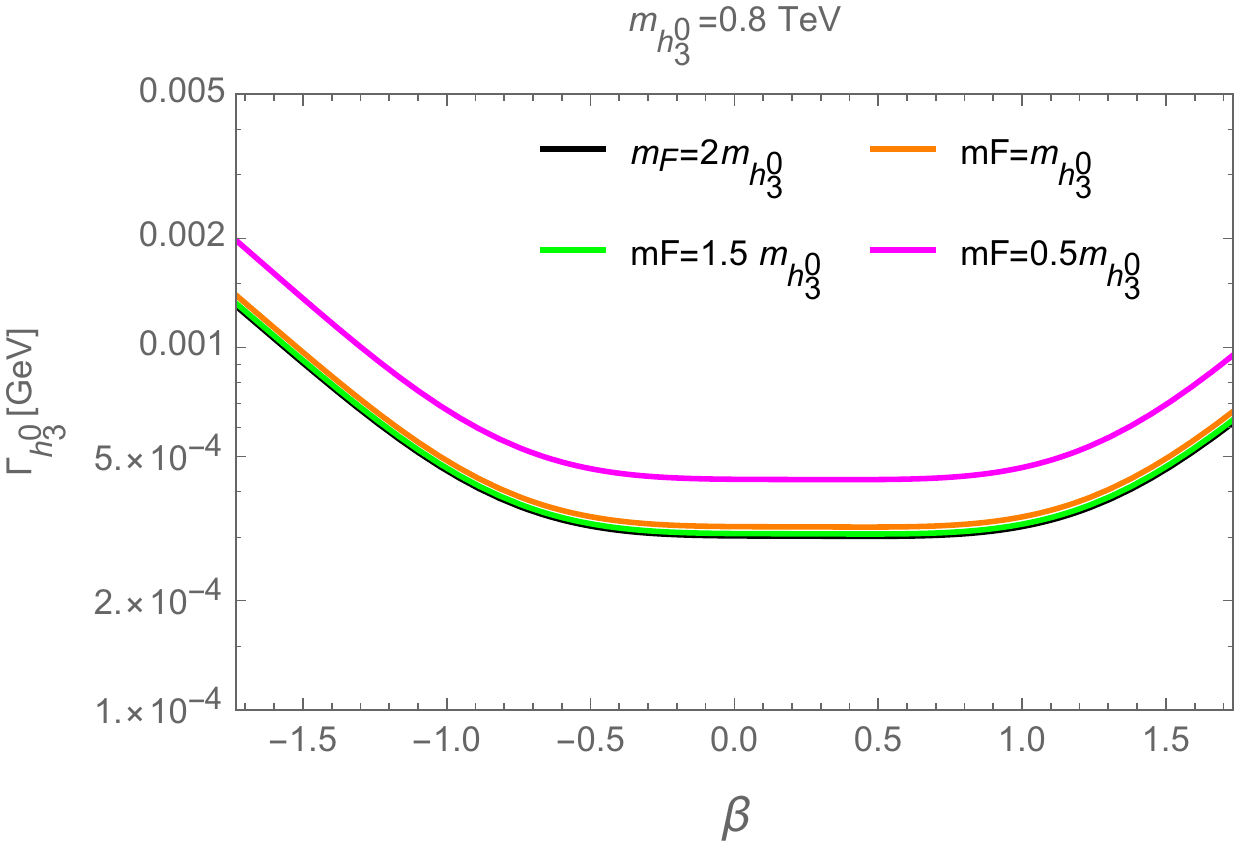} \\
	\end{tabular}%
	\caption{Total decay width of $h^0_3$ as functions of $\beta$, where decays to exotic particle pairs do not included.}
	\label{fig_Gah03}
\end{figure}

For $m_F>m_{h^0_3}$ the total decay width of the $h^0_3$ gets dominant contribution from two gluons decay channel, hence it is sensitive with only $m_{h^0_3}$ and $v_3$, as given in Eq.~\eqref{eq_h03togg}. It is a bit sensitive with $\beta$, see illustrations  in Fig.~\ref{fig_Gah03}.

\section{Conclusions}
\label{conclusion}
The signals of new physics predicted by the 3-3-1 models from the loop-induced neutral Higgs decays $h,h^0_3\rightarrow\gamma \gamma, Z\gamma$ have been discussed. For the general case with arbitrary $\beta$, we have derived that these decays of the SM-like Higgs boson do not depend on the $\beta$, i.e., they cannot be used to distinguish different models corresponding to particular $\beta$ values. This is because of the very large $v_3$ with values around $10$ TeV, leading to the suppressed one-loop contributions from heavy gauge  and charged Higgs bosons, except the $H^{\pm}$, which are also predicted by the 2HDM and are irrelevant with $\beta$. Hence, the large deviations $\delta\mu_{Z\gamma,\gamma\gamma}$  originate from the one-loop contribution of the $H^{\pm}$ and large $|s_{\delta}|$. In the region resulting in large $\delta\mu_{Z\gamma}$, the recent constraint on the $\mu_{\gamma\gamma}$ always gives more strict upper bound on $\mu_{Z\gamma}$ than that obtained from recent experiments. In particular, our numerical investigation predicts  $|\delta\mu_{Z\gamma}|\le |\delta\mu_{\gamma\gamma}|< 0.23$, which is the sensitivity of $\mu_{Z\gamma}$ given in HC-HL project. 

On the other hand, in a model with $\beta=\sqrt{3}$, where  $v_3\simeq 3$ TeV is still valid \cite{Coriano:2018coq},  $\delta\mu_{Z\gamma}$ may be large in the allowed region $\mu_{\gamma\gamma}=0.99\pm 0.14$.  For the near future HC-HL project, where  the experimental sensitivity for the decay $h\rightarrow \gamma\gamma$ may reach $|\delta\mu_{\gamma\gamma}|=0.04$, this model still allows $|\delta\mu_{Z\gamma}|$ to be close to 0.1. But it cannot reach the near future sensitivity $|\delta\mu_{Z\gamma}|=0.23$.  

 Theoretically, we have found two very interesting properties. First,  $F^{331}_{21,sv}$ may have order of $F^{331}_{21,v}$ in allowed regions of the parameter space. This happens also in the 3-3-1 model with $\beta=\sqrt{3}$, where loop contributions from gauge and Higgs bosons may be large and have the same order. Hence, $F^{331}_{21,sv}$ should not be ignored in previous treatments for simplicity ~\cite{Yue:2013qba,Cao:2016uur}.  Second, in the model with $\beta=\sqrt{3}$,  one-loop contributions from gauge  bosons can reach  the order of charged Higgs contributions,  leading to that there appear regions where different contributions to the amplitude $h\rightarrow \gamma\gamma$ are destructive, while they are constructive in contributing to the decay amplitude  $h\rightarrow\,Z\gamma$. This suggests that there  may exist  recent gauge extensions of the SM that allow large $|\delta\mu_{Z\gamma}|$ while still satisfy the future experimental data including $|\delta\mu_{\gamma\gamma}|\leq 0.04$.

Finally, the $h^0_3$ being the $CP$-enven neutral Higgs boson predicted by the $SU(3)_L$ symmetry, not appear in the effective 2HDM. This Higgs boson couples to  only SM-like Higgs through the Higgs self-couplings, while decoulpes to all other SM-like particles . If $h^0_3$ is the lightest among new particles, loop-induced decays $h^0_3\rightarrow \gamma \gamma,Z\gamma,gg$ are still allowed. Our investigation shows that the Br$(h\rightarrow \gamma\gamma)$ is very sensitive with the parameter $\beta$, hence it is a promising channel to distinguish different 3-3-1 models. Because  of the strong Yukawa couplings with new heavy fermions, $h^0_3$ can be produced through the gluon fusion  in the future project HL-LHC.

\section*{Acknowledgments}
This research is funded by the Ministry of Education and Training of Vietnam under grant number: B.2018-SP2-12.

\appendix
\section{\label{app_coupling} Heavy neutral Higgs couplings}
From the Higgs potential and the aligned limit \eqref{eq_alignH0}, the  triple Higgs couplings  containing one heavy neutral Higgs boson $h^0_3$ are listed in table~\ref{table_3H}.  We only mention  the couplings relating with discussion on the decays $h^0_3\rightarrow \gamma\gamma, Z\gamma$.
\begin{table}[ht]
	\centering 
	\begin{tabular}{|c|c|}
		\hline 
		Vertex		& Coupling: $-i\lambda_{S_iS_jS_k}$ \\ 
		\hline 
		$h^0_3 H^+ H^-$	&$-i\left[\left( 1+ s^2_{12}\right) \lambda_{13} + s^2_{12}\lambda_{23} \right]v_3$ \\ 
		\hline 
		$h^0_3 H^{A} H^{-A}$	&  $-i\left[ 2s^2_{13} \lambda_3 +c^2_{13} \lambda_{13}  +\tilde{\lambda}_{13}\right]v_3$\\
		
		\hline 
		$h^0_3 H^{B} H^{-B}$	& $-i\left[ 2s^2_{23} \lambda_3 +c^2_{23} \lambda_{23}  +\tilde{\lambda}_{23}\right]v_3$\\
		\hline 
		\end{tabular}
	\caption{Triple  Higgs couplings of $h^0_3$ involving to the decay $h^0_3\rightarrow \gamma\gamma,Z\gamma$} \label{table_3H}
\end{table}

The nonzero couplings of heavy neutral Higgs bosons with gauge bosons are listed in  table~\ref{table_h023Gcoupling}.  They are derived from the Lagrangian given in~\eqref{eq_lkHiggs},  exactly the same way used to calculate the similar couplings of the SM-like Higgs boson.  Hence,  the notations for the couplings of  these heavy Higgs bosons are the replacements $h \rightarrow h^0_2,h^0_3$ in those given in Eq.~\eqref{eq_lkHiggs}.
\begin{table}[ht]
	\centering 
	\begin{tabular}{|c|c|c|c|}
		\hline 
		Vertex	& Coupling & Vertex&Coupling\\ 
		\hline 
		$g_{h^0_2 W^+W^-}$	&$ g \,m_W \,s_{\delta}$& 	& \\
		\hline 
		$g_{h^0_2Y^{+A}Y^{-A}}$	& $g\,m_W\, s_{12} s_{\alpha}$ & $g_{h^0_3Y^{+A}Y^{-A}}$	& $\frac{g^2v_3}{2}$ \\
		\hline 
		$g_{h^0_2V^{+B}V^{-B}}$&$g\,m_W \,c_{12} c_{\alpha}$   & $g_{h^0_3V^{+B}V^{-B}}$& $\frac{g^2v_3}{2}$ \\
		\hline 
		$g_{h^0_2 HW} $ & $\frac{g\, c_{\delta}}{2}$ & $g_{h^0_3 HW} $ &  0\\
		\hline 
		$g_{h^0_2 H^{-A}Y^{A}}$ & $-\frac{g\, c_{13} s_{\alpha}}{2}$  & $g_{h^0_3 H^{-A}Y^{A}}$ &  $\frac{g\, s_{13}}{2}$\\
		\hline 	
		$g_{h^0_2 H^{-B}Y^{B}}$ & $-\frac{g\, c_{23} c_{\alpha}}{2}$ & $g_{h^0_3 H^{-B}Y^{B}}$ &  $\frac{g\, s_{23}}{2}$ \\
		\hline
	\end{tabular}
	\caption{Heavy neutral Higgs boson couplings   to charged Higgs and  gauge bosons.  }\label{table_h023Gcoupling}
\end{table}
\begin{table}[ht]
	\centering 
	\begin{tabular}{|c|c|}
		\hline 
		Vertex & coupling $g_{h^0_iZZ}$ \\
		\hline 
		$hZZ$&$\frac{g\,m_W}{c^2_W}\left[ c_{\delta}\left(1+ \frac{2\sqrt{3}s_{\theta} c_{\theta}c_W\left(1 -2 s^2_{12} -\sqrt{3}t^2_W\beta\right)}{3\sqrt{1-\beta^2t^2_W}}\right) - \frac{4s_{\delta}c_Ws_{\theta} c_{\theta} s_{12}c_{12} }{\sqrt{3(1-\beta^2t^2_W)}} \right]$\\
		\hline 
		$h^0_2ZZ$& $\frac{g\,m_W}{c^2_W}\left[ s_{\delta}\left(1+ \frac{2\sqrt{3}s_{\theta} c_{\theta}c_W\left(1 -2 s^2_{12} -\sqrt{3}t^2_W\beta\right)}{3\sqrt{1-\beta^2t^2_W}}\right) + \frac{4c_{\delta}c_Ws_{\theta} c_{\theta} s_{12}c_{12} }{\sqrt{3(1-\beta^2t^2_W)}} \right]$\\
		\hline 
	\end{tabular}
	\caption{$h^0_iZZ$ couplings in the limit $s^2_{\theta}=0,\,c^2_{\theta}=1$. }
\end{table}

The couplings of $Z$ to two  exotic fermions are given in table~\ref{table_Z1FF}.
\begin{table}[ht]
	\centering
	\begin{tabular}{|c|c|c|}
		\hline
		$F$ & $g^F_L$&  $g^F_R$\\
		\hline 
		$E_a$ &  $ g^{E_a}_R-\frac{t_{\theta} c_W}{\sqrt{3(1-\beta^2t^2_W)}}$ & $-\frac{(-1 +\sqrt{3}\beta) s^2_W}{2} \left( 1- \frac{t_{\theta}\beta }{c_W \sqrt{1-\beta^2t^2_W}}\right) $  \\
		\hline 
		$J_i$ &  $ g^{J_i}_R +\frac{t_{\theta} c_W}{\sqrt{3(1-\beta^2t^2_W)}}$ & $\frac{(-1 +3\sqrt{3}\beta) s^2_W}{6}\left( 1- \frac{t_{\theta}\beta }{c_W\sqrt{1-\beta^2t^2_W}}\right) $  \\
		\hline 
			$J_3$ &  $ g^{J_3}_R -\frac{t_{\theta} c_W}{\sqrt{3(1-\beta^2t^2_W)}}$ & $-\frac{(1 +3\sqrt{3}\beta) s^2_W}{6}\left( 1- \frac{t_{\theta}\beta }{c_W\sqrt{1-\beta^2t^2_W}}\right) $  \\
		\hline 
	\end{tabular}
	\caption{ Couplings of $Z$ with exotic fermions}\label{table_Z1FF}
\end{table}
\section{\label{app_hzga1loop} Form factors to one-loop amplitudes of the neutral Higgs decays $h,h^0_3\rightarrow Z\gamma,\gamma\gamma$}
\label{appen_loops}

In the  $331\beta$ model, the explicit analytic formulas of one-loop contributions to the anplitudes of the decay $h\rightarrow \gamma\gamma,Z\gamma$ will be presented in terms of the Passarino-Veltmann (PV) functions~\cite{Passarino:1978jh}, namely the one-loop three point PV functions denoted as  $C_{i}$ and $C_{ij}$ with $i,j=0,1,2$. The particular forms for one-loop contributions to the decay amplitudes  $h\rightarrow Z\gamma,\gamma\gamma$ were given in Ref.~\cite{Hue:2017cph}, which are consistent with the previous formulas~\cite{Degrande:2017naf}. We have used the LoopTools~\cite{Hahn:1998yk} to evaluate numerical results.

For the loop-induced decays of the heavy neutral Higgs bosons  $h^0_3$,  the calculation is  the same way as those for the SM-like Higgs boson $h$. Correspondingly, the mass and couplings of $h$ are replaced with those relating with $h^0_3$.  The $h^0_2$ properties were discussed in ref.~\cite{Okada:2016whh}, we do not repeat again.

The contributions from the SM fermions corresponding to the diagram 1 in Fig.~\ref{fig_hzgaDiagram} are 
\begin{align}\label{eq_F331f21}
F^{331}_{21,f}&=- \frac{e\,Q_f\,N_c}{4\pi^2}\left[ m_f Y_{h\bar{f}fL}\frac{gc_{\theta}}{c_W} \left(\, g^f_{L}  + g^f_{R}\right)\right]  \left[4  \left(C_{12} +C_{22} +C_2\right) +C_0\right], 
\end{align} 
where $C_{0,i,ij} \equiv C_{0,i,ij}(m_Z^2, 0,m_h^2; m_f^2,m_f^2,m_f^2)$; $Q_f$ $N_c$ and $m_f$ are respectively  the electric charge, color factor and mass of the SM fermions. The factors $Y_{h\bar{f}fL}$ and $g^f_{L,R}$ are listed in tables~\ref{table_h0coupling} and \ref{table_Z1ff}, respectively. 

The contributions from the charged Higgs  bosons  $s=H^{\pm}, H^{\pm A}, H^{\pm B}$ corresponding to the diagram 2  in Fig.~\ref{fig_hzgaDiagram} are 
\begin{align}
F^{331}_{21,s}= \frac{e\,Q_s \lambda_{hss}g_{Zss} }{2\pi^2} \left[C_{12}+C_{22} +C_2\right],\label{F21sa}
\end{align} 
where $s=H^{\pm},H^{\pm A}, H^{\pm B}$,  $C_{0,i,ij} \equiv C_{0,i,ij}(m_Z^2, 0,m_h^2; m_s^2,m_s^2,m_s^2)$,  and the couplings $\lambda_{hss},g_{Zss}$ are listed in table~\ref{table_h0coupling} and~\ref{table_Z1A}.

The contributions from the diagrams containing both charged Higgs and gauge bosons  $\{v,s\}=\{ W^{\pm},H^{\pm}\}, \{Y^{\pm A},H^{\pm A}\} , \{V^{\pm B},H^{\pm B}\}$ corresponding to the two diagrams 3  and 4 in Fig.~\ref{fig_hzgaDiagram} are 
\begin{align}
F^{331}_{21,vss}&= \frac{e\,Q_s\, g_{hvs}g_{Zvs} }{4\pi^2} \left[ \left(1+\frac{-m_s^2+m_h^2}{m_v^2}\right)(C_{12}+C_{22} +C_2) +2(C_1+C_2 +C_0)\right],\label{F21VSSa}\\
F^{331}_{21,svv}&= \frac{e\,Q_v\, g_{hvs}g_{Zvs}}{4\pi^2}   \left[ \left(1+\frac{-m_s^2+m_h^2}{m_v^2}\right)(C_{12}+C_{22} +C_2) -2(C_1+C_2)\right], \label{eq_F21SVVa} 
\end{align} 
where $C_{0,i,ij} \equiv C_{0,i,ij}(m_Z^2, 0,m_h^2; m_V^2,m_s^2,m_s^2)$  or $ C_{0,i,ij}(m_Z^2, 0,m_h^2; m_s^2,m_V^2,m_V^2)$ corresponding to Eqs.~\eqref{F21VSSa} or~\eqref{eq_F21SVVa}.  The vertex factors are listed in table~\ref{table_HGcoupling} and \ref{table_Z1A}.

The contributions from the charged gauge  bosons  $v=W^{\pm}, Y^{\pm A}, V^{\pm B}$ corresponding to the diagram 5  in Fig.~\ref{fig_hzgaDiagram} are 
\begin{align}
F^{331}_{21,v}&= \frac{e\,Q_v\,g_{hvv}\,g_{Zvv}}{8\pi^2} \crn  &\times\left\{ \left[8+ \left( 2+ \frac{m_h^2}{m_v^2}\right)\left( 2- \frac{m_Z^2}{m_v^2}\right)\right] \left(C_{12}+C_{22}+C_{2}\right)   +  2\left( 4 - \frac{m_Z^2}{m_v^2}\right)C_0 \right\},\label{eq_F21Va}
\end{align} 
where $v=W^{\pm}, Y^{\pm A}, V^{\pm B}$, $C_{0,i,ij} \equiv C_{0,i,ij}(m_Z^2, 0,m_h^2; m_v^2,m_v^2,m_v^2)$.  The vertex factors are listed in table~\ref{table_HGcoupling} and \ref{table_3gaugcoupling}. 

For the decay $h\rightarrow \gamma \gamma$, analytic formulas of $F^{331}_{\gamma}$  can be derived from the $F^{331}_{21}$ by taking replacements $g_{Zvv}, g_{Zss}, \frac{gc_{\theta}}{c_W}  g^f_{L,R}  \rightarrow  eQ_v ,  eQ_s, eQ_f$ and the respective PV functions, name ly: 
\begin{align}
\label{eq_F331gaga}
F^{331}_{\gamma \gamma,f}&=- \frac{e^2\,Q^2_f\,N_c}{2\pi^2}\left( m_f Y_{h\bar{f}fL}\right)  \left[4  \left(C_{12} +C_{22} +C_2\right) +C_0\right], \crn 
F^{331}_{\gamma \gamma,s}&= \frac{e^2\,Q^2_s \lambda_{hss}}{2\pi^2} \left[C_{12}+C_{22} +C_2\right], \crn
F^{331}_{\gamma \gamma,v}&=\frac{e^2\,Q^2_V\,g_{hvv}}{4\pi^2} \times\left\{ \left( 6+ \frac{m_h^2}{m_V^2}\right) \left(C_{12}+C_{22}+C_{2}\right)   +  4C_0 \right\},
\end{align}
where $C_{0,i,ij}\equiv C_{0,i,ij}(0,0,m_h^2; m_x^2, m_x^2, m_x^2)$ with $x=f,s,v$ corresponding to the contribution from fermions, charged Higgs and gauge bosons. 

Regarding to $h^0_3$, we emphasize again that the only nonzero coupling with SM particle is the triple couplings with two SM-like Higgs bosons. Hence the fermion contributions to the decay amplitudes $h^0_3\rightarrow \gamma\gamma, Z\gamma,  gg$ are only exotic fermions $F=E_a,J_a$. These contributions are denoted as  $F^{331,h^0_3}_{\gamma\gamma,F},F^{331,h^0_3}_{21,F},F^{331,h^0_3}_{gg,F}$. They are derived base on Eq.~\eqref{eq_F21331be} with  the following replacement,
\begin{align}\label{eq_F12h03}
F^{331}_{21,F}(h^0_3\rightarrow Z\gamma)&=F^{331}_{21,f}(f\rightarrow F, h\rightarrow h^0_3), \crn 
F^{331}_{\gamma\gamma,F}(h^0_3\rightarrow Z\gamma)&=F^{331}_{\gamma\gamma,f}(f\rightarrow F, h\rightarrow h^0_3).
\end{align}
The other contributions to the mentioned $h^0_3$ decays are calculated  by simple replacements the mass and couplings of the SM-like Higgs bosons with those of the $h^0_3$. We note that the $W$ bosons are not included in these amplitudes.
\section{\label{app_numerical}More numerical illustrations discussed in  section \ref{numerical}}
\begin{figure}[ht]
	\centering
	\begin{tabular}{cc}
		\includegraphics[width=7.5cm]{fconla15} &
		\includegraphics[width=7.5cm]{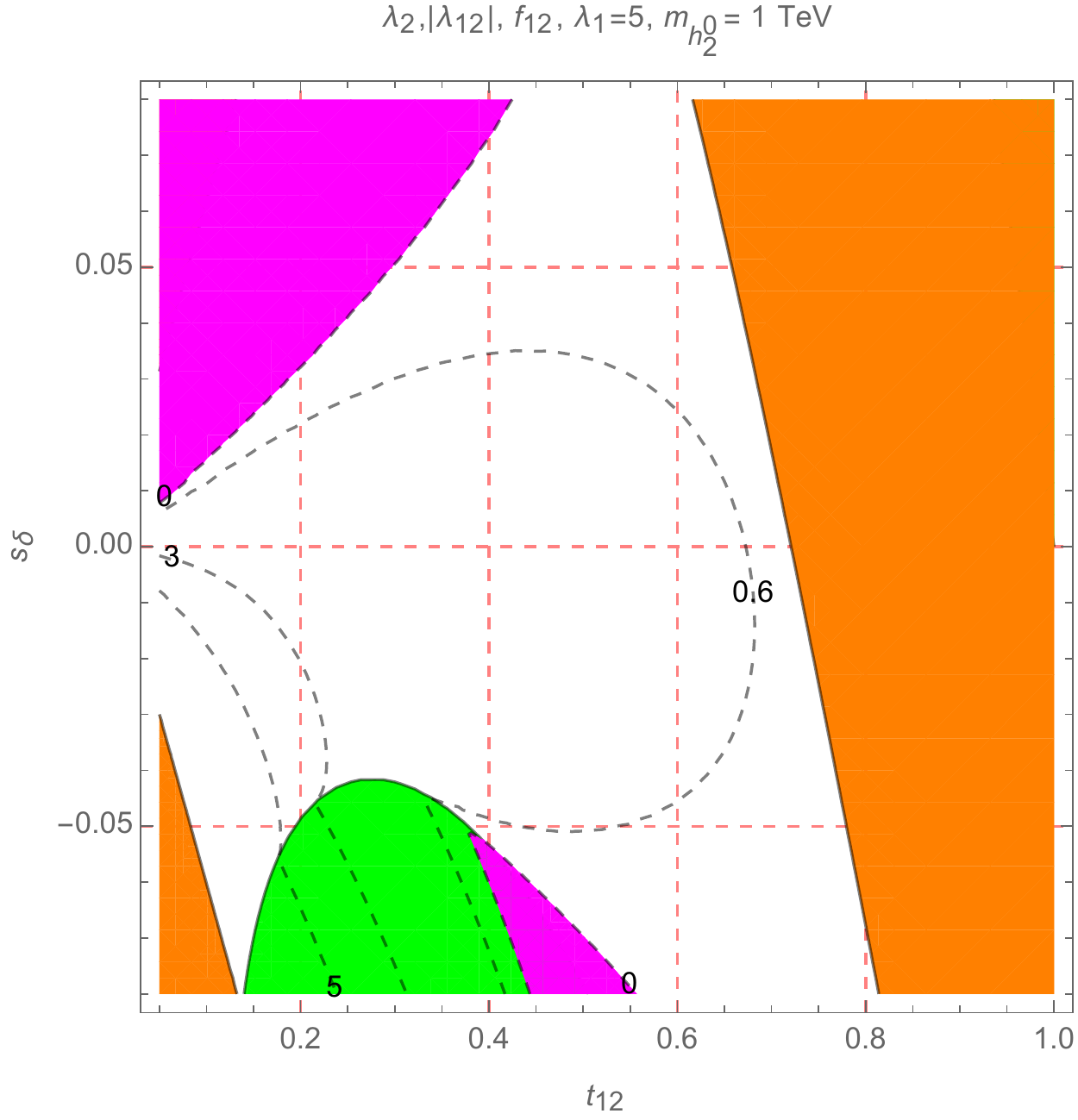} \\
	\end{tabular}%
	\caption{ Contour plots of $\lambda_{2}$,  $|\lambda_{12}|$ and $f_{12}$ as functions of $s_{\delta}$ and $t_{12}$. The green, orange, magenta regions are excluded by requirements that $0<\lambda_2<10$, $|\lambda_{12}|<10$, and $f_{12}>0$, respectively}
	\label{fig_la1contour1}
\end{figure}

\begin{figure}[ht]
	\centering
	\begin{tabular}{cc}
		\includegraphics[width=7.8cm]{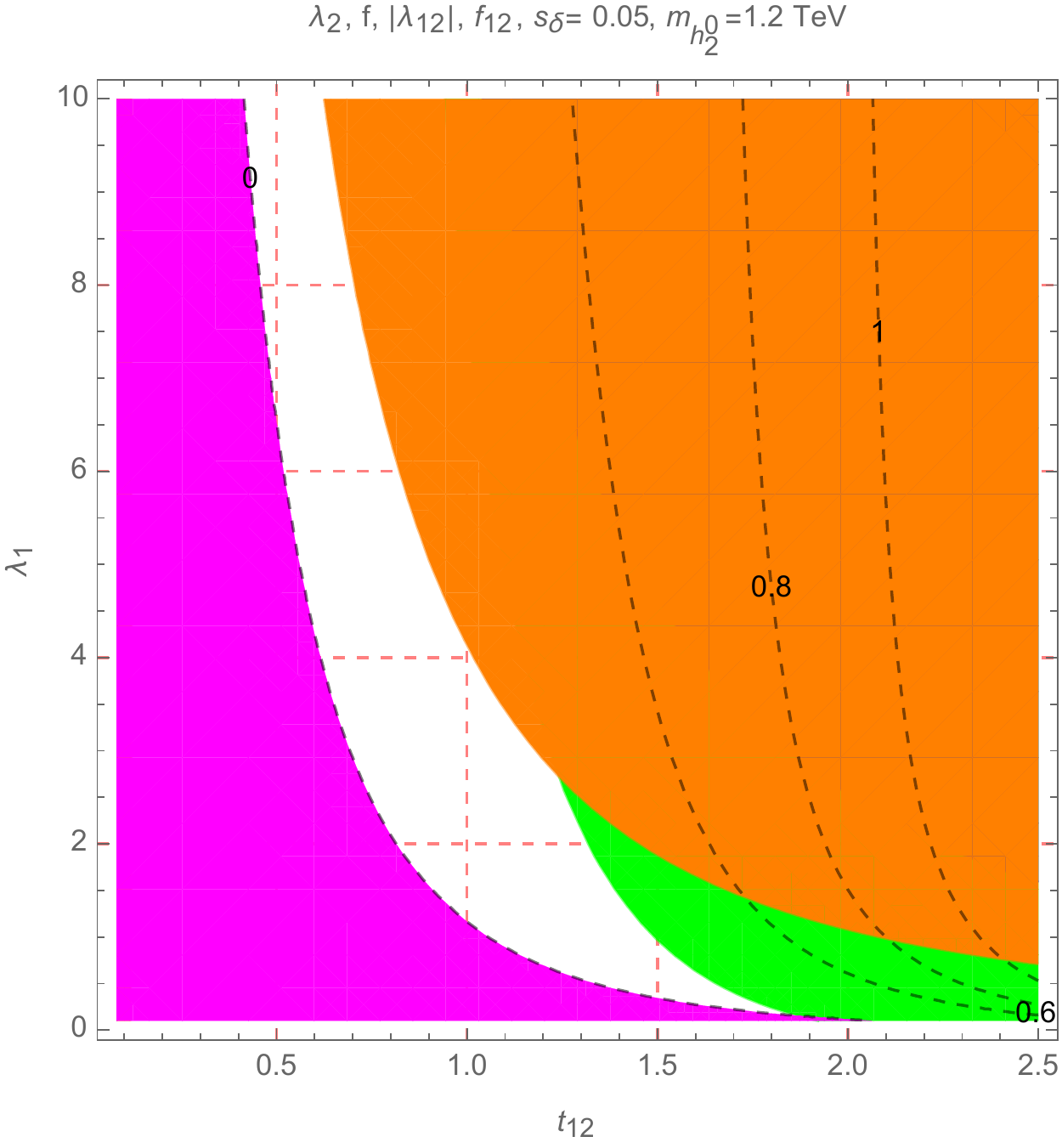} &
		\includegraphics[width=7.8cm]{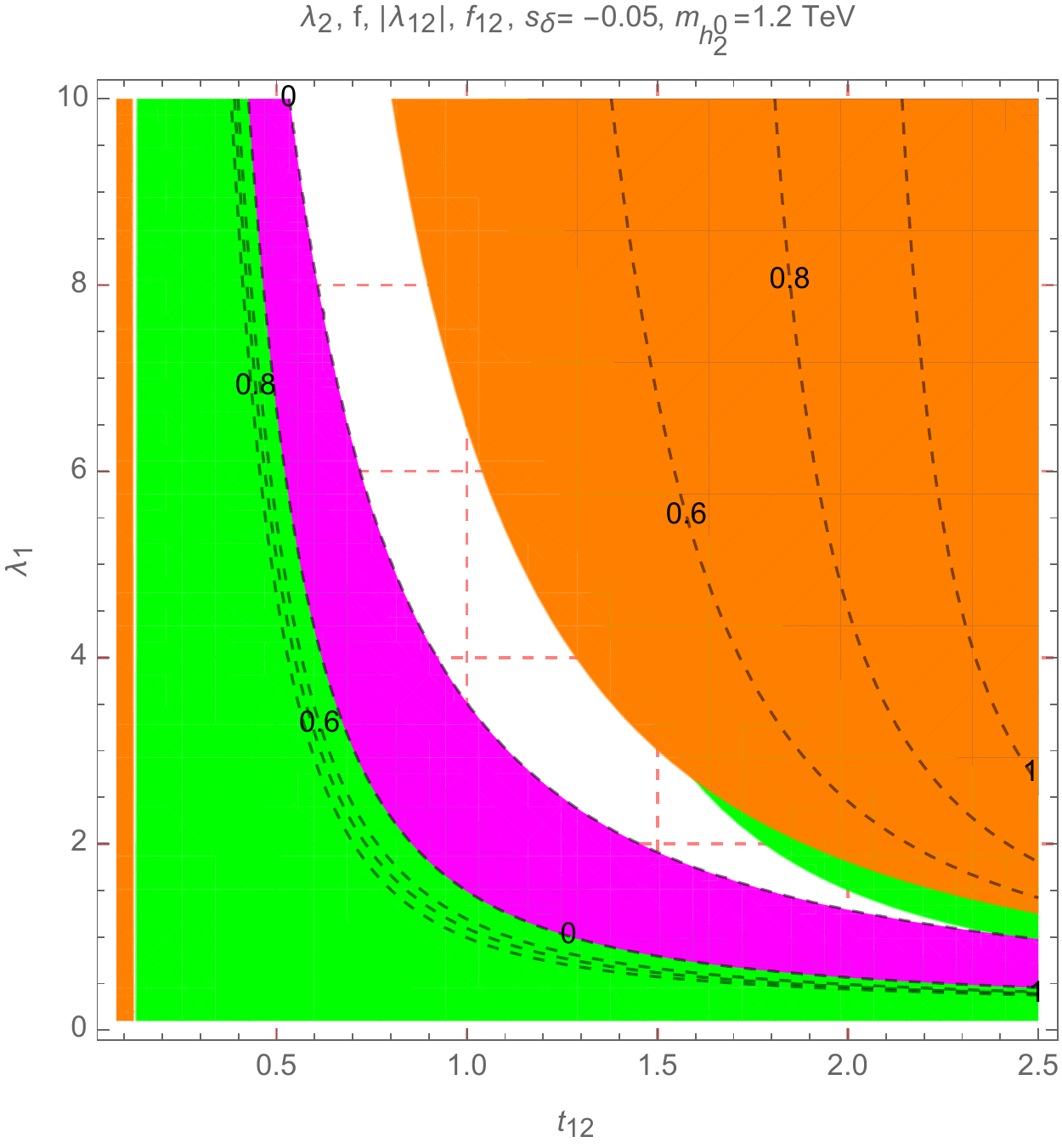} \\
	\end{tabular}%
	\caption{ Contour plots of $\lambda_{2}$,  $|\lambda_{12}|$ and $f_{12}$ as functions of $\lambda_1$ and $t_{12}$ with some fixed $m_{h^0_2}$ . The green, orange, magenta regions are excluded by requirements that $0<\lambda_2<10$, $|\lambda_{12}|<10$, and $f_{12}>0$, respectively}
	\label{fig_contourft12la1p}
\end{figure}

Contour plots with other numerical values of $\lambda_1$ (Fig.~\ref{fig_la1contour1}) and $|s_{\theta}=0.05|$ (figure~\ref{fig_contourft12la1p}).


\end{document}